\documentclass[11pt,a4paper]{article}
\usepackage{graphicx}
\usepackage{jheppub}
\usepackage{tikz-feynman}
\usepackage{caption} 
\usepackage{subcaption} 
\usepackage{cleveref}

  \usepackage{amsmath}
  \usepackage{amsfonts}
  \usepackage{amssymb}

  \usepackage{graphicx}
  \usepackage{float}
  \DeclareGraphicsExtensions{.ps}

\title{Quantum tachyonic preheating, revisited}

	\author[a]{Anders Tranberg,}
    \author[a]{Gerhard Ungersb\"ack.}
	\affiliation[a]{Faculty of Science and Technology, University of Stavanger, 4036 Stavanger, Norway}
	\emailAdd{anders.tranberg@uis.no}
	\emailAdd{gerhard.ungersback@uis.no}

\abstract{In certain models of inflation, the postinflationary reheating of the Universe is not primarily due to perturbative decay of the inflaton field into particles, but proceeds through a tachyonic instability. In the process, long-wavelength modes of an unstable field, which is often distinct from the inflaton itself, acquire very large occupation numbers, which are subsequently redistributed into a thermal equilibrium state. We investigate this process numerically through quantum real-time lattice simulations of the Kadanoff-Baym equation, using a $1/N$-NLO truncation of the 2PI-effective action. We identify the early-time maximum occupation number, the "classical" momentum range, the validity of the classical approximation and the effective IR temperature, and study the kinetic equilibration of the system and the equation of state.}

\begin{document}

\maketitle

\section{Introduction}
\label{sec:Intro}

At the end of inflation \cite{PhysRevD.23.347,Liddle:2000cg}, most of the energy in the Universe resided in the potential energy of an inflaton field. This energy is expected to have been transferred to the Standard Model degrees of freedom through one or a combination of processes collectively referred to as reheating \cite{PhysRevLett.73.3195}. One of these is {\it tachyonic preheating} \cite{PhysRevLett.87.011601}, which occurs if at the end of inflation a field (the inflaton itself, but more commonly another field coupled to it) becomes unstable and undergoes a spinodal transition. In the process, the unstable field acquires very large occupation numbers in the long-wavelength (IR) modes.

The most common realisation of this scenario is a two-field model
\cite{PhysRevD.49.748}, where one field $\sigma$ plays the role of the inflaton and a second field\footnote{The "waterfall" field.} $\phi$ undergoes spontaneous a symmetry breaking transition at zero temperature triggered by the (slow-)rolling of the inflaton. A generic model for this follows from the action
\begin{eqnarray}
S=\int d^4x \left[\frac{1}{2}\partial_\mu\sigma\partial^\mu\sigma -V(\sigma) + \frac{1}{2}\partial_\mu\phi\,\partial^\mu\phi -\frac{1}{2}(g^2\sigma^2-\mu^2)\phi^2-\frac{1}{24}\lambda\phi^4-V_0
\right].
\end{eqnarray}
where $\sigma$ is real-valued and $\phi$ is here written as a real singlet, but could have any number of components. 
The inflaton potential $V(\sigma)$ is not specified but must have the property that the potential minimum is at $\sigma=0$ and allows for slow-roll inflation at values of $\sigma$ larger than $\mu/g$. 

As the inflation rolls towards $\sigma=0$, the effective mass of the second field $\phi$ changes sign
\begin{eqnarray}
m^2(t) = g^2\sigma^2(t)-\mu^2,
\end{eqnarray}
and $\phi$ undergoes a symmetry breaking transition, as the potential acquires new minima at non-zero field value 
\begin{eqnarray}
\phi^2\rightarrow v^2 = 6 \mu^2/\lambda, 
\end{eqnarray}
with a speed determined by the evolution of $\sigma$, 
\begin{eqnarray}
u=\frac{1}{\mu^3}\frac{dm^2(t)}{dt}_{\sigma=\mu/g}=\frac{2g\dot{\sigma}}{\mu^2}.
\end{eqnarray}
We may fix the constant $V_0=3\mu^4/2\lambda$ in such a way that the potential is zero at $\phi=v$, $\sigma=0$.

In hybrid inflation models \cite{PhysRevD.49.748}, the end of inflation is itself triggered by the symmetry breaking transition, as the increase of $\phi^2$ spoils the flatness of the $\sigma$-potential. In other implementations, inflation has ended well before the transition. Either way, the potential initially stored in $V_0$ is transferred to all the degrees of freedom, and the system eventually equilibrates to some reheating temperature $T_{\rm reh}$ near the potential minimum.

The process of (p)reheating is often quite complicated and model-dependent, since in addition to triggering symmetry breaking, the $\sigma$ field itself may also oscillate around $\sigma=0$ so that the sign of $m^2(t)$ flips multiple times. Such oscillations may also lead to resonant preheating \cite{PhysRevLett.73.3195}, where certain momentum modes in resonance with $\sigma$ grow exponentially. A substantial body of analytic and in particular numerical work has been done over the past 25 years, exploring many model of preheating after inflation \cite{Garcia-Bellido:2002fsq, Smit:2002yg, Arrizabalaga:2004iw, Aarts:1997kp, Dux:2022kuk, Garcia:2023eol, Mahbub:2023faw}. A review can be found in \cite{Amin:2014eta}. 

Several aspects of a tachyonic reheating transition deserve detailed scrutiny. One is the transition itself, where the field dynamics may be solved numerically in real-time and the spectrum of fluctuations computed. Because occupation numbers are large, at least for some modes, for some of the time, most numerical treatments rely on the classical approximation to the dynamics \cite{Aarts:1997kp, Aarts:2001yn}. Yet, the initial state of the system is the quantum vacuum, and the unstable modes are seeded by the vacuum fluctuations. Also, at late times classical dynamics is notoriously incomplete, in that the classical equilibrium is badly defined and does not match the quantum equilibrium state. To quantify the validity of the classical approximation and follow the thermalisation process using quantum dynamics seems worthwhile \cite{Arrizabalaga:2004iw}. 

The approach to equilibrium (kinetic and chemical) also has a set of characteristic timescales and an effective equation of state which enters in the cosmological evolution.
An out-of-equilibrium regime with very large occupation numbers may be suitable for other interesting processes to take place. These include the creation of heavy exotic particles, primordial black holes, gravitational waves, topological defects and a baryon asymmetry. When model building in these contexts, it is useful to known how large occupation numbers become, in which momentum range and over what timescales. 

In the simplest analysis, one might assume that the potential energy is instantaneously redistributed onto $g^*$ non-interacting relativistic degrees of freedom, so that the final reheating temperature is simply
\begin{equation}
\label{eq:simplestT}
V_0 = \frac{\pi^2}{30}g^* T_{\rm reh}^4 \rightarrow \frac{T_{\rm reh}}{\mu}= \left(\frac{45}{\pi^2 g^*\lambda}\right)^{1/4}.
\end{equation}
In the Standard Model below $T_{\rm reh}=100$ GeV, one expects\footnote{With the appropriate normalisation of $V_0$ in the SM, $T_{\rm reh}\simeq 42$ GeV.} $g^*\simeq 107$, while for the O(N) model to be discussed below, the expectation is $g^*=N-1$. We will see that at intermediate times, this naive estimate has limitations. 

We may also quantify under what conditions we are allowed to ignore cosmological expansion in our simulations. The longest timescale in the problem is the chemical equilibration time $\tau_{ch}$, and so we can ignore expansion as long as $H\tau_{ch}<1$, where $H$ is the Hubble rate. Since
\begin{eqnarray}
H^2=\frac{V_0}{3M_{\rm pl}^2}\rightarrow \frac{H}{\mu}= \frac{1}{\sqrt{2\lambda}}\frac{\mu}{M_{\rm pl}},
\end{eqnarray}
as long as $M_{\rm pl}/\mu\gg\mu \tau_{\rm ch}$ and the coupling is not too small, expansion is negligible. If not, the evolution equations have to be restated and solved in a Friedmann-Robertson-Walker metric as in \cite{Tranberg:2008ae}. 

In the present work, we will ignore expansion and compute the out-of-equilibrium quantum dynamics of a O(N) scalar field system under a mass quench, solving the Kadanoff-Baym equations truncated at NLO in a 1/N-2PI expansion \cite{Aarts:2001qa,Berges:2000ur,Aarts:2002dj}. Early work considering such a transition include \cite{Guth, PhysRevD.36.2474, Calzetta}, with a substantial body of work based on the Hartree approximation (see for instance \cite{Boyanovsky,Boyanovsky2,Boyanovsky3}), which in the 2PI-language is LO in a coupling expansion. Truncations at LO are very useful for the early stages of the evolution, and are numerically straightforward to implement. However, including only local (in time) self-energies, they are unable to capture the quantum equilibration and ultimate thermalisation of the system (see however \cite{Salle,Salle2} for inhomogeneous systems).

The paper is organised as follows: In section \ref{sec:Model} we set up our simplified model of spinodal symmetry breaking. In section \ref{sec:eoms} we introduce the 2PI formalism, the quantum evolution equations and their classical limit. We present our numerical results in sections \ref{sec:EarlyTimes}, \ref{sec:IntResults} and \ref{sec:finitetQ}, where the first section focuses on classical aspects of tachyonic preheating and the applicability of classical approximations, and the second addresses questions outside the reach of classical approximations, where quantum equations of motions are essential. We briefly consider the early time dynamics in finite time mass quenches in section \ref{sec:finitetQ} and conclude in section \ref{sec:conclusion}. Some technical aspects of the simulations are placed in an appendix \ref{app:kernel}. Some of the present work can be seen as a refinement and continuation of \cite{Arrizabalaga:2004iw}, and shares much of the notation.

\section{Symmetry and symmetry breaking in the O(N) scalar field model}
\label{sec:Model}

We will consider a N-component scalar field with an $O(N)$-symmetric potential, described by the following action (sum over components $a=1,...,N$ implied)
\begin{eqnarray}
S=\int d^4x \left[
\frac{1}{2}\partial_\mu\phi_a\partial^\mu\phi_a-\frac{m^2(t)}{2}\phi_a\phi_a-\frac{\lambda}{24 N}(\phi_a\phi_a)^2.
\right]
\end{eqnarray}
This is a generalisation of the single real field $\phi$ in the preceding section. For $N=4$ we have a model reminiscent of the Standard Model complex Higgs doublet. Keeping $N$ general allows us to mimic an arbitrary number of spinodally unstable degrees of freedom to be reheated at the end of inflation. We will also use $1/N$ as the expansion parameter to organise our 2PI diagram series below. 

When $m^2(t) =-\mu^2$, the potential has minima at $\phi_a\phi_a=6N\mu^2/\lambda\equiv v^2$. A common prescription is to choose a global rotation to fix this in the $\phi_1$-direction, $\phi_a=\delta_{a1}v$. Then the mass spectrum becomes
\begin{eqnarray}
M^2_{ab}= \left(m^2+\frac{\lambda}{6N}\phi_c\phi_c\right)\delta_{ab}+\frac{\lambda}{3N}\phi_a\phi_b=\textrm{diag}(2\mu^2,0,0,...),
\end{eqnarray}
with a massive Higgs mode with mass $m_H^2=2\mu^2$, corresponding to fluctuations in the $\phi_1$ direction around $v$, and $N-1$ massless Goldstone modes, corresponding to fluctuations perpendicular to $\phi_1$. These two modes are often explicitly encoded in a longitudinal (along the 1-direction) and transverse (to the 1-direction) propagator. 

On the other hand, the system has O(N) symmetry, and so the equations of motion will conserve $\langle \phi_a\rangle=0$ for all $a$. We may therefore choose to study the dynamics keeping this symmetry manifest, and consider only one "compound" correlator encoding both massive and massless modes \cite{Arrizabalaga:2004iw}. Close to equilibrium, the compound propagator contains $N-1$ Goldstone modes and one Higgs mode, and at equal times it may be decomposed as\footnote{This is in contrast to the common procedure of defining the connected correlator $G_c=\langle\phi\phi\rangle-\langle\phi\rangle^2$, and from there the transverse and longitudinal components of this correlator. Here, symmetry ensures $\langle\phi\rangle=0$, but the correlator still solves $dV/d\phi=0=(\lambda \langle\phi^2\rangle-\mu^2)\phi$.} 
\begin{align}
    \langle\phi_a({\bf x},t)\phi_b({\bf y},t)\rangle =G({\bf x-y},t)\delta_{ab} = \left[\frac{v^2}{N} + \frac{N-1}{N} F^G({\bf x-y},t) + \frac{1}{N} F^H({\bf x-y},t)\right]\delta_{ab}.
    \label{eq:compdecomp}
\end{align}
It follows that the expectation value $v$ enters as a contribution to the zero momentum mode
\begin{equation}
  \frac{G(k=0)}{L^3}\simeq \frac{v^2}{N}.
  \label{eq:vev}
\end{equation}
We will return to this point below. 

\subsection{Triggering the symmetry breaking transition}
\label{sec:triggering}

The transition is triggered as the function $m^2(t)$ changes sign. This may be parametrised in terms of a finite quench time $\tau_Q$, as
\begin{align}
m^2(t) &= \mu_0^2,\quad t<0,\\
&=\mu_0^2\left[1-\left(1+\frac{\mu^2}{\mu_0^2}\right)\frac{t}{\tau_Q}\right], \quad 0<t<\tau_Q,\\
&=-\mu^2,\quad t>\tau_Q.
\label{eq:linquench}
\end{align}
and the corresponding quench speed is given by $u=-\frac{1}{\mu \tau_Q} (1+ \frac{\mu_0^2}{\mu^2})$. 
An instantaneous quench amounts to $\tau_Q=0$, cutting out the transition region altogether
\begin{align}
m^2(t)&= \mu_0^2, \quad t<0,\\
m^2(t)&=-\mu^2, \quad t>0.
\end{align}
In the general case of a dynamical inflaton $m^2(t)=g^2\sigma^2-\mu^2$, $\mu_0^2$ corresponds to some initial value of $\sigma$. For simplicity, we will consider "symmetric" quenches where $\mu^2=\mu_0^2$. In the example of an instantaneous quench, modes with momentum $k<\mu$ evolve as
\begin{eqnarray}
\phi_k(t)\propto e^{\pm i\omega_kt}\rightarrow e^{\pm \sqrt{\mu^2-k^2}t},
\end{eqnarray}
which is the tachyonic (or spinodal) instability leading to exponential growth of the long-wavelength modes. Performing a more detailed computation \cite{Guth,Calzetta,Boyanovsky,Smit:2002yg}, using as initial condition the quantum vacuum prior to the quench ($\pi_k=\partial_t\phi_k$) 
\begin{eqnarray}
\langle \phi_k^\dagger\phi_k\rangle=\frac{1}{2\omega_k^+},\quad \langle \pi_k^\dagger\pi_k\rangle= \frac{\omega_k^+}{2},
\end{eqnarray}
one finds
\begin{align}
\label{eq:linQ1}
\langle \phi_k^\dagger(t)\phi_k(t')\rangle_{t=t'}=F_k(t,t^{\prime}) |_{t=t^{\prime}} =
\frac{1}{2 \omega^+_k} 
\Biggl[  1 - \biggl( \frac{\omega_k^{+ \, 2}}{\omega_k^{- \, 2}} - 1 \biggr) \sinh^2 (\omega_k^{-} t) \Biggr], \\
\langle \pi_k^\dagger(t)\pi_k(t')\rangle_{t=t'}=\partial_t \partial_{t^{\prime}} F_k(t,t^{\prime}) |_{t=t^{\prime}} =
\frac{\omega^{- \, 2}_k}{2 \omega^+_k} 
\Biggl[  1 + \biggl( \frac{\omega_k^{+ \, 2}}{\omega_k^{- \, 2}} - 1 \biggr) \cosh^2 (\omega_k^{-} t) \Biggr],
\label{eq:linQ2}
\end{align}
where  $\omega_k^{\pm 2} = \pm \mu^2 + k^2$. For a linear quench, the corresponding expression may be found in \cite{Garcia-Bellido:2002fsq}. The exponential growth continues until the self-interactions become important (see below). 

\section{Quantum evolution equations}
\label{sec:eoms}
\subsection{The 2PI effective action}
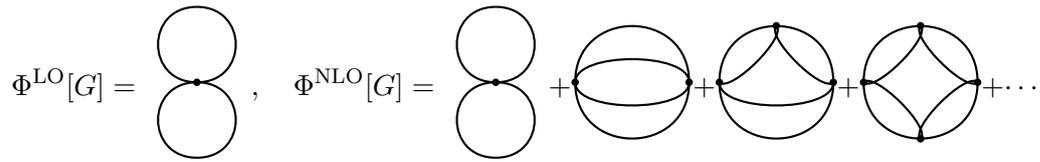
\begin{figure}
\begin{subfigure}[b]{0.2\textwidth}
    \begin{equation*}
        \Phi^{\rm LO}[G] =
        \begin{gathered}
        \begin{tikzpicture}[line width=.8pt]
        \begin{feynman}
                \vertex (v);
                \vertex[above=1.0cm of v](t);
                \vertex[below=1.0cm of v](b);
                \diagram*{
                (v)  -- [out=180,in=180, looseness=1.75] (t) -- [ out=0,in=0, looseness=1.75] (v)
                     -- [out=-180,in=180, looseness=1.75] (b) --[ out=0,in=0, looseness=1.75] (v)
                };
                \draw[fill=black] (v) circle (1pt);
        \end{feynman}
        \end{tikzpicture}
        \end{gathered}
    \end{equation*}
\end{subfigure}
\begin{subfigure}[b]{0.3\textwidth}
    \begin{equation*}
        , \quad \Phi^{\rm NLO}[G] =
        \begin{gathered}
        \begin{tikzpicture}[line width=.8pt]
        \begin{feynman}
                \vertex (v);
                \vertex[above=1.0cm of v](t);
                \vertex[below=1.0cm of v](b);
                \diagram*{
                (v)  -- [ out=180,in=180, looseness=1.75] (t) --[ out=0,in=0, looseness=1.75] (v)
                     -- [ out=-180,in=180, looseness=1.75] (b) --[  out=0,in=0, looseness=1.75] (v)
                };
                \draw[fill=black] (v) circle (1pt);
        \end{feynman}
        \end{tikzpicture}
        \end{gathered}
    +        
        \begin{gathered}
        \begin{tikzpicture}[line width=.8pt]
        \begin{feynman}
            \vertex (v1);
            \vertex[right=1.5cm of v1] (v3);
            \diagram* {
                (v1) -- [ half left, looseness=1.7] (v3) -- [  half left, looseness=1.7] (v1)
                };
            \diagram* {
                (v1) -- [ half left, looseness=0.7] (v3) -- [  half left, looseness=0.7] (v1)
                };
            \draw[fill=black] (v1) circle (1pt);
            \draw[fill=black] (v3) circle (1pt);
        \end{feynman}
        \end{tikzpicture}
        \end{gathered}
    +
        \begin{gathered}
        \begin{tikzpicture}[line width=.8pt]
        \begin{feynman}
            \vertex (v1);
            \vertex[right=1.5cm of v1] (v3);
            \vertex[above right=1.06cm of v1] (v2); 
            \diagram* { 
                (v1) -- [ half left, looseness=1.7] (v3) -- [  half left, looseness=1.7] (v1)
                };
            \diagram* { 
                (v1) -- [ half right, looseness=0.5] (v2) -- [ half right, looseness=0.5] (v3) -- [ half left, looseness=0.7] (v1)
                };
            \draw[fill=black] (v1) circle (1pt);
            \draw[fill=black] (v2) circle (1pt);
            \draw[fill=black] (v3) circle (1pt);
        \end{feynman}
        \end{tikzpicture}
        \end{gathered}
    +
        \begin{gathered}
        \begin{tikzpicture}[line width=.8pt]
        \begin{feynman}
            \vertex (v1);
            \vertex[right=1.5cm of v1] (v3);
            \vertex[above right=1.06cm of v1] (v2); 
            \vertex[below right=1.06cm of v1] (v4); 
            \diagram* { 
                (v1) -- [ half left, looseness=1.7] (v3) -- [ half left, looseness=1.7] (v1)
                };
            \diagram* { 
                (v1) -- [ half right, looseness=0.5] (v2) -- [ half right, looseness=0.5] (v3) -- [ half right, looseness=0.5] (v4) -- [ half right, looseness=0.5] (v1)
                };
            \draw[fill=black] (v1) circle (1pt);
            \draw[fill=black] (v3) circle (1pt);
            \draw[fill=black] (v2) circle (1pt);
            \draw[fill=black] (v4) circle (1pt);
        \end{feynman}
        \end{tikzpicture}
        \end{gathered}
    + \cdots
    \end{equation*}
\end{subfigure}
    \caption{Feynman diagrams corresponding to LO (left) and NLO (right) in an 1/N expansion. In contrast to standard perturbation theory, the diagrams are built from self-consistently dressed propagators, resumming an infinite set of perturbative diagrams.}
    \label{fig:diagrams}
\end{figure}
For a system with large occupation numbers, classical dynamics may be a good approximation to the full quantum evolution. In that case, variation of the action provides classical equations of motion for the field variables. Solving these equations straightforwardly in real time and averaging observables over a suitable ensemble of initial realisations constitutes the classical statistical approximation to quantum field theory (see, for instance \cite{Tranberg:2022noe,Berges:2013lsa,Aarts:2001yn}). By construction, all quantum effects are then discarded.

Simulating quantum dynamics in real time is highly nontrivial. One very successful approach is to derive evolution equations\footnote{Kadanoff-Baym equations or real-time Schwinger-Dyson equations} for the propagator and the mean field from the 2PI effective action \cite{Berges:2000ur}. We define
\begin{eqnarray}
\bar{\phi}_a(t)=\langle\phi_a({\bf x},t)\rangle,\qquad G_{ab}({\bf x-y},t,t')= \langle T_\mathcal{C}\phi_a({\bf x},t)\phi_b({\bf y},t)\rangle,
\end{eqnarray}
where we have explicitly assumed a homogeneous state, but have not yet explicitly imposed that the mean field vanishes. The two-point function is time-ordered along the Keldysh contour $\mathcal{C}$. The 2PI effective action may then be written in the form \cite{Cornwall:1974vz}:
\begin{equation}
    \Gamma[\bar{\phi},G]=S[\bar{\phi}]-\frac{i}{2} \rm Tr \ln{G}+\frac{i}{2} \rm Tr \, G_{0}^{-1}[\bar{\phi}]G+\Phi[\bar{\phi},G],
\end{equation}
where for our model,
\begin{align}
    iG^{-1}_{0,\,ab}(x,y)=
    \frac{\delta^{2}\,S[\bar{\phi}]}{\delta\bar{\phi}_{a}(x)\,\bar{\phi}_{b}(y)}=
    \left(\partial^{2}_{x}\,
    \delta_{ab}+\mu^2\,
    \delta_{ab}-\frac{\lambda}{6N}\,\bar\phi_c\bar\phi_c \,
    \delta_{ab}
    - \frac{\lambda}{3N}\,\bar\phi_a\bar\phi_b
    \right) \delta(x-y).
\end{align}
The functional $\Phi[\bar{\phi},G]$ may be written as an expansion in terms of 2PI skeleton diagrams  with full propagators and mean fields and bare vertices.
The equations of motion are obtained from the stationarity conditions
\begin{align}
\label{eq:varEOM}
    \frac{\delta \Gamma[\bar{\phi},G]}{\delta G_{ab}(x,y)}=0, 
    \quad 
    \text{and}
    \quad 
    \frac{\delta \Gamma[\bar{\phi},G]}{\delta \bar\phi_{a}(x)}=0.
\end{align}
In what follows we consider a system initially in the symmetric phase $\bar\phi_a=0$, for all a.
Because of the O(N) symmetry of the equations of motion, if $\bar\phi_a=0$ at the initial time it will remain zero for all time.
The first equation is formally solved by
\begin{align}
    \frac{\delta \Gamma[\bar{\phi},G]}{\delta G_{ab}(x,y)}=0
    \to
    G^{-1}_{ab}(x,y)=G_{0,ab}^{-1}(x,y)+i\Sigma_{ab}(x,y),
    \label{eq:G1}
\end{align}
in terms of the self-energy
\begin{align}
    \Sigma_{ab}(x,y) =
    -2\frac{\delta \Phi[\bar{\phi},G]}{\delta G_{ab}(x,y)}.
\end{align}
Multiplying (\ref{eq:G1}) by $G_{bc}(y,z)$ turns it into
an integro-differential equation,
\begin{align}
    \delta_{ac}\delta(x,z)=\int d^{4}{\bf y}\,\left[
    G^{-1}_{0,\, ab}(x,y)+i\Sigma_{ab}(x,y)\right]\,G_{bc}(y,z).
    \label{eq:G2}
\end{align}
At leading order (LO) in a $1/N$ expansion we have \cite{Aarts:2002dj}
\begin{align}
\Phi^{\rm LO}[G] &= - \frac{\lambda}{4! N} 
  \int d^{4}x\, G_{aa}(x,x) G_{bb}(x,x), 
\label{eq:PhiLO}
\end{align}
Since $G_{aa} \propto N$, the entire contribution is $\mathcal{O}(N)$ (see left-most diagram in \Cref{fig:diagrams}). 
At NLO we have
\begin{align}
\Phi^{\rm NLO}[G] &= \frac{i}{2} \int d^{4}x\, \ln \textrm{B}(G) (x,x),
\label{eq:PhiNLO}
\end{align}
where 
\begin{align}
    \textrm{B}(x,y; G) = \delta_{\mathcal{C}}(x-y) + i \frac{\lambda}{6N} G_{ab}(x,y) G_{ab}(x,y).
\end{align}
The expression (\ref{eq:PhiNLO}) resums an infinite set of diagrams of $\mathcal{O}(N^0)$ (see \Cref{fig:diagrams}, \cite{Aarts:2002dj}). We note that the figure-8 diagram appears at both LO and NLO, providing local contributions to the self-energy at both orders. We will work mostly at NLO, sometimes comparing to LO.  Including diagrams at NNLO is numerically highly challenging \cite{Aarts:2008wz}, and we will not do so here. 

At this stage, it is useful to further decompose the propagator in terms of the statistical ($F$) and spectral ($\rho$) components as
\begin{align}
G_{ab}(x,y)= F_{ab}(x,y) - \frac{i}{2} \rho_{ab}(x,y) \textrm{sign}_{\mathcal{C}}(x^0-y^0),
\label{eq:Gdecomp}
\end{align}
where
\begin{align}
F_{ab}(x,y) = \frac{1}{2} \langle \{\phi_a(x), \phi_b(y)\} \rangle
,\qquad \rho_{ab}(x,y) = i \langle [\phi_a(x), \phi_b(y)] \rangle,
\end{align}
and the sign-function again refers to ordering along the Keldysh contour in the complex time plane\footnote{This decomposition is different from, but completely equivalent to, the +/- formalism, often stated to be doubling the number of degrees of freedom. In both formalisms, there are two independent propagators components, here $F$ and $\rho$.}.
It follows that $F$($\rho$) is symmetric (antisymmetric) with respect to space-time indices $(x,y)$. In particular, $G_{ab}(x=y)=F_{ab}(x=y)$. An $O(N)$ symmetric state ($ \bar{\phi_a} = 0$) and spatial homogeneity allows to further simplify $F_{ab}(x,y)=\delta_{ab} F({\bf x-y},t,t')$ and $\rho_{ab}(x,y)=\delta_{ab} \rho({\bf x-y},t,t')$. We may similarly decompose the self-energy into two components
\begin{eqnarray}
\Sigma_{ab}(x,y)= \Sigma_{F}\delta_{ab}(x,y) - \frac{i}{2} \Sigma_{\rho}\delta_{ab}(x,y) \textrm{sign}_{\mathcal{C}}(x^0-y^0), 
\label{eq:Sigmadecomp}
\end{eqnarray}
Inserting this decomposition into \cref{eq:G2}, one obtains the real-time equations of motion for $F$ and $\rho$:
\begin{align}
        \Big( \partial_t^2 - \partial_x^2 + M^2(x) \Big) F(x,y) =
        &- \int_0^{x^0} dz^0 \int d^3z \Sigma^{\rho}(x,z) F(z,y) \nonumber \\ 
        &+ \int_0^{y^0} dz^0 \int d^3z \Sigma^{F}(x,z) \rho(z,y),\\
        \Big( \partial_t^2 - \partial_x^2 + M^2(x) \Big) \rho(x,y) =
        &- \int_{y^0}^{x^0} dz^0 \int d^3z \Sigma^{\rho}(x,z) \rho(z,y).
\label{eq:eom}
\end{align}
We note that the evolution equations are explicit, and depend on the entire past history of the evolution through the time integrals over the self-energy and propagator components.
The local parts of the self-energies are accounted for in the effective mass 
\begin{align}
M_{\rm LO}^2(t) = m^2 + \frac{\lambda}{6} F({\bf 0},t,t), \quad M_{\rm NLO}^2(t) = M_{\rm LO}^2(t) + \frac{\lambda }{3 N} F({\bf 0},t,t).
\label{eq:mass}
\end{align}
At NLO the non-local self energies are given by   
\begin{align}
\label{eq:selfenNLO1}
    \Sigma^{\rho}(x,y) &= - \frac{\lambda}{3 N} \Big( F(x,y) I_{\rho}(x,y)     + \rho(x,y) I_{F}(x,y) \Big), \\
    \label{eq:selfenNLO2}
    \Sigma^{F}(x,y)    &= - \frac{\lambda}{3 N} \Big( F(x,y) I_{F}(x,y) - \frac{1}{4} \rho(x,y) I_{\rho}(x,y) \Big),
\end{align}
where the $1/N$ resummation is performed through the objects $I_{F,\rho}$, that in turn satisfy\footnote{We note in passing that not iterating the $I_{F,\rho}$, but simply inserting the local (no time integration) part of the expression into (\ref{eq:selfenNLO1}, \ref{eq:selfenNLO2}) reduces the evolution to NLO in a coupling expansion \cite{Berges:2000ur,Arrizabalaga:2005tf}.}
\begin{align}
\label{eq:Ieom1}
    I_{\rho}(x,y) = &- \frac{\lambda}{3} \int_{y^0}^{x^0} d^3z I_{\rho}(x,z)  F(z,y)\rho(z,y) 
    + \frac{\lambda}{3} F(x,y) \rho(x,y), \\
    \label{eq:Ieom2}
    I_F(x,y) = &- \frac{\lambda}{6}  
    \int_0^{x^0} d^3z I_{\rho}(x,z) \big( F^2(z,y) -\frac{1}{4} \rho^2(z,y) \big)
    + \frac{\lambda}{3} \int_0^{y^0} d^3z I_{F}(x,z) F(z,y) \rho(z,y) \nonumber \\
    &+ \frac{\lambda}{6} \Big( F^2(x,y) -\frac{1}{4} \rho^2(x,y) \Big).
\end{align}

\subsection{Numerical implementation}
\label{sec:Numerics}

The non-linear integro-differential equations of motion may be evaluated numerically, in terms of a discrete set of degrees of freedom. This may be done either through discretising the system on a space-time lattice at the level of the action, or through discretising momentum space and time at the level of the equations of motion. We will here opt for the lattice implementation. 

\subsubsection{Discretization and observables}
\label{sec:disc}

We introduce a cubic lattice of $N_x^3=32^3$ sites with lattice spacing $a\mu=0.7$, and a time discretization $a_t=a dt$, $dt=0.1$. Although in principle, the memory integrals over the non-local self-energies should stretch all the way to the initial time, in practice we keep only the last $n_t$ timesteps for a memory range $t_K=n_t a_t$. Further discussion on this point may be found in Appendix \ref{app:kernel}.
The lattice momentum operator is
\begin{align}
   (a k)_{\textrm{lat}}^2 = \sum_{i=1}^3 \left(2 - 2 \cos  k_i \right), \quad  
    k_i = \frac{2 \pi}{N_x} n_i, \quad n_i \in \Big(-\frac{N_x}{2}+1, \frac{N_x}{2}\Big).
\end{align}
The initial conditions for the spectral propagator follow from the basic commutation relations
\begin{align}
        \rho_k(t,t) = 0, \quad \partial_t \rho_k(t,t^{\prime})|_{t=t^{\prime}} = 1,
\end{align}
while for the statistical propagator we write
\begin{align}
        F_k(t,t^{\prime}) |_{t=t^{\prime}=0} &= \frac{1}{\omega_k} \Big(n_k + \frac{1}{2}\Big), \\ 
        \partial_t F_k(t,t^{\prime}) |_{t=t^{\prime}=0} &= 0 , \\
        \partial_t \partial_{t^{\prime}} F_k(t,t^{\prime}) |_{t=t^{\prime}=0} &= \omega_k \Big(n_k + \frac{1}{2}\Big). 
\end{align}
Choosing $n_k=0$ and $\omega_k = k_{\textrm{lat}}^2 + m^2(t<0)$ amounts to selecting the free-field vacuum state prior to the mass quench. 
Conversely, for all later times, we may extract a particle number and a dispersion relation from \footnote{On the lattice, we further introduce a correction $n_k(t) + \frac{1}{2} \rightarrow c_k(t)(n_k(t) + \frac{1}{2})$, where $c_k(t) = \sqrt{ 1 - \frac{1}{4} dt^2 \omega_k^2(t) }$, to account for time-discretization effects (see, for instance \cite{Salle}).}
\begin{align}
    \omega_k(t) &= \sqrt{ \frac{\partial_t \partial_{t^{\prime}} F_k(t,t^{\prime})|_{t^{\prime}=t}}{F_k(t,t)} } 
   \label{eq:disprelation},  \\
    n_k(t) + \frac{1}{2} &=  \sqrt{ \partial_t \partial_{t^{\prime}} F_k(t,t^{\prime})|_{t^{\prime}=t} F_k(t,t)},
\end{align}
Very far from equilibrium, the physical interpretation of these quantities is not clear, but once the system settles down enough that the field excitations are quasi-particle like, they are good representations of the occupation number and frequency of a given mode (see for instance \cite{Salle}).

Real time evolution derived from the 2PI effective action have the advantage of conserving an energy functional. We monitor the energy and pressure densities during the simulations, from which we compute the equation of state parameter $\omega = \frac{P}{\rho}$. The expressions are given by
\begin{align}
\label{eq:Eden}
    \frac{\rho(t)}{N} = 
    & \frac{1}{2} \int d^3 k \partial_t \partial_{t^{\prime}} F_k(t,t) 
     +\frac{1}{2} \int d^3 k  \Big( k^2 + m^2 + \frac{\lambda}{2} \frac{N+2}{6N} \int \frac{d^3k^{\prime}}{(2\pi)^3} F_{k^{\prime}}(t,t) \Big) F_k(t,t) \nonumber\\
    & +\frac{1}{4} \int_0^t dt^{\prime} \int d^3k \Big(\Sigma^{\rho}_k(t,t^{\prime}) F_k(t^{\prime},t) - \Sigma^{F}_k(t,t^{\prime}) \rho_k(t^{\prime},t) \Big) +V_0,
\end{align}
and
\begin{align}
\label{eq:Pden}
    \frac{P(t)}{N} = 
    & \frac{1}{2} \int d^3k \partial_t \partial_{t^{\prime}} F_k(t,t) 
     -\frac{1}{2} \int d^3k  \Big( k^2 +m^2 + \frac{\lambda}{2} \frac{N+2}{6N} \int \frac{d^3k^{\prime}}{(2\pi)^3} F_{k^{\prime}}(t,t) \Big) F_k(t,t)\nonumber\\
    & -\frac{1}{4} \int_0^t dt^{\prime} \int d^3k \Big(\Sigma^{\rho}_k(t,t^{\prime}) F_k(t^{\prime},t) - \Sigma^{F}_k(t,t^{\prime}) \rho_k(t^{\prime},t) \Big)
    + \frac{1}{3} \int d^3 k \; k^2 F_k(t,t)-V_0.
\end{align}
Initially, the system is in a free-field vacuum state with energy $\rho_0/N=V_0$ and pressure $P_0/N=-V_0$, where $V_0=\frac{3 \mu^4}{2 \lambda}$. Hence the initial equation of state parameter is $\omega=-1$. Energy density and pressure are divergent quantities and subject to renormalisation as described in the following. 

\subsubsection{Renormalisation}
\label{sec:renorm}

The evolution equations are UV divergent, and although the lattice provides a regularization, a substantial cut-off dependence is present, calling for renormalisation. 2PI-truncated evolution equations are renormalisable \cite{Berges:2004hn,Berges:2005hc}, but because of the infinite resummation of diagrams, a precise cancellation of all divergences comes at a substantial numerical cost \cite{Borsanyi}. An approximate, but much simpler renormalisation procedure addresses only the local self-energies and quadratic divergences. In effect, given an input renormalised $a\mu=0.7$, we must perform the simulation with a bare mass parameter given by
\begin{align}
    -(a\mu_0)^2  = -(a\mu)^2 - \delta (a\mu)^2,
\end{align}
where the counterterm on the lattice at NLO is given by:
\begin{align}
    \delta (a\mu)^2 = \lambda \frac{N+2}{6N} \frac{1}{N_x^3} \sum_{k_{\rm lat}} \frac{1}{2 \sqrt{(ak_{\rm lat})^2 + (a\mu)^2}} \simeq \Big|_{a\mu=0.7}  \lambda \frac{N+2}{6N} 0.212.
\end{align}
At LO the prefactor changes $\frac{N+2}{6N} \rightarrow \frac{1}{6} $ to account for only the LO local self-energy \cref{eq:mass}.
This approximate subtraction works well\footnote{In the sense that the measured output mass of the system matches the input mass of the simulation, and that physical quantities do not change appreciably when varying the cut-off.} for small-to-moderate values of the coupling, and not too close to the continuum limit (as discussed for instance in \cite{Arrizabalaga:2004iw}). 

The energy density and pressure density (\ref{eq:Eden}), (\ref{eq:Pden}) are also divergent quantities. After implementing the mass renormalisation, we can remove the remaining divergences by adding an overall constant counterterm (one for energy, one for pressure). As renormalisation condition, we choose for the initial renormalised energy and pressure density to be $+V_0$ and $-V_0$, respectively. 

\subsubsection{The classical approximation}
\label{sec:classicallimit}

The classical limit is intuitively when the occupation numbers are large, $n_k\gg 1$, and indeed one may extend classical perturbation theory\footnote{Omitting certain vertices in the Keldysh field basis at the level of the action.}  to this resummed non-equilibrium framework \cite{Aarts:1997kp,Aarts:2001yn}, in which case the classical limit amounts to
\begin{eqnarray}
F^2\gg \rho^2.
\end{eqnarray}
This requirement is more general than $n_k\gg 1$, but since in the quasi-particle limit, $F\propto n_k$ and $\rho\propto 1$, the two are consistent when both apply.

The 2PI-classical approximation consequently follows from discarding from the evolution equations (\ref{eq:selfenNLO1}, \ref{eq:selfenNLO2}, \ref{eq:Ieom1}, \ref{eq:Ieom2}) all instances of $\rho^2$ (and $\rho I_\rho$) when added to $F^2$ (or $F I_F$).
For the system and truncation considered here, we find
\begin{align}
    \Sigma_{\rm cl}^{F}(x,y)    &= - \frac{\lambda}{3 N} F(x,y) I_{F}(x,y), \\
     I^{\rm cl}_F(x,y) &=  \frac{\lambda}{6} F^2(x,y) \nonumber \\
    &- \frac{\lambda}{6}
    \int_0^{x^0} d^3z I_{\rho}(x,z) F^2(z,y) 
    + \frac{\lambda}{3} \int_0^{y^0} d^3z I_{F}(x,z) F(z,y) \rho(z,y).  
\end{align}
We note that at LO, the classical and the quantum evolution are the same, and so quantum effects only enter at NLO\footnote{This is true for Gaussian approximations, although beware of subtleties such as \cite{Millington:2020vkg}.}. It is also only at NLO that non-local effects such at equilibration and thermalisation enter, which is why the LO truncation (as well as the Hartree approximation) are insufficent for simulating preheating  beyond the initial stages.

\section{Early time dynamics}
\label{sec:EarlyTimes}

\subsection{Linear regime}
\label{sec:Qlinear}

\begin{figure}[ht]
  \includegraphics[width=0.5\textwidth]{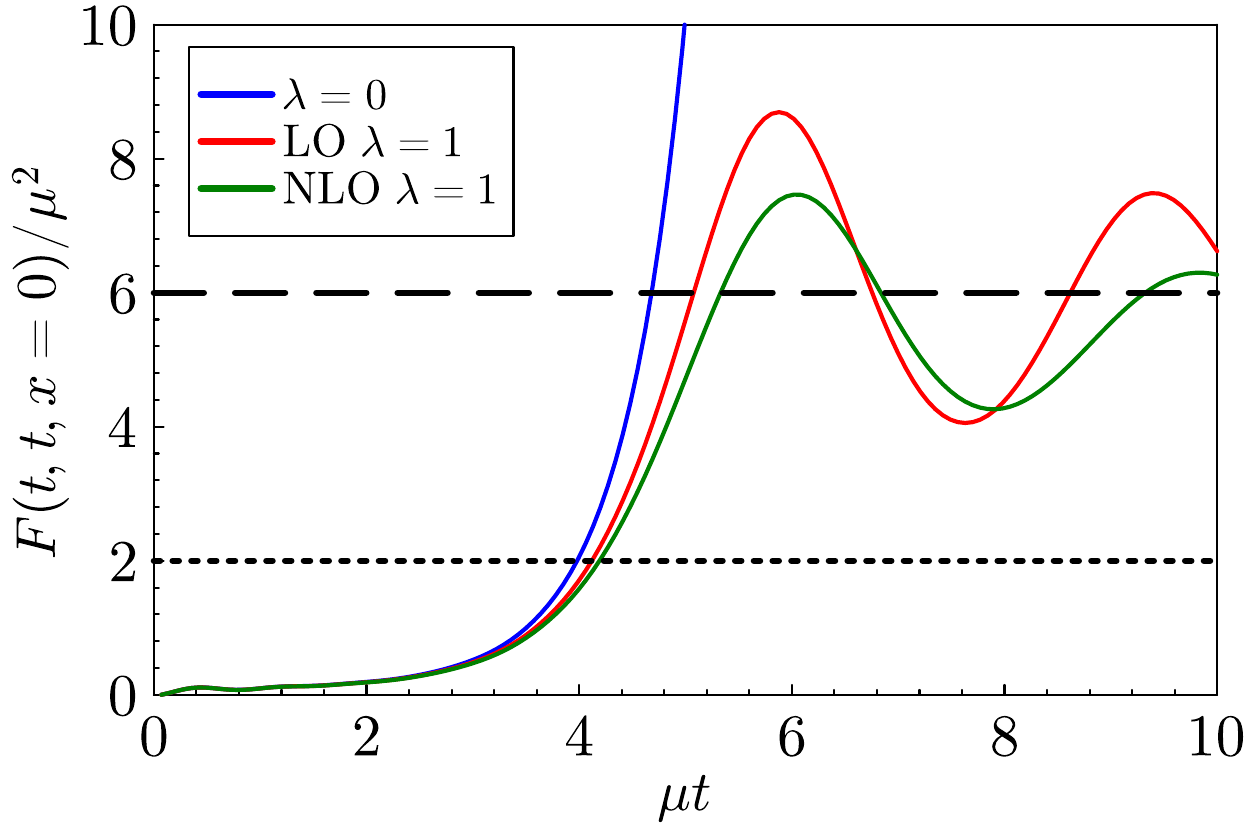}
  \includegraphics[width=0.5\textwidth]{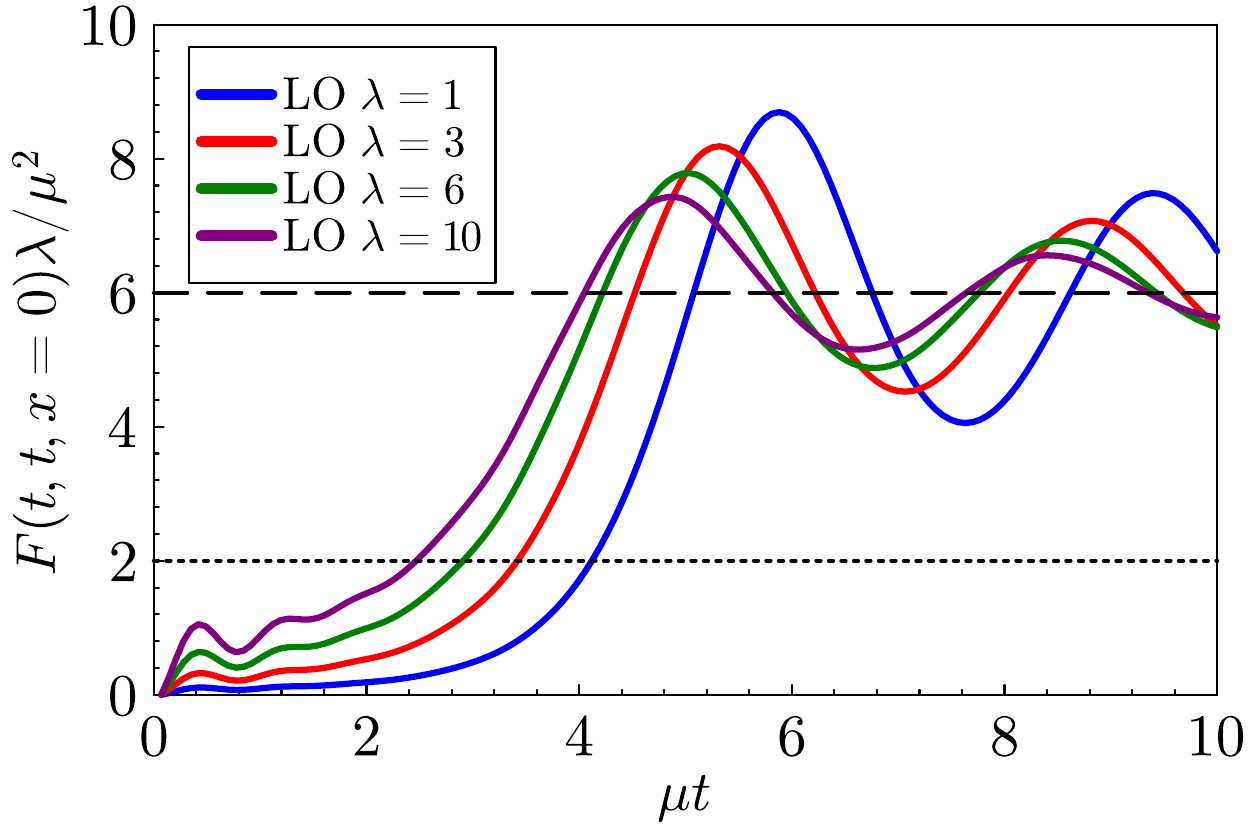}
  \caption{Left: Time evolution of the statistical propagator $F(t,t,x=0)$ in the free theory, LO and NLO approximation for $N=4$ and $\lambda=1$. The dashed (dotted) horizontal line corresponds to the minimum ($\alpha=1$) and inflection point ($\alpha=1/3$) of the potential, respectively. Right: Comparing the free field to the LO evolution for different $\lambda$. Note that the result is independent of the number of fields $N$. Also note the different normalisation of $F$ in the left vs. righthand plot. 
  }
  \label{fig:linear}
\end{figure}
To get us started, we will first consider the very early stages of the spinodal transition. The initial state for the transition is the quantum vacuum prior to the mass quench, and since occupation numbers are not large, the classical approximation would seem to not apply. However, because the quantum and classical evolution of the propagators coincide for free fields and in the LO approximation, at small coupling the classical and quantum evolution will agree for some time.  

When the field self-interaction may be neglected, $\lambda\simeq 0$, 
the evolution equation becomes linear
\begin{align}
        \Big( \partial_t^2 + \omega_k^2(t) \Big) F_k(t,t^{\prime}) &= 0,
        \quad \text{with} \quad \omega_k^2(t) = m^2(t) + k^2 ,
\end{align}
and closed-form solutions exist for the instantaneous (eqs. \ref{eq:linQ1}, \ref{eq:linQ2}, \cite{Smit:2002yg}) and linear quench \cite{Boyanovsky,Garcia-Bellido:2002fsq}. 

For $\lambda=0$, there is no notion yet of $v$ (the bottom of the potential) and also no separation into Higgs and Goldstone modes. The free field approximation to the interacting dynamics is reliable until the non-linear term (order $\lambda$, at LO or NLO) becomes comparable to the mass parameter,
\begin{equation}
    F(x,x)=\alpha\frac{6 \mu^2}{\lambda}.
    \label{eq:lincrit}
\end{equation}
with some number $\alpha$. Often, the criterion is taken to be either $\alpha=1$ (the minimum of potential is reached) or $\alpha=1/3$, (the correlator $F$ passes the inflection point of the potential).
Figure \ref{fig:linear} (left) shows the equal-time statistical propagator in the free field approximation, and with LO and NLO corrections. The dashed line is the minimum of the potential and the dotted line the inflection point. We see that the latter is perhaps a better criterion for the onset of non-linearities, as the LO (and NLO) depart from the pure exponential growth around $\mu t=4$. The non-linear evolution does still overshoot the potential minimum before settling down to an equilibrium state. We will discuss the differences between LO and NLO below.
\begin{figure}[ht]
  \includegraphics[width=0.5\textwidth]{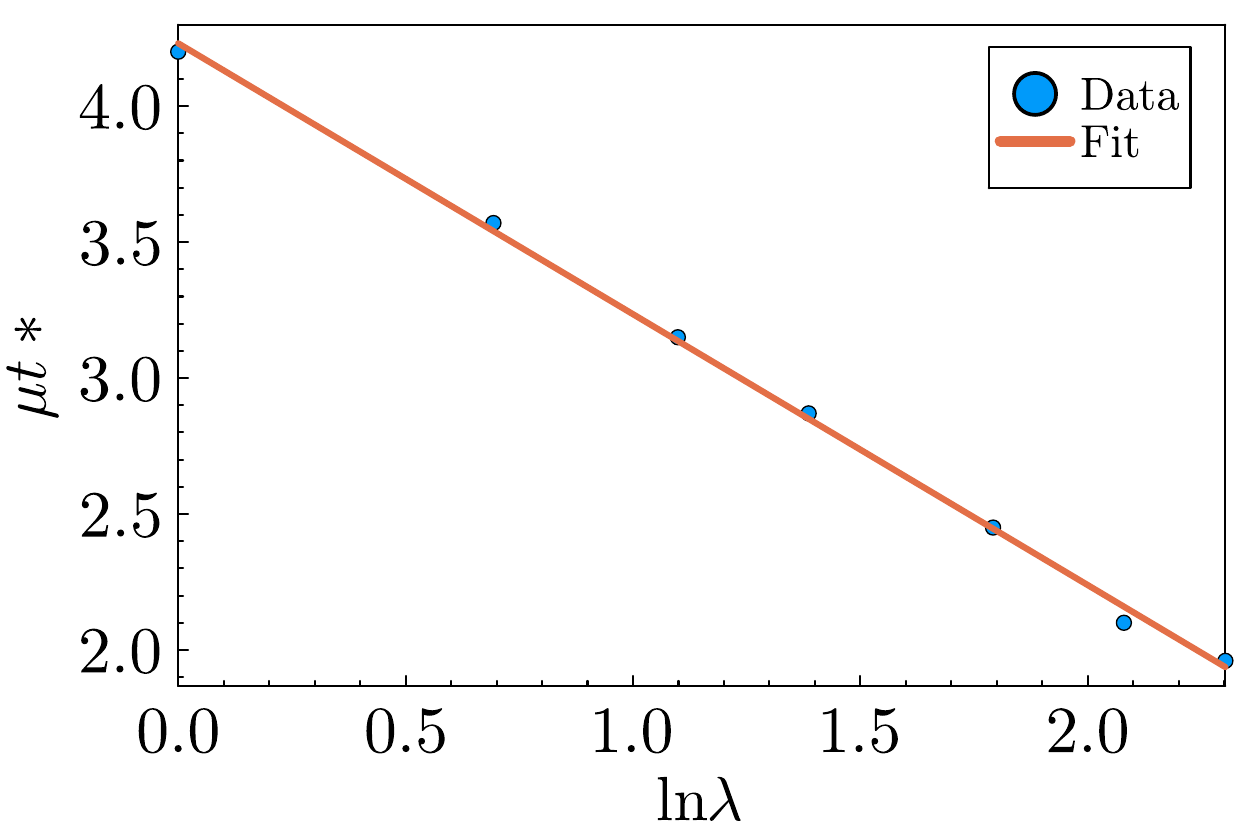}
  \includegraphics[width=0.5\textwidth]{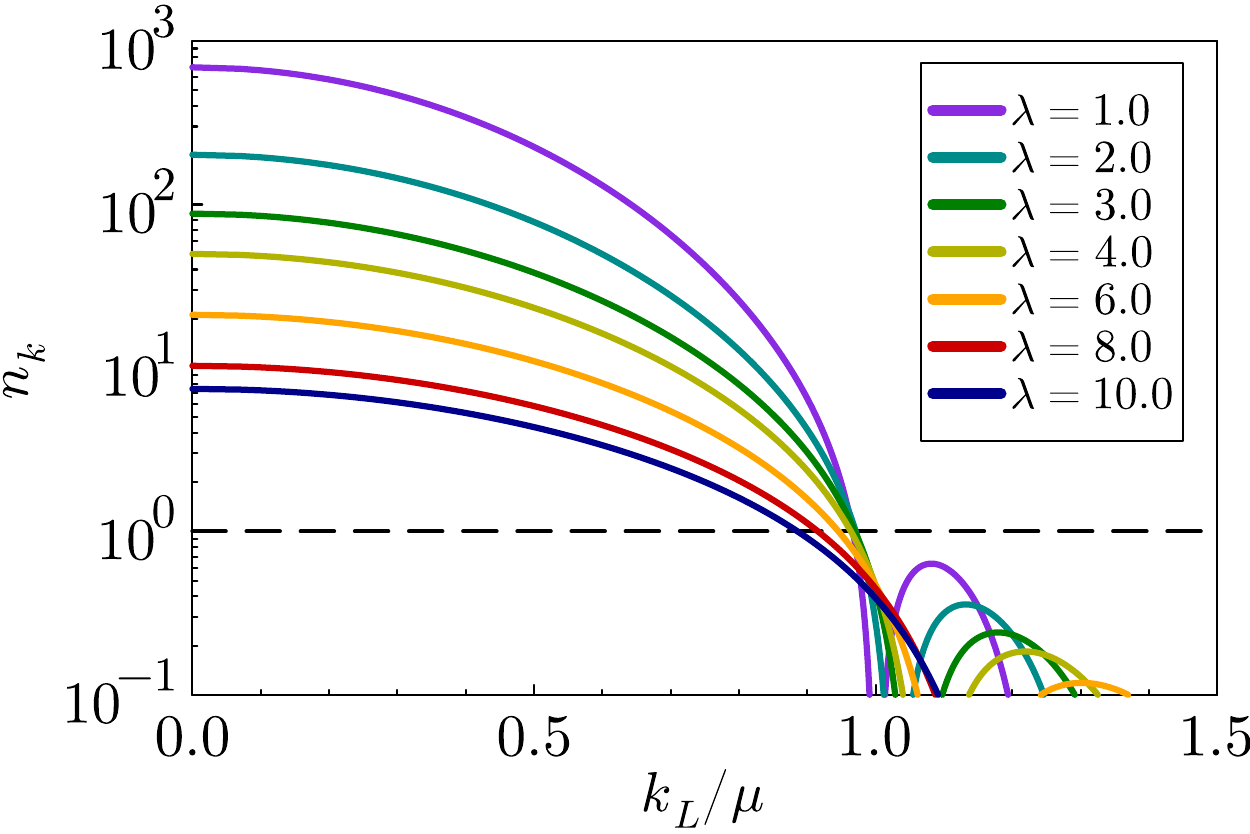}
  \caption{Left: Time at which free theory approximation deviates from the LO result by 20 percent ($\Delta F(t,t,x=0) = 0.2$). Right: Particle spectrum in the LO approximation for $N=4$ at the time when the free field approximation becomes unreliable for different values of $\lambda$.
  Even relatively high values of $\lambda$ produce highly occupied modes at this very early stage. The lattice has the size $N_x^3=128^3$.}
  \label{fig:linearLOtimefit}
\end{figure}

In figure \ref{fig:linear} (right) we see that the inflection point of the potential (equivalently the minimum of the potential) is reached earlier with higher $\lambda$ in the LO approximation, thus limiting the applicability for the free field approximation to earlier times. We can make this statement more precise by defining a time at which the equal time statistical propagator in the free field approximation differs from the LO approximation by 20 percent:
\begin{equation}
    \Delta F(t) = \Big| \frac{F_{free} - F_{\lambda}}{F_{free}} \Big|(t) = 0.2 .
\end{equation}
Figure \ref{fig:linearLOtimefit} (left) shows the relation between this time and $\lambda$. Based on a simple fit, we conclude that free theory is a good approximation until a time
\begin{equation}
    \mu t^* = a - b \ln \lambda \quad \text{where} \quad a = 4.13 \pm 0.20, \quad b = 1.00 \pm 0.03.
\end{equation}
The quoted error of the mean corresponds to varying the tolerance over an interval $\Delta F \in (0.1,0.3)$.
We can also examine the particle spectrum at the time when the free theory approximation breaks down. Figure \ref{fig:linearLOtimefit} (right) shows the occupation numbers in the IR for various values of $\lambda$. Although the free field regime only lasts for a time $\mu t \simeq 2 - 4$ we see that the unstable modes are already highly populated $n_k \gg 1$, even for large values of $\lambda$. At least for these modes, as we leave the early time, small-coupling regime, the classical approximation should be robust.

\subsection{The classical regime}
\label{sec:Qclassical}

\begin{figure}[ht]
  \includegraphics[width=0.5\textwidth]{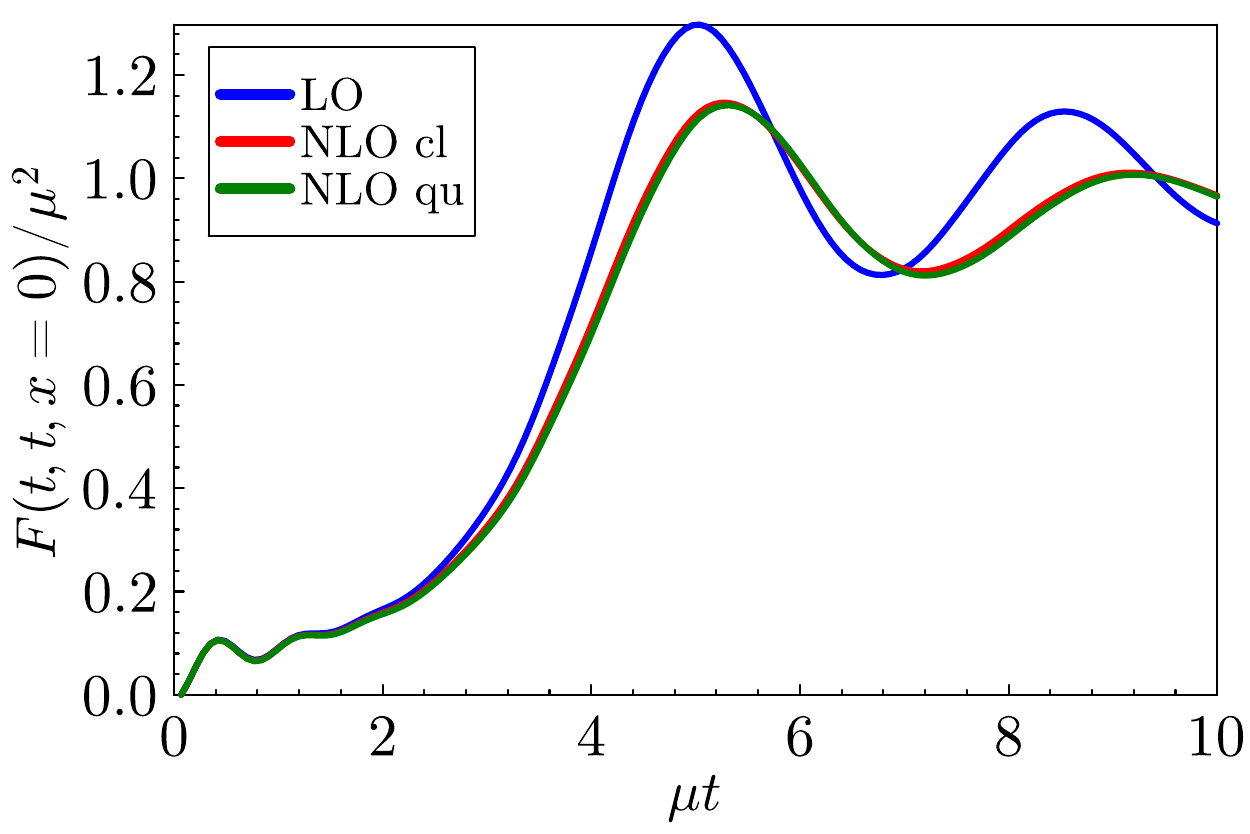}
  \includegraphics[width=0.5\textwidth]{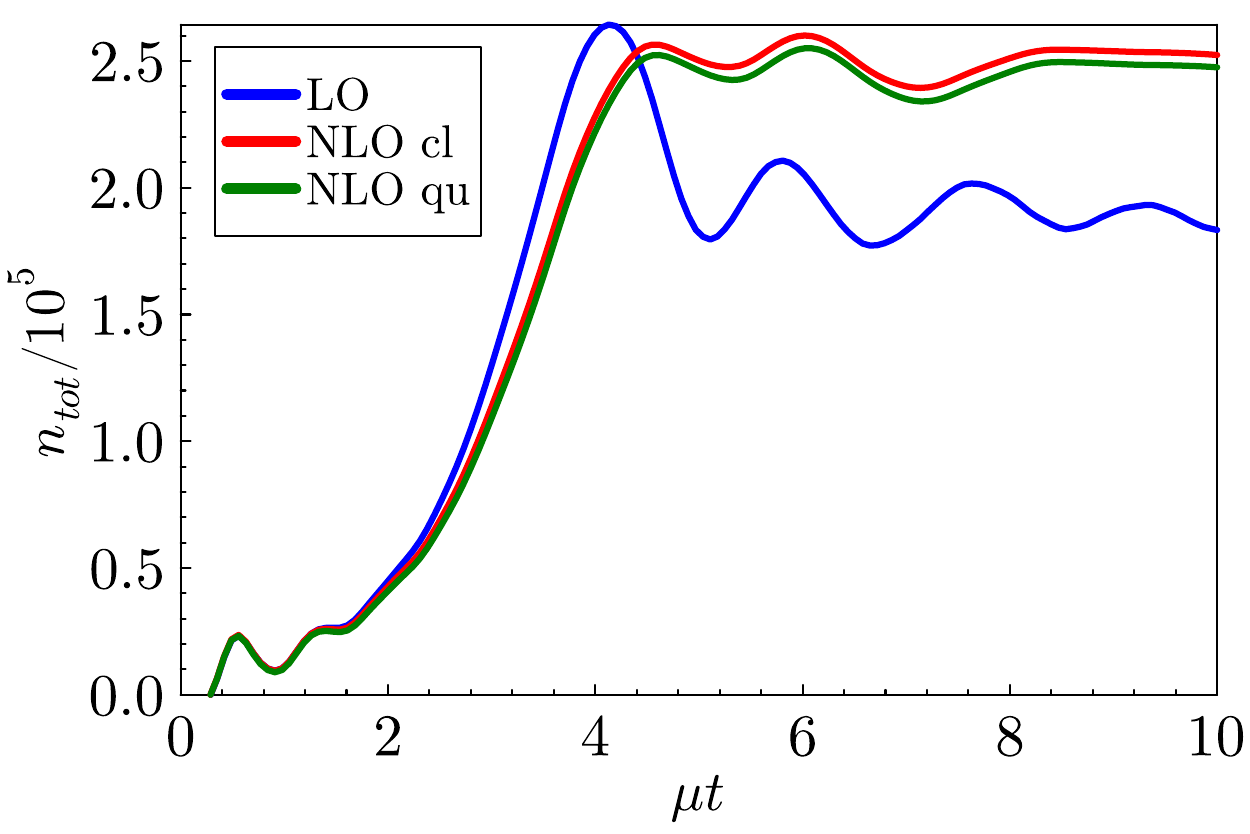}
  \caption{Left: Time evolution of the statistical correlator $F(t,t,x=0)$ at LO, quantum-NLO and classical-NLO. Apparently, occupation numbers are large enough for the system to act effectively classically. Right: The time evolution of the total particle number, summed over all momentum modes. $N=4$, $\lambda=6$. }
  \label{fig:classicalityFtt}
\end{figure}

\begin{figure}[ht]
  \centering
  \includegraphics[width=0.5\textwidth]{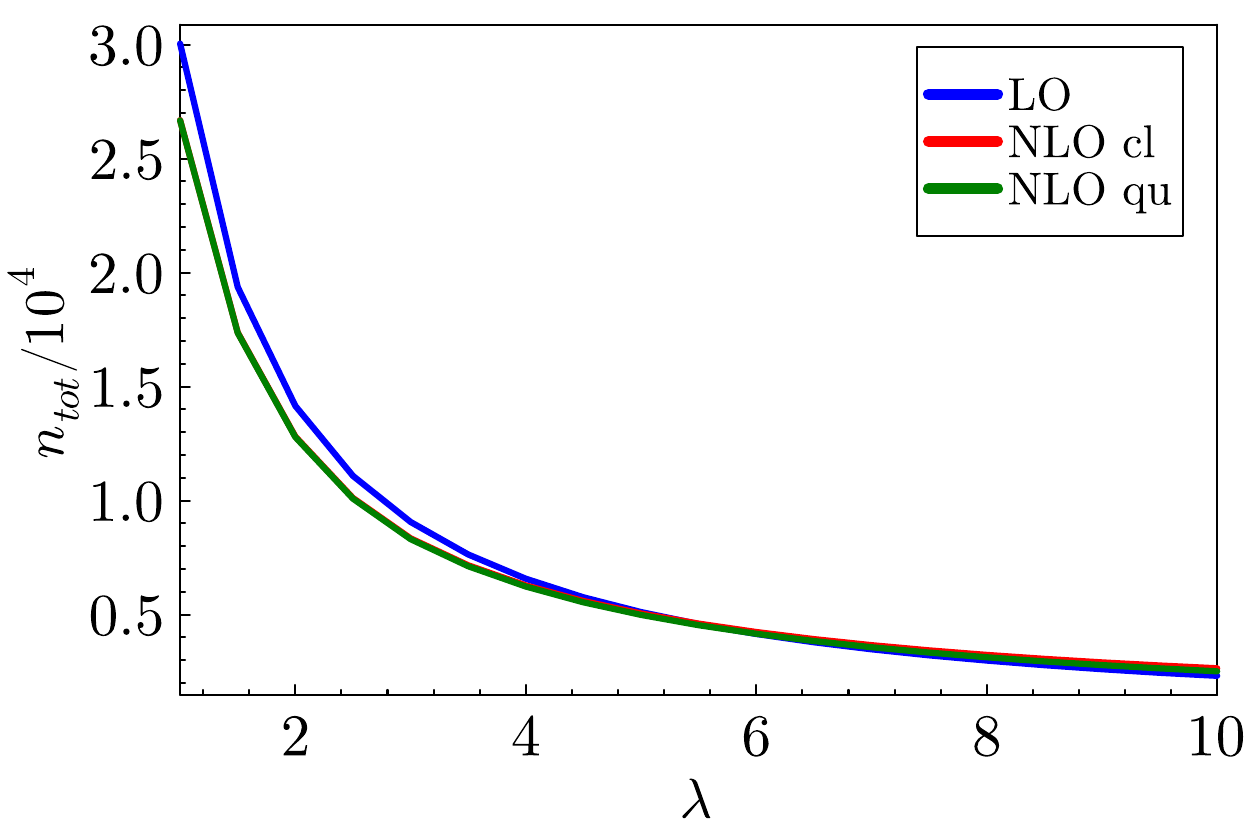}
  \caption{Highest value for total particle numbers $n_{tot}$ for the LO and NLO truncation. Even for large coupling, $n_{\rm tot}$ reaches large values. $N=4$.}
  \label{fig:ntot_lambdas}
\end{figure}

As we have seen, in the language of correlators, classicality follows in the limit where $F^2\gg \rho^2$ (see also \cite{Aarts:1997kp} for a discussion of the classical-statistical approximation). 
Often, $n_k>1$ is applied as a minimal requirement, and often it is stated that as long as this is fulfilled for the modes directly relevant to the phenomenon of interest, it is sensible to proceed classically for all modes. 
That requirement is fulfilled during a spinodal transition, at least in the range $k_L/\mu<0.8$ and we may further test the classical nature of the tachyonic transition by computing the evolution of $F(t,t,x=0)$ in the LO, the quantum-NLO and the classical-NLO approximation, shown in figure \ref{fig:classicalityFtt} (left).
 We see that the classical and quantum evolution agree very well, even though only unstable modes are highly occupied. Figure \ref{fig:classicalityFtt} (right) shows the same agreement in terms of the total particle number. 
Since self-interactions stop the instability when \cref{eq:lincrit} is satisfied, we can expect the individual occupation numbers to scale as $ n_k \propto 1/ \lambda $. During the evolution, the total particle number (summed over all the modes) grows to a maximum value and then decreases. In figure \ref{fig:ntot_lambdas} we show this maximal value as a function of $\lambda$, at LO and NLO.
We see that the LO dynamics describes the early time particle production process very well, and we find that for all $\lambda$, the maximal particle number occurs at times $\mu t < 15$.

\section{Intermediate time dynamics}
\label{sec:IntResults}
\subsection{The classical momentum range}
\label{sec:Qearly}

While the early-time dynamics of the tachyonic transition is well described by first a free field and then the LO approximation, at times $\mu t>6$, the non-local dynamics first entering at NLO becomes important. 
\begin{figure}[ht]
  \includegraphics[width=0.5\textwidth]{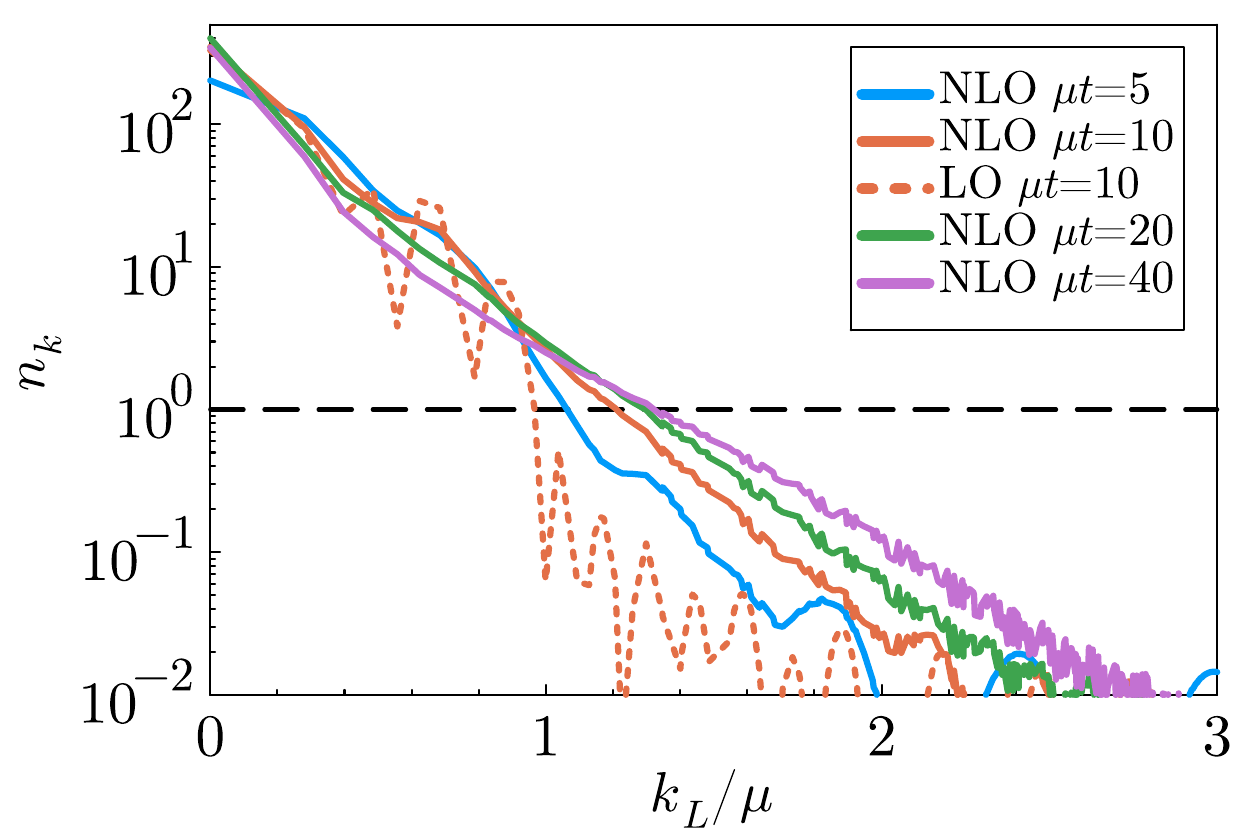}
  \includegraphics[width=0.5\textwidth]{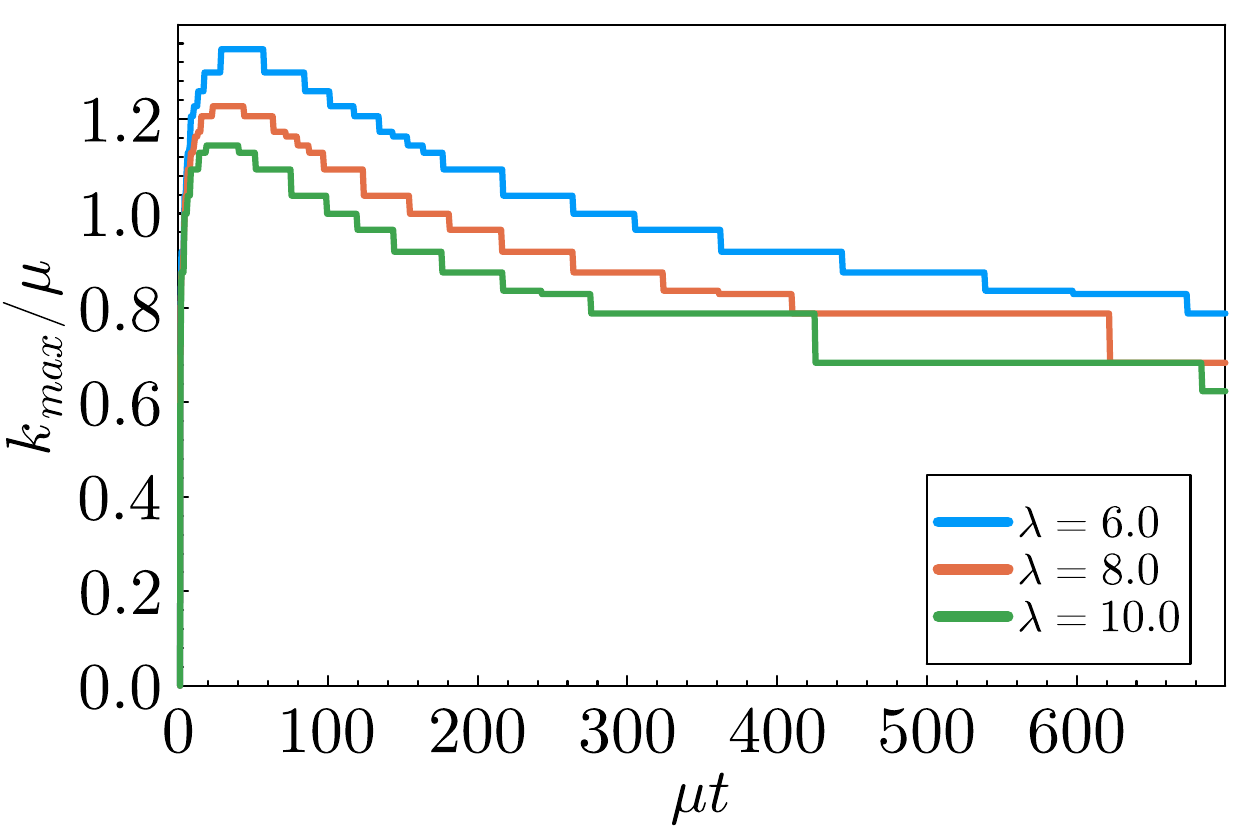}
  \caption{Left: The particle spectrum at the early stage of NLO evolution. $N=4$, $\lambda=6$. Right: The largest momentum wih $n_k>1$ in time, for different values of $\lambda$. $N=4$.}
  \label{fig:Earlytimespectra}
\end{figure}

\begin{figure}[ht]
  \centering
  \includegraphics[width=0.5\textwidth]{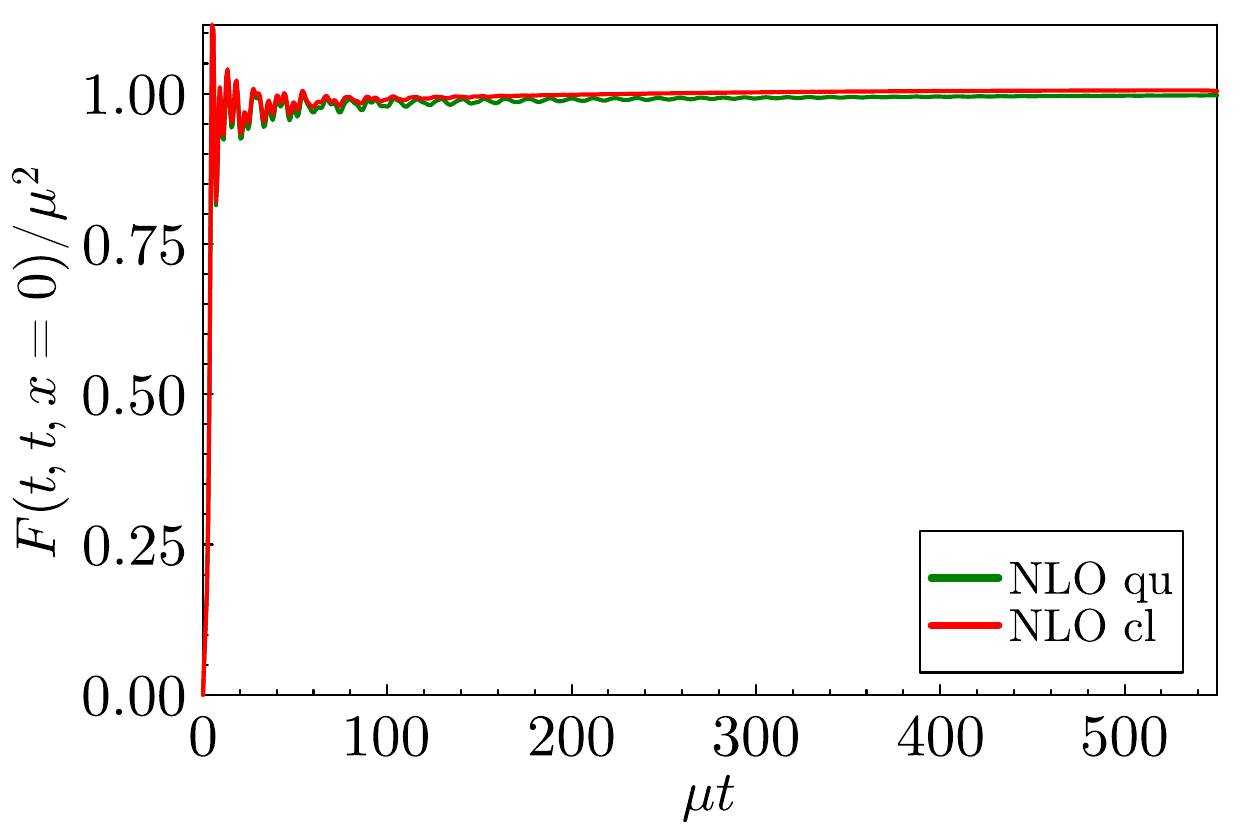}
  \caption{Time evolution of $F(t,t,x=0)$ at NLO, in the quantum system and the classical approximation.}
  \label{fig:FttzeroNLOquvscl}
\end{figure}

Figure \ref{fig:Earlytimespectra} (left) shows the particle spectrum at several times during the evolution, and we see that for $\mu t \lesssim 40 $ the range 
of momenta with $n_k>1$ grows, as the high occupancy in the IR is redistributed towards the UV. In figure \ref{fig:Earlytimespectra} (right) we show the value of the largest momentum with $n_k>1$ ($k_{max}$), which at first grows and subsequently shrinks on a time-scale of $\mu t=500$. Judging from the local correlator $F(t,t)$ only (figure \ref{fig:FttzeroNLOquvscl}), the classical approximation for the whole system is valid throughout.

\begin{figure}[ht]
  \centering
  \includegraphics[width=0.5\textwidth]{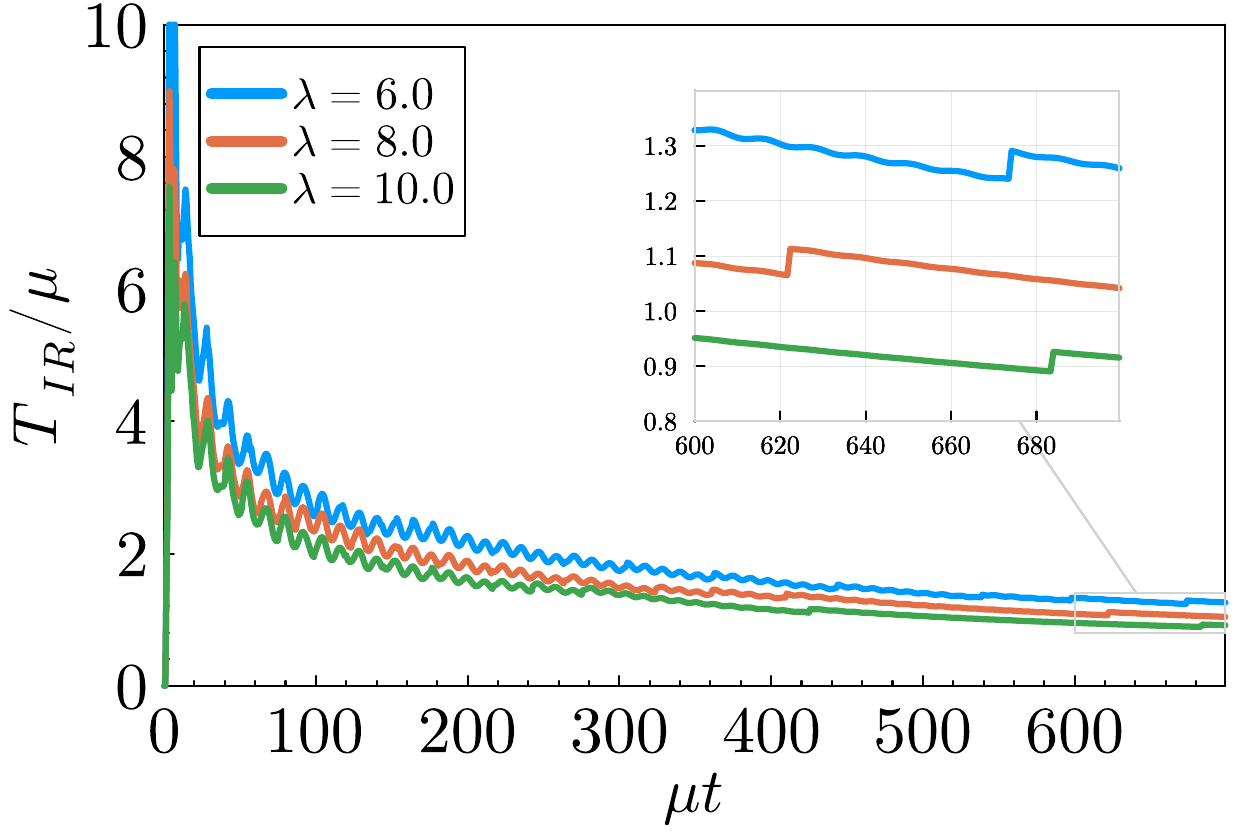}
  \caption{Time evolution of the effective temperature in the IR according to \cref{eq:TempIR}. The inset shows the values of the temperature at the end of the evolution. The discontinuities are caused by modes falling out of the classicality region and are therefore a discretisation effect. $N=4$.
  }
  \label{fig:InttimeTIR}
\end{figure}

Several out-of-equilibrium processes may have taken place during the initial violent stage of tachyonic preheating. Exotic heavy particles, topological defects, sphalerons and other relics such as primordial black holes may have been created in this environment of large IR field fluctuations. 
We can attempt to describe the out-of-equilibrium stage through an effective infrared temperature $\overline{T}_{\rm IR}$, which we define to be the average temperature in the classical range, had the spectrum been classical equilibrium 
\begin{equation}
    \overline{T}_{\rm IR}=\frac{\sum_{n_k>1}\omega_kn_k}{\sum_{n_k>1}1}.
    \label{eq:TempIR}
\end{equation}
This is shown in figure \ref{fig:InttimeTIR} as a function of time, and we see that although the asymptotic equilibrium temperature is $T_{\rm reh}\simeq \mu$ or less, at intermediate times the effective temperature may be twice as large or more. Provided the out-of-equilibrium process of interest takes place over a time-scale $\mu t\simeq\mathcal{O}(100-1000)$, this effective temperature should likely enter estimates of particle and relic production.

\subsection{Kinetic equilibration}
\begin{figure}[ht]
  \includegraphics[width=0.5\textwidth]{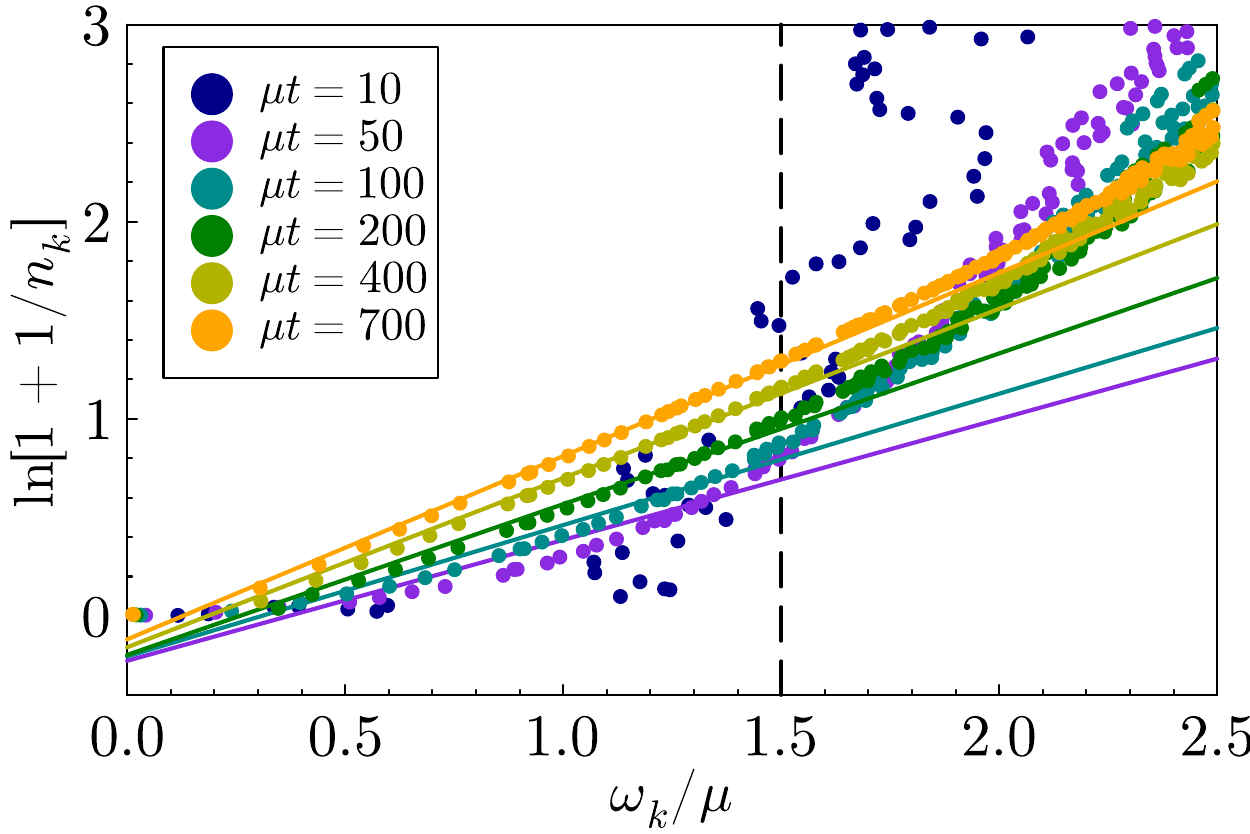}
  \includegraphics[width=0.5\textwidth]{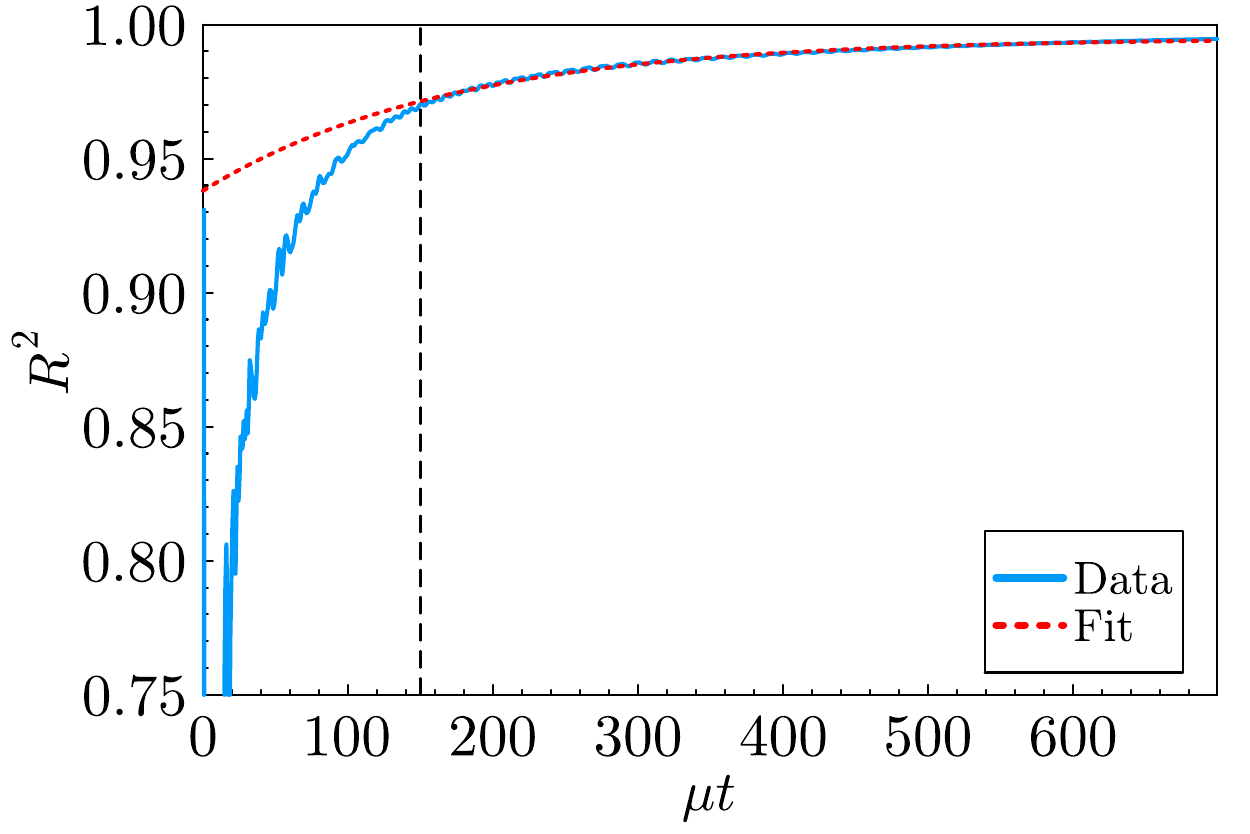}
  \caption{Left: Particle spectra for various stages of equilibration and the corresponding fits to Bose-Einstein distributions (straight lines). 
  Right: The $R^2$ value of the fit quantifies how close the particle spectrum is to a thermal distribution. $N=4$, $\lambda=6$.}
  \label{fig:kineticequilibrationBEandR2}
\end{figure}
The (quantum-)NLO evolution equations provide correct thermalization, in the sense that at late times, the occupation numbers approach a quantum thermal equilibrium state \cite{Berges:2000ur,Arrizabalaga:2005tf,Shen:2020jya}. There are two thermalization stages: kinetic equilibration, whereby particles are redistributed into a thermal distribution with a non-zero chemical potential; followed by chemical equilibration, where the overall number of particles changes to eventually relax the system to zero chemical potential and its final temperature. These processes may be identified with certain constituent perturbative diagrams (2-to-2 scattering, 2-to-4 scattering, ...), and so for small coupling, the timescales are expected to be well separated. 
We can quantify the equilibration process by performing a fit of the particle spectrum with a Bose-Einstein distribution:\footnote{We trust that the standard notation using $\mu_{ch}$ as the chemical potential is not confused with the mass parameter $\mu^2$.}
\begin{equation}
    n_k= \frac{1}{e^{(\omega_k-\mu_{ch})/T}-1},
\label{eq:BEdist}
\end{equation}
where $\omega_k$ is the derived dispersion relation (\ref{eq:disprelation}). 
\begin{figure}[ht]
  \includegraphics[width=0.5\textwidth]{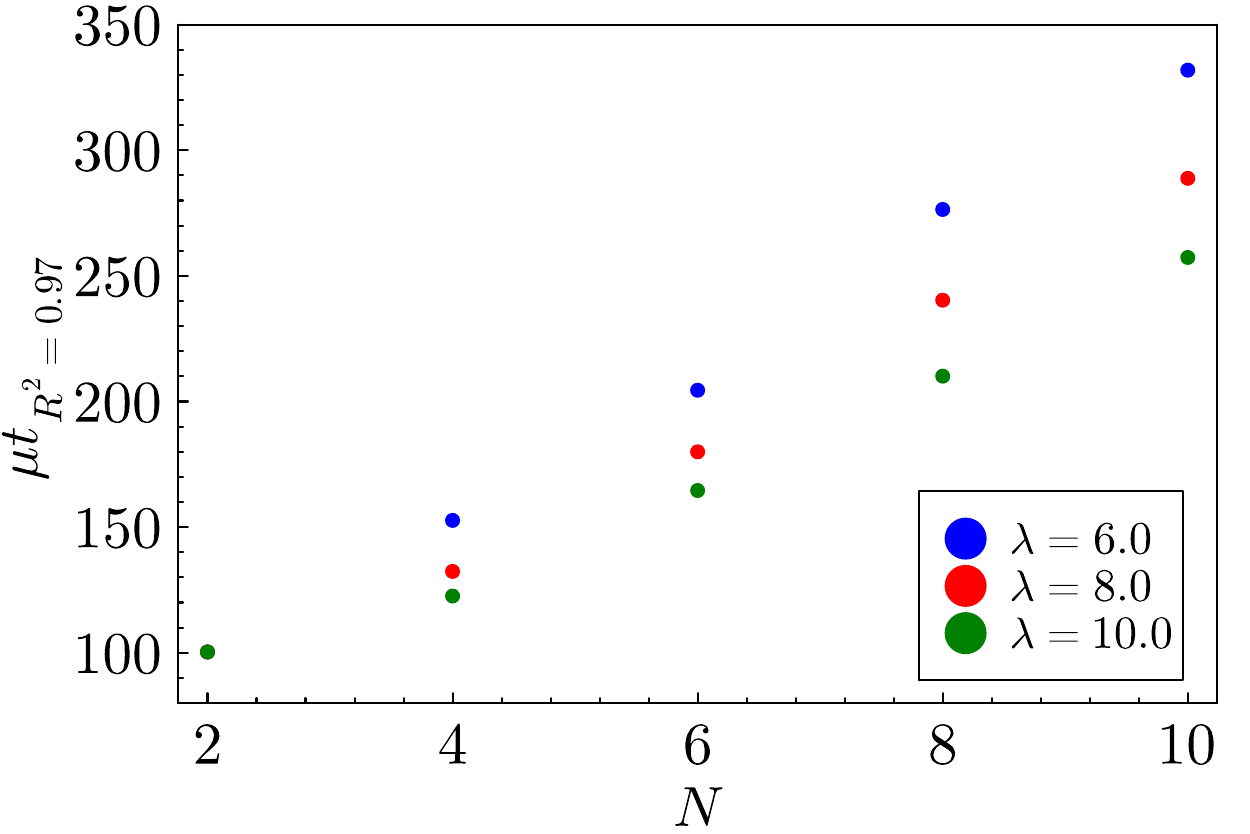}
  \includegraphics[width=0.5\textwidth]{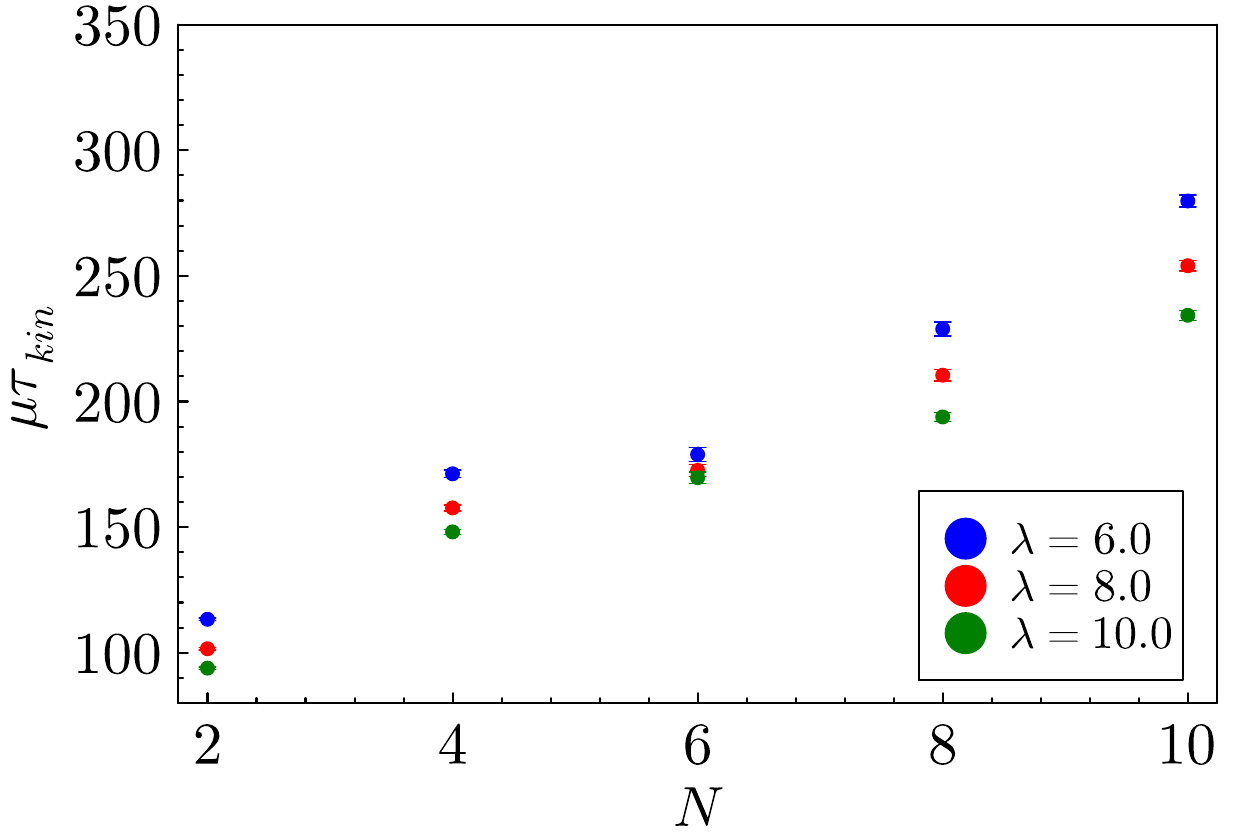}
  \caption{Time scale of kinetic equilibration: Time when $R^2$ of Bose-Einstein distribution fit reaches $0.97$ (left) and time scale defined by \cref{eq:R2fit} (right).}
  \label{fig:kineticequilibrationtimes}
\end{figure}
We perform a least square fit with
\begin{equation}
    \ln \big(1 + 1/n_k \big) = (\omega_k-\mu_{ch})/T,
\label{eq:BEdistlinear}
\end{equation}
and quantify kinetic equilibration by calculating the $R^2$ value of this fit \footnote{We use the standard definition
\begin{equation}
    R^2 = 1 - \frac{ \sum \big( y_i - f_i \big)^2}{ \sum \big( y_i - \bar{y} \big)^2},
\end{equation}
where $y_i$ refers to the data points, $\bar{y}$ to the mean and $f_i$ the predictions of the model with best fit parameters. A $R^2$ value of 1 corresponds to a perfect fit where all data points are matched by the model.The best fit parameters $T$ and $\mu$ in \ref{eq:BEdistlinear} can be associated with an uncertainty from the fit procedure, quantifying a range of parameters in agreement with data. In the following these uncertainties are propagated to observables as $T_{\rm reh}$ and time scales. We stress that they are systematic rather than statistical.}

In figure \ref{fig:kineticequilibrationBEandR2} we show the process of kinetic equilibration, for $N=4$, $\lambda=6$. On the left we see the particle spectrum at various stages of equilibration and the corresponding fit to a Bose-Einstein distribution. We restrict our fits to the IR defined by $\omega_k / \mu < 1.5$, indicated by the dashed vertical line. 
In figure \ref{fig:kineticequilibrationBEandR2} (right) we see that the $R^2$ gradually approaches 1, and that although after the initial growth of particle numbers the spectrum is far from thermal, at intermediate times $\mu t \approx 150$ the particles are reorganised into a thermal-like distribution. At this stage the distribution is well approximated by \cref{eq:BEdist}.  
We define kinetic equilibration to be completed when $R^2$ reaches $0.97$, which happens at $\mu t =150$ in figure \ref{fig:kineticequilibrationBEandR2}. Furthermore we can extract a kinetic equilibration time scale by fitting the $R^2$ evolution to the form
\begin{equation}
    R^2(t) = R^2_{\rm final}+(R^2_{\rm init}-R^2_{\rm final})e^{-t/\tau_{\rm kin}}.
    \label{eq:R2fit}
\end{equation}
where we discard $R^2$ values $<0.97$. The fit is also shown in figure \ref{fig:kineticequilibrationBEandR2} (right).

\begin{figure}[ht]
  \centering
  \includegraphics[width=0.5\textwidth]{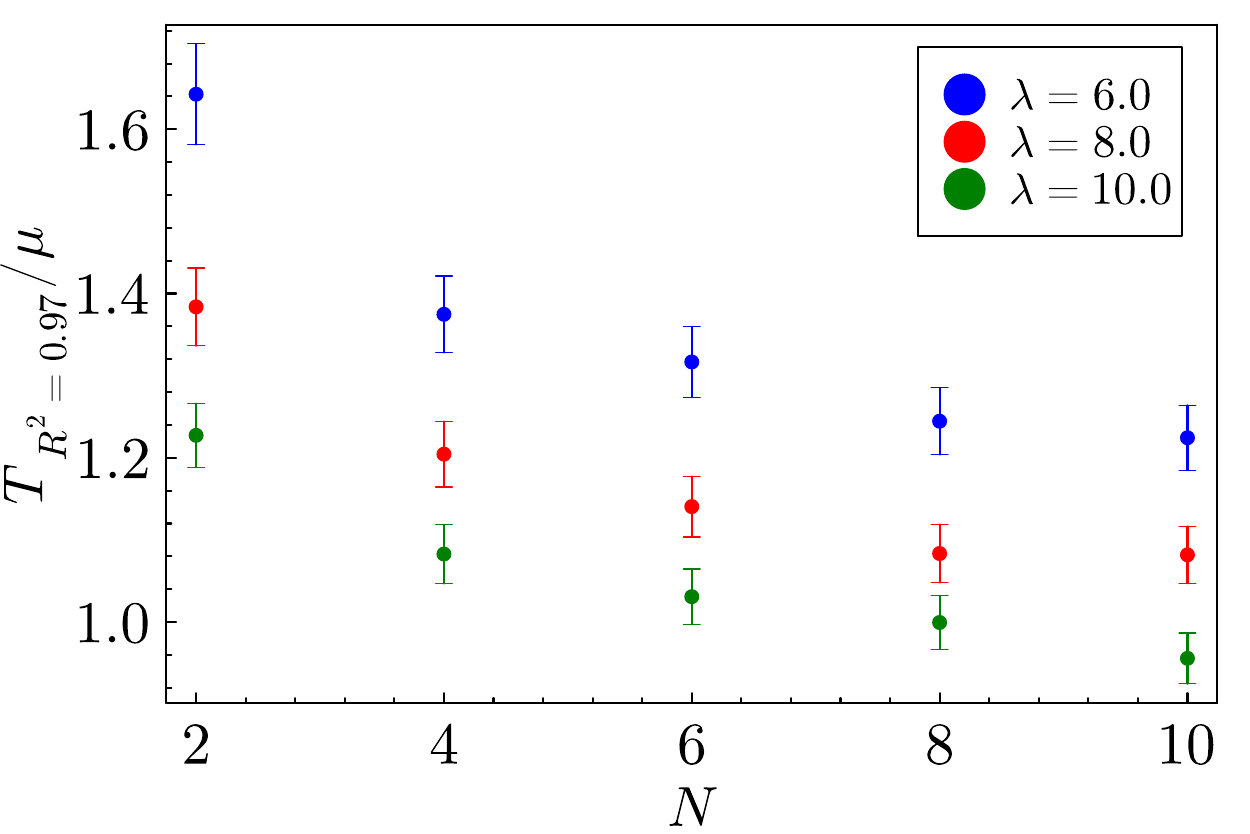}
  \caption{Temperature at the time when kinetic equilibrium is reached.}
  \label{fig:kineticequilibrationtemperatures}
\end{figure}

Figure \ref{fig:kineticequilibrationtimes} shows the time when $R^2$ reaches $0.97$ (left) and the kinetic equilibration time defined by \cref{eq:R2fit} (right). The two are consistent, and reveal a linear dependence on $N$ and that larger coupling produces faster kinetic equilibration. 
At the onset of kinetic equilibration we extract a temperature, shown in figure \ref{fig:kineticequilibrationtemperatures}, with a chemical potential in the range $\mu_{ch}\mu=0.2-0.33$. In the following stage of chemical equilibration the temperature approaches its final value and the chemical potential goes to zero.
\begin{figure}[ht]
  \includegraphics[width=0.5\textwidth]{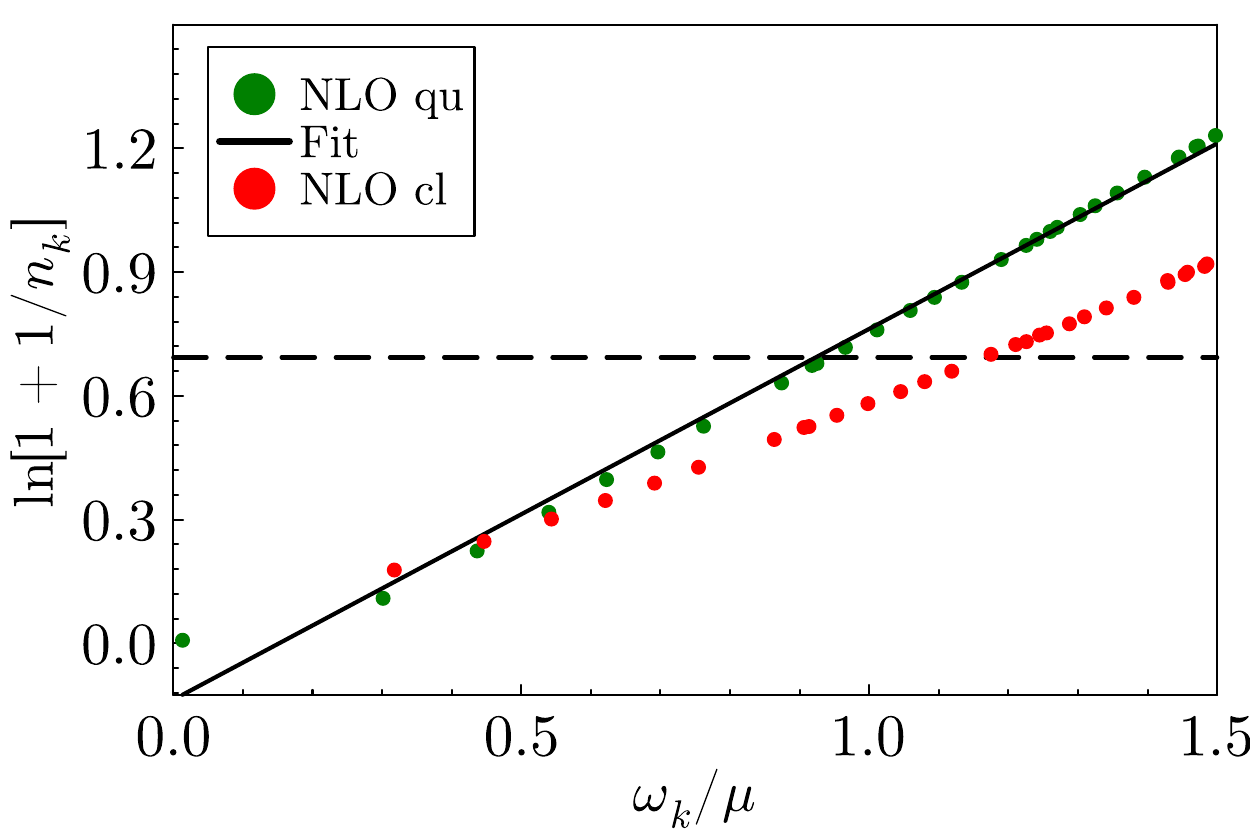}
  \includegraphics[width=0.5\textwidth]{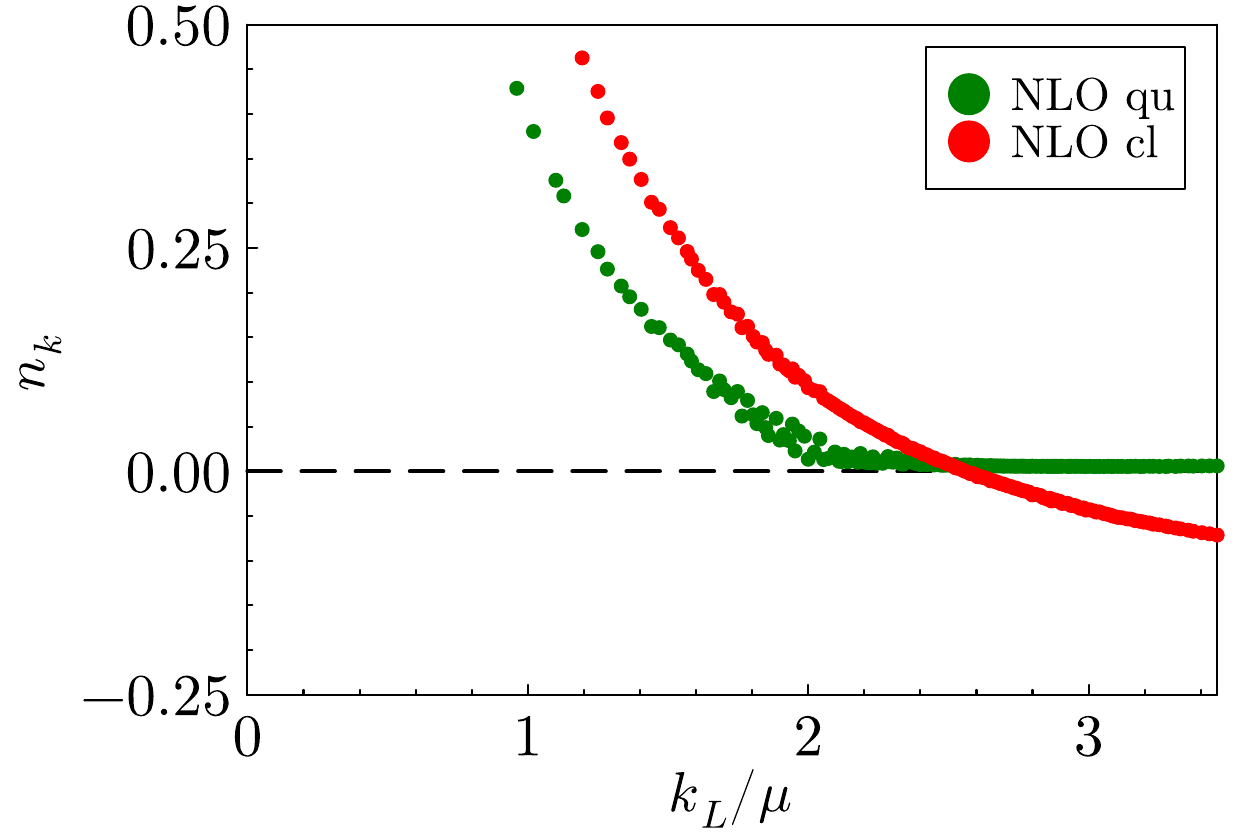}
  \caption{Left: The particle spectrum at $\mu t = 550$ for $N=4$, $\lambda=6$ at NLO and a fit to a Bose-Einstein distribution, which in this variable corresponds to a straight line $(\omega_k-\mu_{ch})/T$. The red dots show the spectrum in the classical approximation.  Right: Comparison of UV part of the particle spectrum at $\mu t =550 $ for quantum evolution at NLO and in the classical approximation. $N=4$, $\lambda=6$.}
  \label{fig:kineticeqBEfit}
\end{figure}

In figure \ref{fig:kineticeqBEfit} (left), we show an advanced stage of kinetic equilibration, for $N=4$, $\lambda=6$. The distribution is very well approximated by \cref{eq:BEdist} ($R^2>0.99$). For comparison, we show the equivalent spectrum in a classical-NLO simulation, which does not agree very well, even for modes with $n_k>1$ (shown by the dashed horizontal line). The classical line is lower than the quantum spectrum, corresponding to larger occupation numbers. Only for $n_k>3$ do the two spectra agree. 
This implies that one should be wary of approximating intermediate to late time quantum evolution by classical dynamics. In figure \ref{fig:kineticeqBEfit} (right), we show the UV part of the spectrum\footnote{But without performing the transformation where a Bose-Einstein distribution becomes a straight line.} at this same time of $\mu t=550$. The quantum dynamics correctly conserves the zero-point fluctuations ($n_k+1/2\geq 1/2$, $n_k\geq 0$), as shown by the green dots. Zero point fluctuations are there to scatter on, but it is not physical to extract net particles or energy from the vacuum. The classical simulation however has no such constraint, and does not distinguish between the $n_k$ and the $1/2$, as shown by the red dots. As such, it allows to deplete the UV vacuum so that $n_k+1/2 < 1/2$, to incorrectly increase the occupation in the IR and ultimately to thermalise to an incorrect, classical equilibrium \cite{Tranberg:2022noe}. In a manifestly classical simulation, one might consider not initialising with the zero-point fluctuations, or only in the unstable modes \cite{Arrizabalaga:2004iw}. This has some advantages, although one would lose some of the physics of scattering on the UV fluctuations, and the late time equilibrium would still be wrong. 

\subsection{Chemical equilibration}
 \begin{figure}[ht]
  \includegraphics[width=0.5\textwidth]{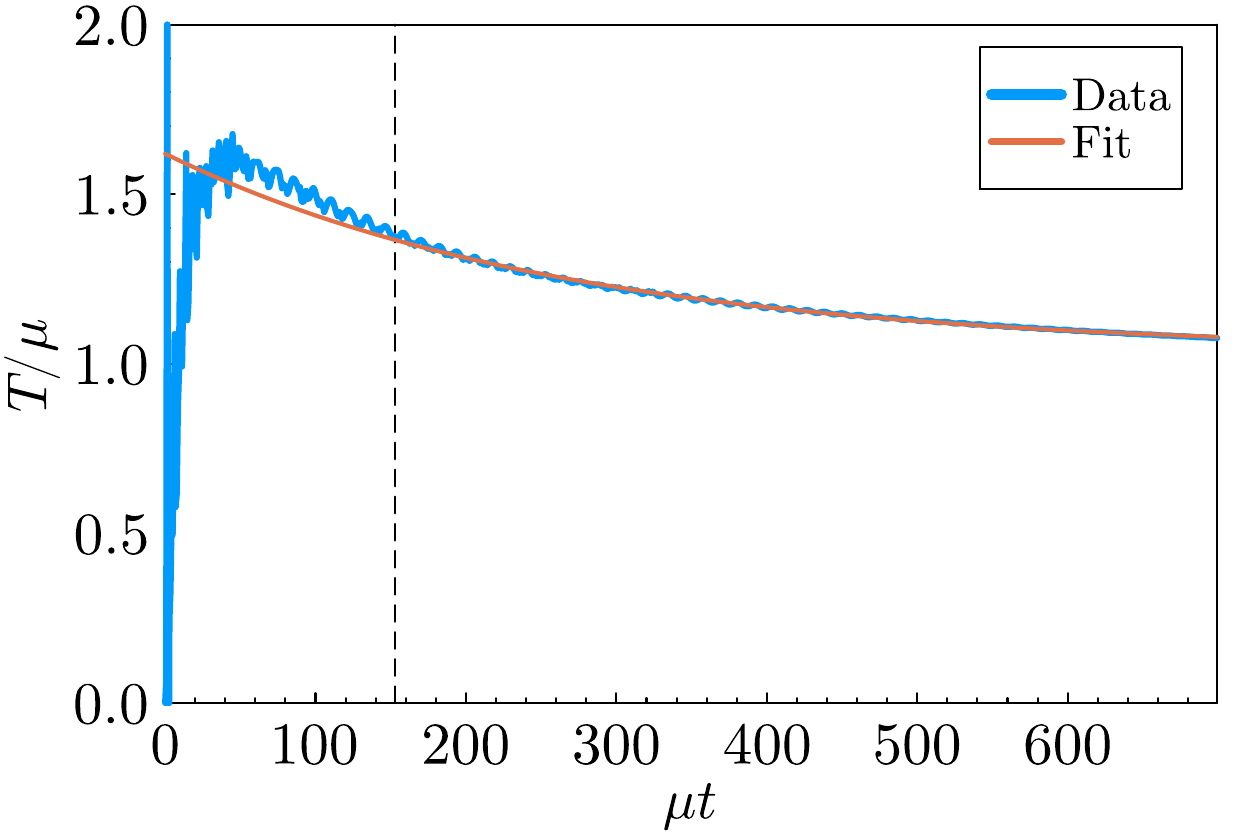}
  \includegraphics[width=0.5\textwidth]{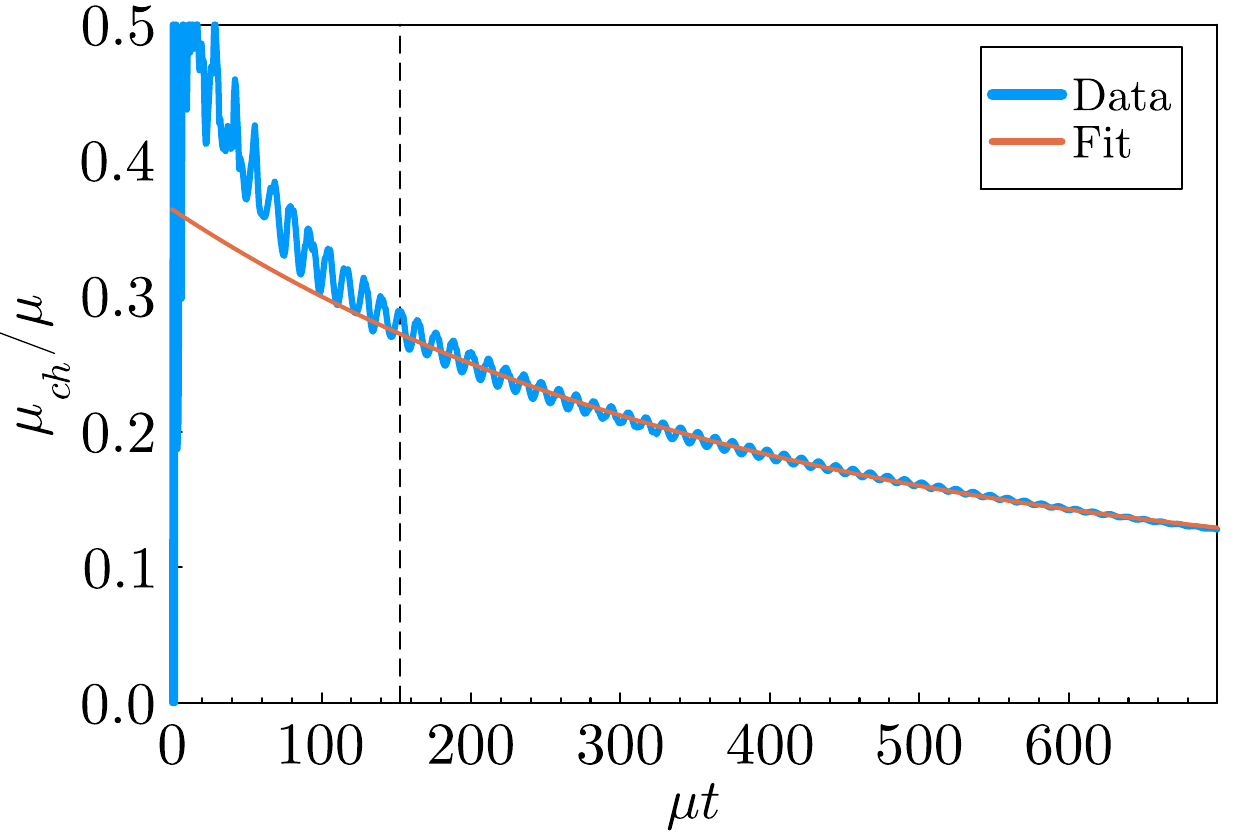}
  \caption{Evolution of the Bose-Einstein fit parameters $T$ and $\mu_{ch}$ in time, and the fit to \cref{eq:Tfit} and \cref{eq:mufit}. The dashed line indicates the beginning of the fit region. $N=4$, $\lambda=6$.}
  \label{fig:kineticeqTmuevol}
\end{figure}
The fit parameters $\mu_{ch}$ and $T$ are time-dependent, and are expected to asymptote to zero and some finite reheating temperature $T_{\rm reh}$, respectively. Figure \ref{fig:kineticeqTmuevol} shows $T$ and $\mu_{ch}$ in time, and we may further extract a final temperature $T_{\rm reh}$ and a chemical equilibration timescale $\tau_{\rm ch}$ by a fit to
\begin{equation}
    T(t) = T_{\rm reh}+(T_{\rm init}-T_{\rm reh})e^{-t/\tau_{\rm ch}}.
    \label{eq:Tfit}
\end{equation}
and
\begin{equation}
    \mu_{ch}(t) = \mu_{ch}^{\rm final}+(\mu_{ch}^{\rm init}-\mu_{ch}^{\rm final})e^{-t/\tau_{\rm ch}}.
    \label{eq:mufit}
\end{equation}
We must note here that we are somewhat limited by the available computational resources. The asymptotic temperature $T_{\rm reh}$ and chemical equilibration time scale $\mu \tau_{\rm ch}$ depend on the available time-range. 
Based on simulations of a smaller lattice size $16^3$ to much late times ($\mu t =2800$), we find that data until $\mu t = 700 $ overestimates $T_{\rm reh}$ by about $20\%$. Furthermore, the chemical equilibration time is underestimated by about $80 \%$ (see Appendix \ref{App:16}).

\begin{figure}[ht]
  \includegraphics[width=0.5\textwidth]{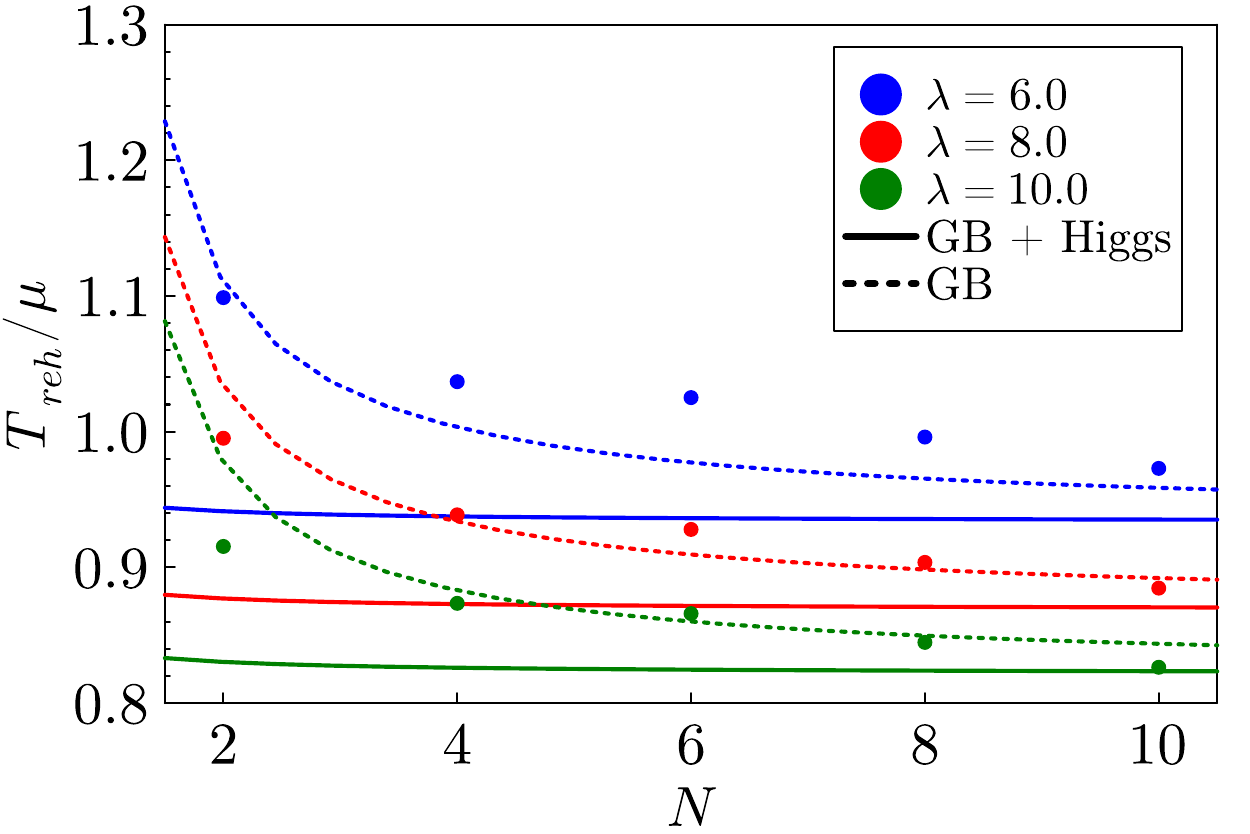}
  \includegraphics[width=0.5\textwidth]{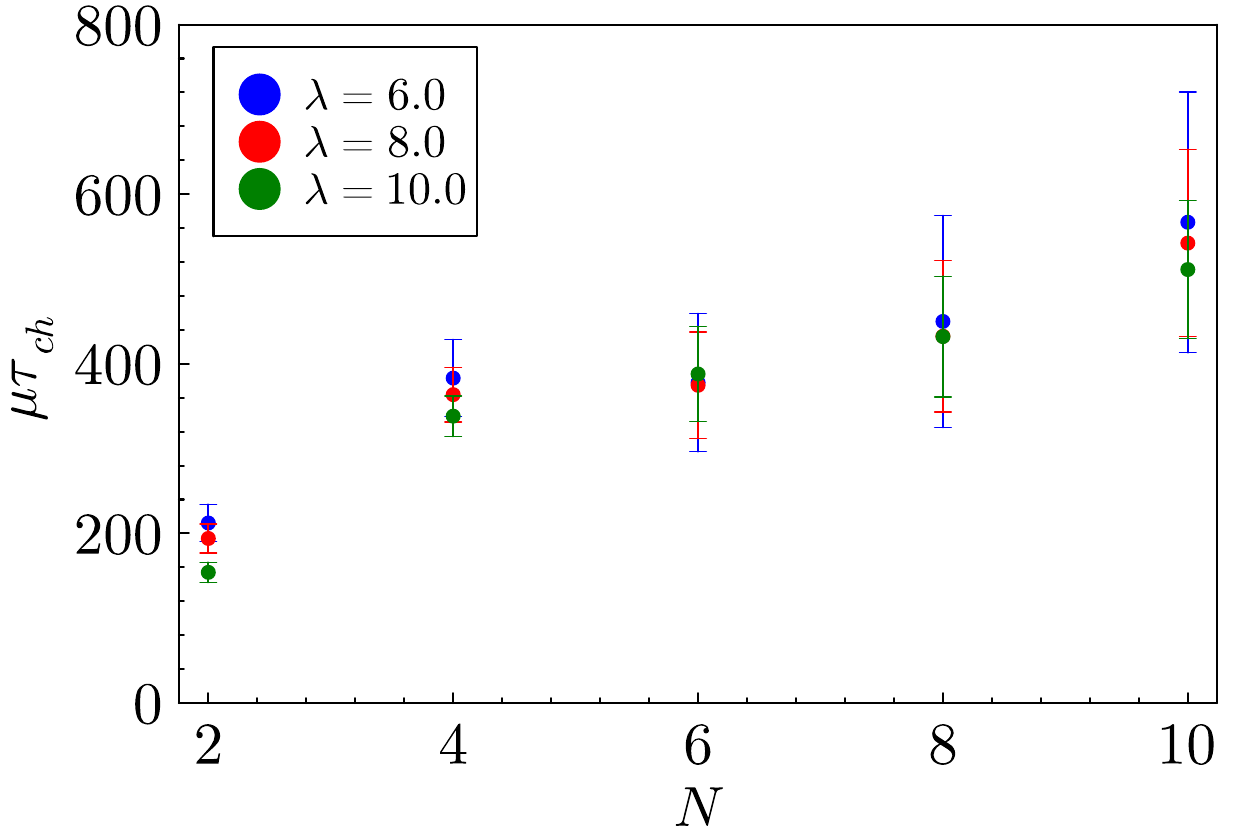}
    \caption{Left: Extrapolated equilibrium temperatures for various values of $N$ and $\lambda$. Shown are values obtained from the simulation (dots) and a simple Higgs-Goldstone model (solid line) and for Goldstone modes only (dotted line). Right: Chemical equilibration time for various values of $N$ and $\lambda$.}
  \label{fig:kinequTeff}
\end{figure}

\begin{figure}[ht]
  \includegraphics[width=0.5\textwidth]{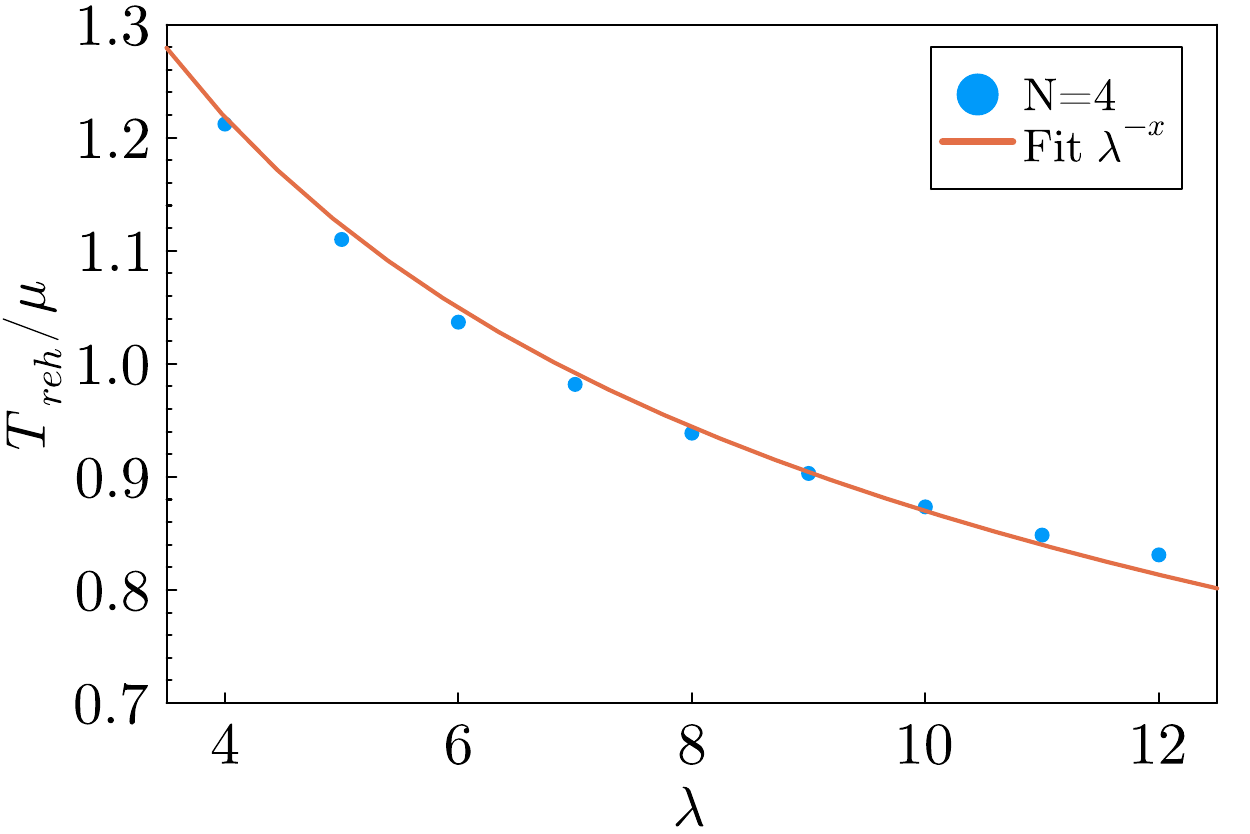}
  \includegraphics[width=0.5\textwidth]{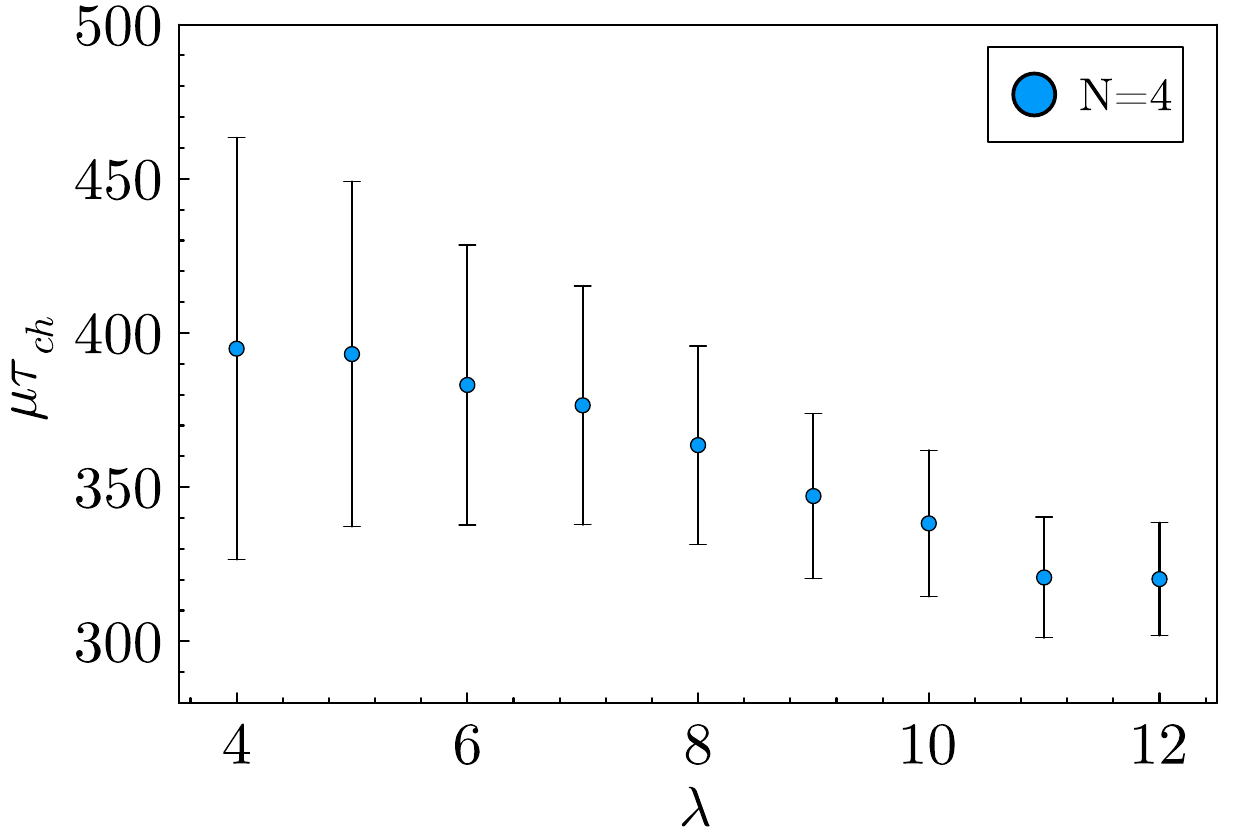}
  \caption{Left: Coupling strength dependence of $T_{\rm reh}$ and power fit $\propto \lambda^{-x}$ for which $x=0.37$. Right: Coupling strength dependence of chemical equilibration time for $N=4$.}
  \label{fig:Lambdadep}
\end{figure}

With this caveat in mind, we show in figure \ref{fig:kinequTeff} the reheating temperature $T_{\rm reh}$ for various $N$ and $\lambda$ values. We also show for comparison temperatures from a simple toy model in which all available energy is distributed among a massive Higgs mode with $m_H=\sqrt{2} \mu$ and $N-1$ massless Goldstone modes. We obtain the temperature in this model by solving
\begin{align}
    \frac{3 N \mu^4}{2 \lambda} = &\int \frac{d^3k}{(2 \pi)^3} \Big( \sqrt{k^2 - m_H^2} \, n_k(m_H,T) + (N-1) \, k \, n_k(0,T) \Big),
\end{align}
for $T$, where we assume equilibrated constituents $n_k(m,T) = ( e^{ \sqrt{k^2 -m^2} / T } - 1)^{-1}$. The temperatures are almost independent of $N$. An even simpler model is to discard the Higgs mode, in which case one returns to (\ref{eq:simplestT}) (with $N-1$ degrees of freedom). In that case the temperature scales with $T \propto \Big( \frac{N}{N-1} \Big)^{1/4} $, with $T/\mu = 0.93$ for $\lambda=6$.
Taking into account the overestimation of $T_{\rm reh}$ due to the time range limitation, we see that the fit temperatures end up between the Higgs-Goldstone and Goldstone-only models. 
Figure \ref{fig:kinequTeff} (right) shows the kinetic equilibration time $\mu \tau_{\rm kin}$ and we identify an approximate linear $N$ dependence. Furthermore, we find as for kinetic equilibration that larger couplings lead to faster equilibration. The systematic errors estimated from the fitting procedure are however substantial. In addition, we note that for these rather large couplings, the chemical equilibration times are only about twice as large as the kinetic equilibration time scale, so that it is hard to disentangle the two processes.
In figure \ref{fig:Lambdadep} we show the coupling strength dependence on $T_{\rm reh}$ and $\tau_{\rm ch}$. 

\subsection{The compound propagator}
\label{sec:Qcompound}

\begin{figure}[ht]
  \includegraphics[width=0.5\textwidth]{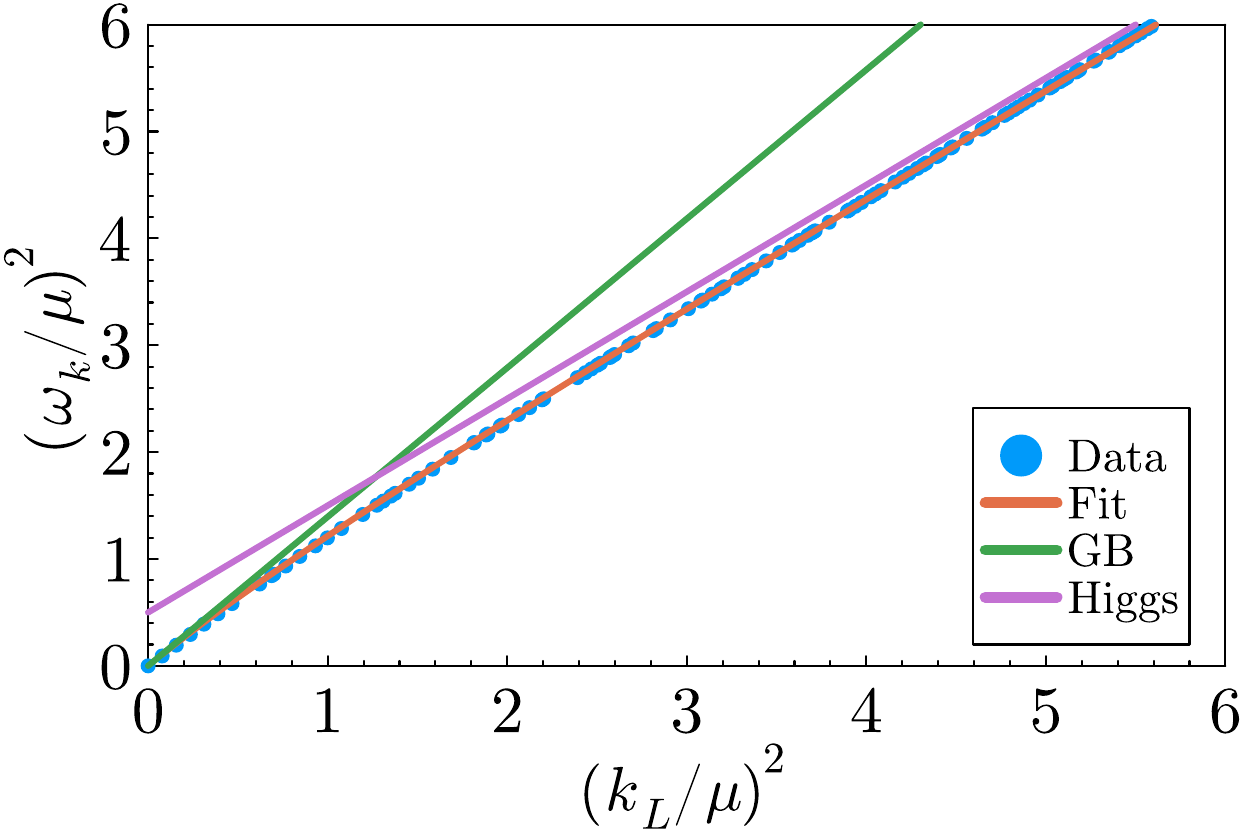}
  \includegraphics[width=0.5\textwidth]{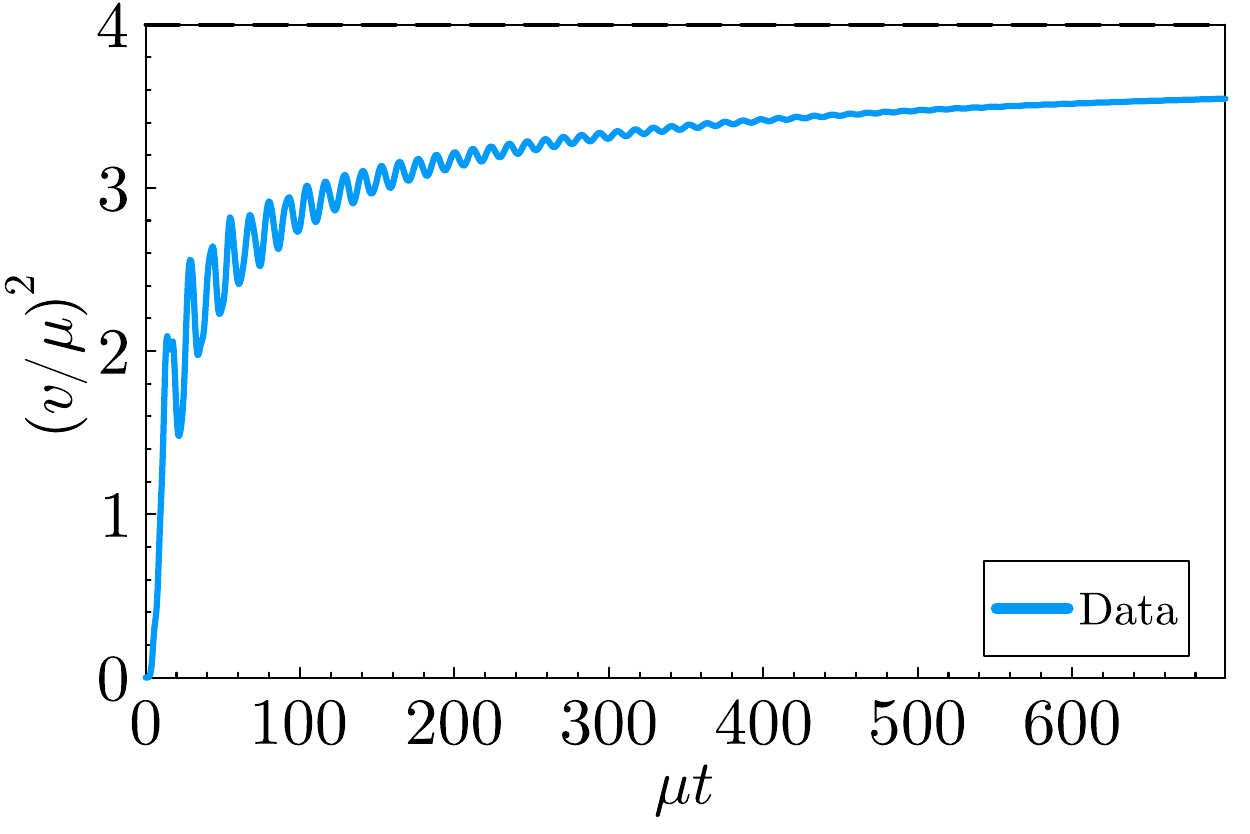}
  \caption{Left: Dispersion relation for $N=4$ and $\lambda=6$ and the fit to the result of the model of the compound propagator. Right: The field expectation value (squared) extracted from the propagator by \cref{eq:vev}. 
  }
  \label{fig:compoundomega}
\end{figure}
As we discussed in section \ref{sec:Model}, the O(N) symmetric correlator is a compound of a Higgs mode and $N-1$ Goldstone modes. Sofar, we have been fitting the particle number with a single Bose-Einstein distribution, in effect assuming that the $N-1$ light modes dominate. In figure \ref{fig:compoundomega} (left), we fit the decomposition \cref{eq:compdecomp} to the dispersion relation, while inserting the temperatures as obtained in the previous chapter. We indeed see that the dispersion relation is a composite of two particle modes, one with non-zero mass (Higgs, purple line) and one massless (Goldstone, green line).
The field expectation value can be recovered from the zero mode of the propagator \cref{eq:vev}, which we show in \cref{fig:compoundomega} (right). We see that although it asymptotes to a finite value, this is not the zero-temperature (vacuum) expectation value, but is the minimum of the finite-temperature effective potential. Also, settling into this minimum happens gradually on a time-scale similar to kinetic equilibration. 

\subsection{The equation of state}

\begin{figure}[ht]
  \includegraphics[width=0.5\textwidth]{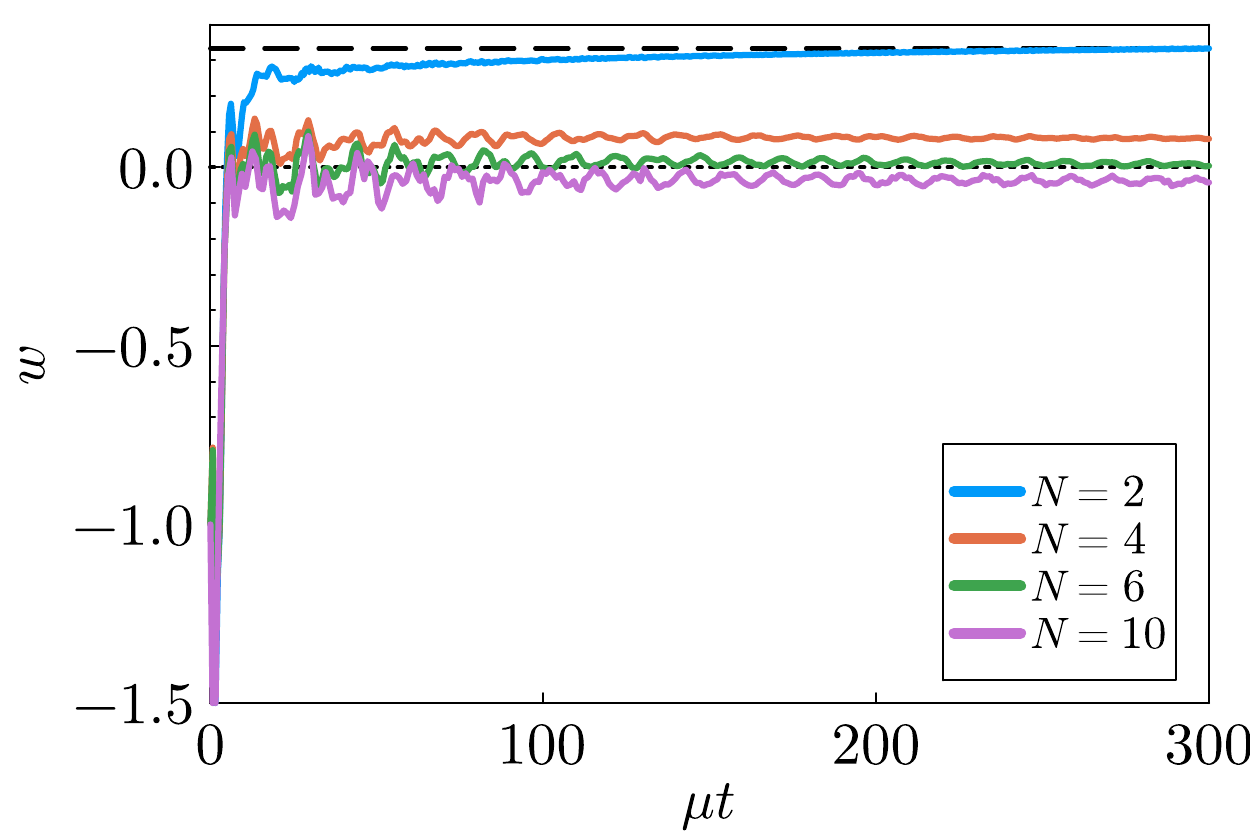}
  \includegraphics[width=0.5\textwidth]{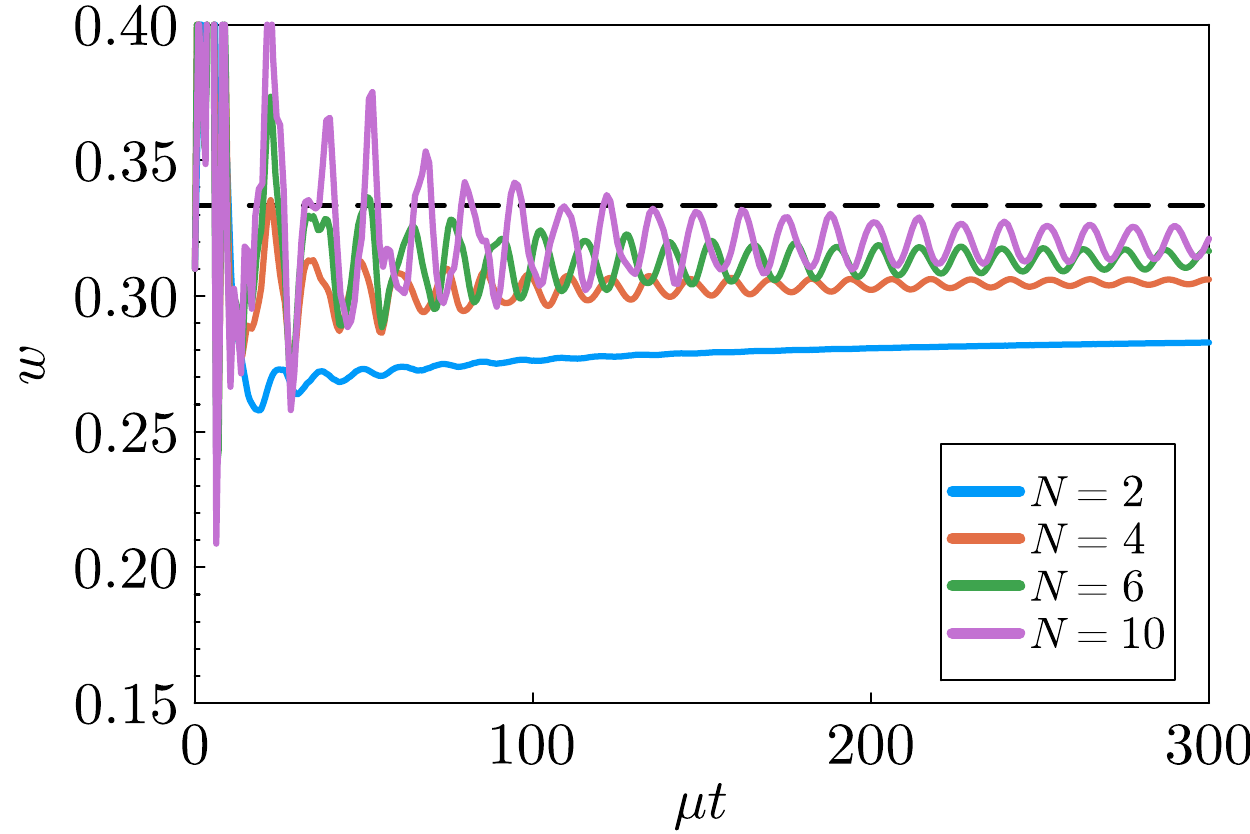}
  \caption{Time evolution of the equation of state for $\lambda=6$. Left: Equation of state derived from the energy density and pressure density functionals. Right: Equation of state from a quasi particle ansatz.}
  \label{fig:eosL6}
\end{figure}
\begin{figure}[ht]
  \includegraphics[width=0.5\textwidth]{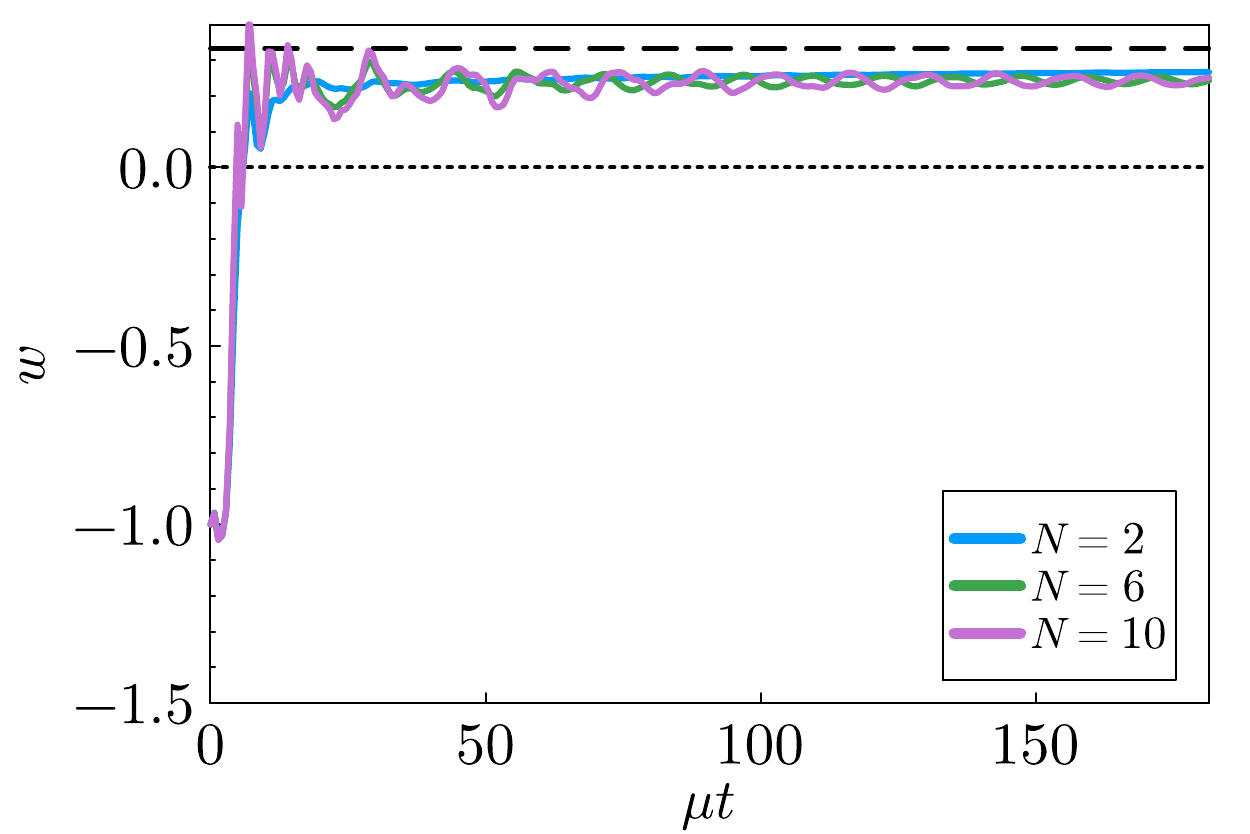}
  \includegraphics[width=0.5\textwidth]{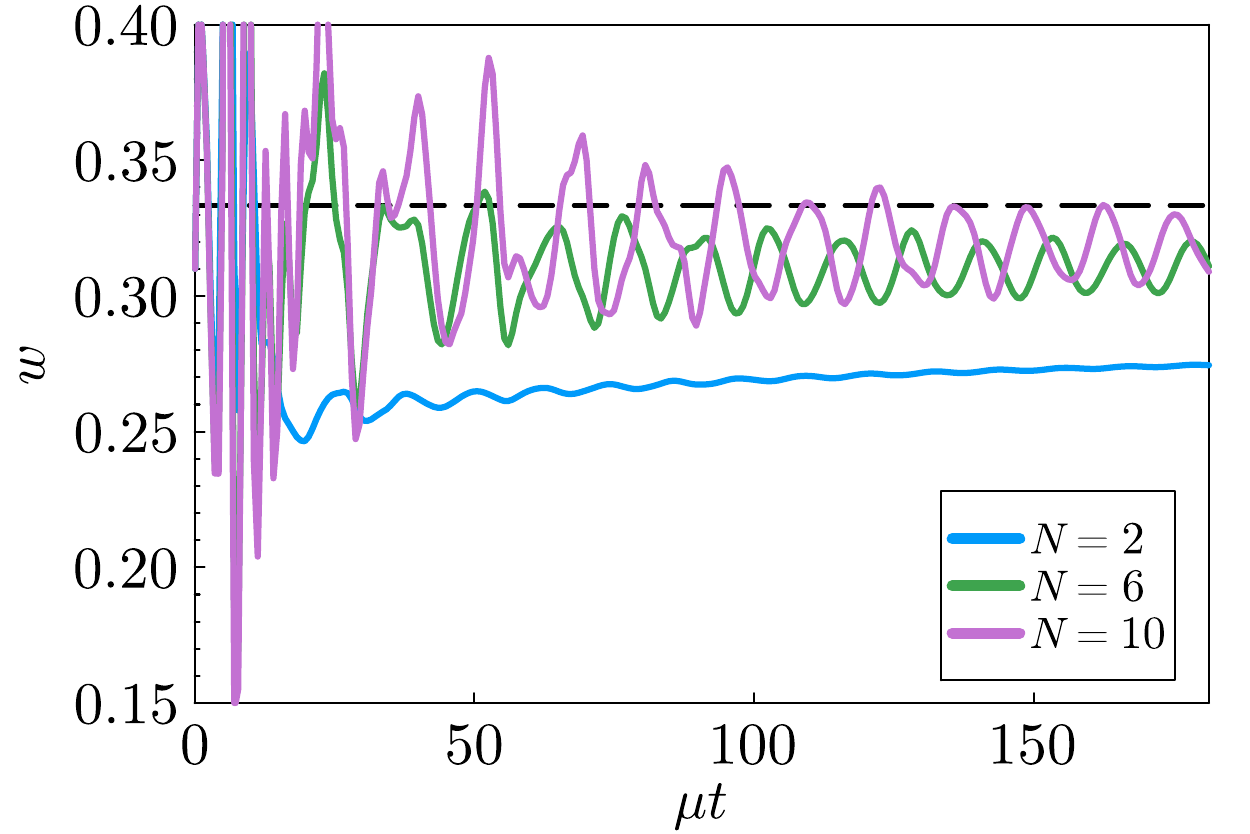}
  \caption{As figure \ref{fig:eosL6} but with $\lambda=1$.}
  \label{fig:eosL1}
\end{figure}
For cosmological purposes, complete equilibration is often less important than the effective equation of state of the system, which often settles much faster \cite{Berges:2004ce}. In figure \ref{fig:eosL6} we show this quantity in time for $\lambda=6$. One might have expected that the state was an equipartitioned mixture of $g^*$ "radiation" Goldstone components with $P/\rho\simeq 1/3$ and $(N-g^*)$ "matter" components with $P/\rho\simeq 0$. In that case, the equation of state should be
\begin{eqnarray}
\omega\simeq\frac{g^*}{3N}.
\end{eqnarray}
At low temperature, one would expect $g^*=N-1$, so that for $N=4$, $\omega=1/4$. But the situation is more complicated. 

On the left of figure \ref{fig:eosL6} we show how the equation of state parameter evolves from the initial $\omega=-1$ to its final value for $\lambda=6$, within a time $\mu t =100-200$. 
The dependence on $N$ and the overall magnitude is not as naively expected. This is due to our renormalisation procedure, which calibrates to a potential minimum of $V_0$, while the finite temperature effective potential has much shallower minima. If we instead compute the equation of state using a quasi-particle ansatz
\begin{align}
    \rho = \int d^3k\, n_k \omega_k,  \qquad
    P = \frac{1}{3} \int d^3k \; k^2 \frac{n_k}{\omega_k},
\end{align}
we find in figure \ref{fig:eosL6} (left) that the equation of state indeed approaches its radiation value of $1/3$ in the case of many Goldstone boson components (large $N$). 
In figure \ref{fig:eosL1} we repeat our simulation, but for $\lambda=1$, where the expectation value is closer to the zero temperature vev. We see that both for the direct (left) and quasi-particle (right) equation of state, the system displays the correct behaviour.

\section{Finite quench time}
\label{sec:finitetQ}
\begin{figure}[ht]
  \centering
  \includegraphics[width=0.5\textwidth]{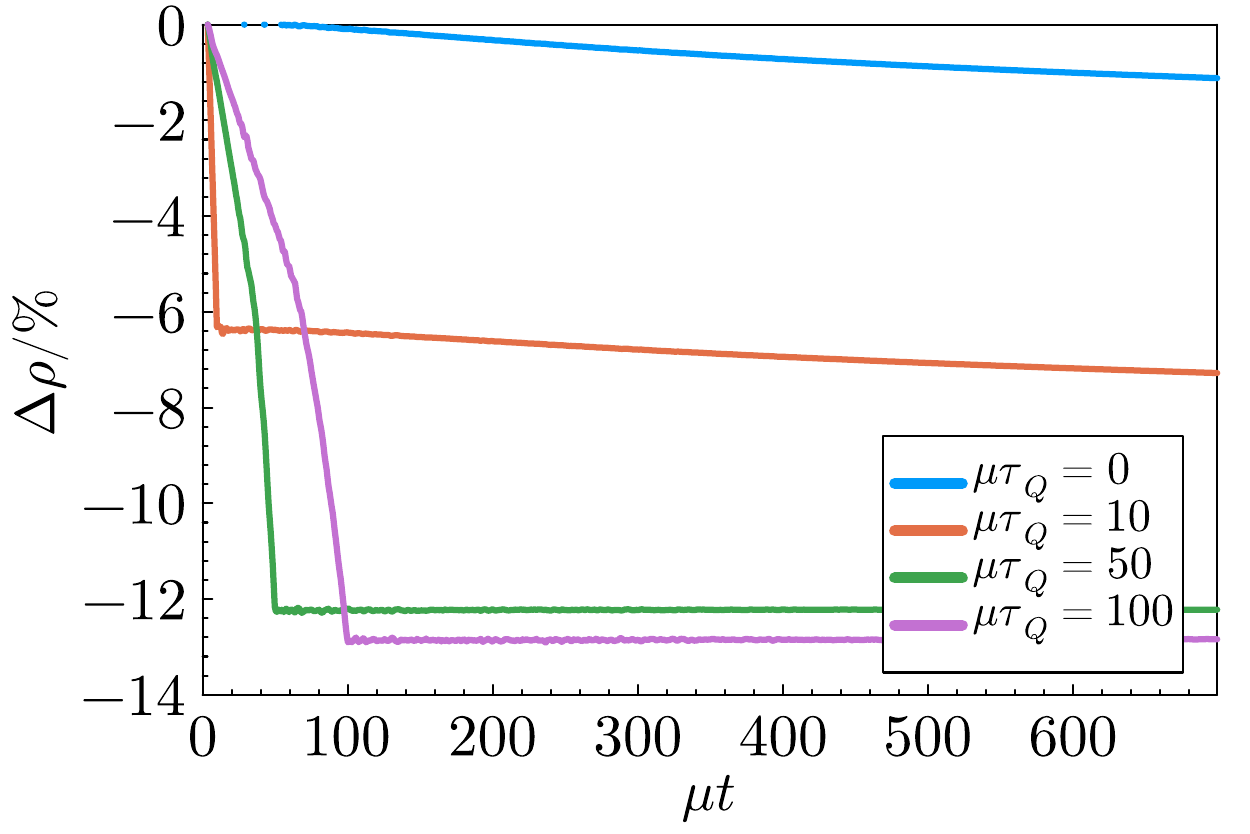}
  \caption{Energy loss for finite time quenches. Note that the post-quench, gradual energy loss is due to a truncation of the memory kernel (see appendix \ref{app:kernel}). $N=4$, $\lambda=6$.}
  \label{fig:Energyquench}
\end{figure}
A natural generalisation of the instantaneous quench it to consider finite-time linear quenches of the form (\ref{eq:linquench}). Implementing this as a by-hand mass function has the drawback that total energy is no longer conserved. We straightforwardly have that
\begin{eqnarray}
\dot{\rho}= \frac{1}{2} \dot{\mu^2(t)} \phi_a^2 = \dot{\mu} \mu \phi_a^2 ,
\end{eqnarray}
so that the total energy loss during the transition is 
\begin{eqnarray}
\Delta \rho= \int_0^t dt' \left[\dot{\mu}(t') \mu(t')\phi_a^2(t')\right].
\end{eqnarray}
For a very fast quench, $\phi_a$ is not able to react to the quench of the mass parameter, and since we start out at $\phi_a=0$, the energy loss is negligible. Conversely, if the flip is very slow (adiabatic), the field is able to track the bottom of the potential $\phi_a^2=\frac{6 N \mu^2}{\lambda}$
in which case
\begin{eqnarray}
\Delta \rho= -\frac{6N}{\lambda}\int_0^t dt' \dot{\mu}\mu^3 = -\frac{3 \mu^4(t)}{2\lambda}.
\label{eq:deltaElambda}
\end{eqnarray}
In this limit, all the potential energy is lost, as the field is slowly deposited in the new potential minimum. In figure \ref{fig:Energyquench} we show the energy loss for different quench times. In a real dynamical realisation of the $\sigma-\phi$ coupling, the energy is strictly conserved and instead transferred to the inflaton $\sigma$. The whole system of $N+1$ degrees of freedom eventually thermalise to some reheating temperature. We postpone this highly model-dependent complication to future work, and only briefly discuss the early time dynamics. 
\begin{figure}[ht]
  \includegraphics[width=0.5\textwidth]{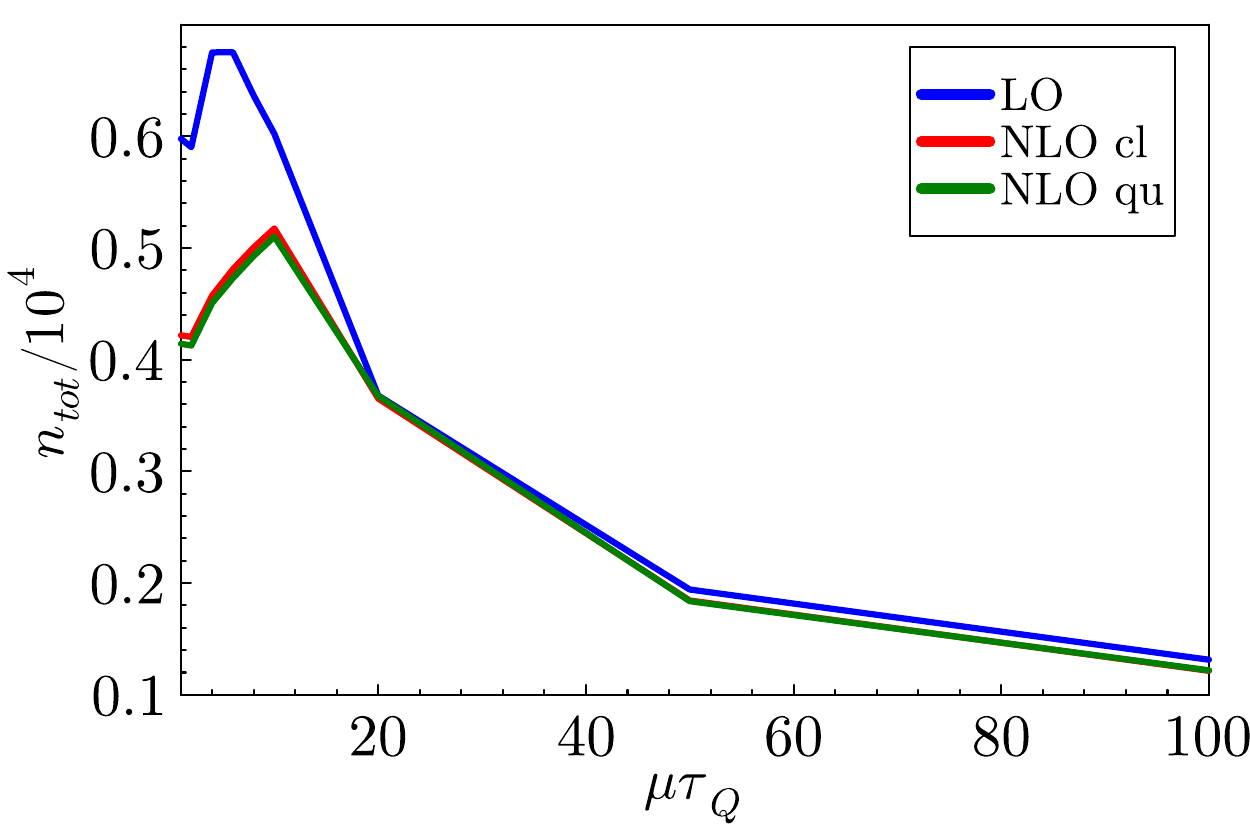}
  \includegraphics[width=0.5\textwidth]{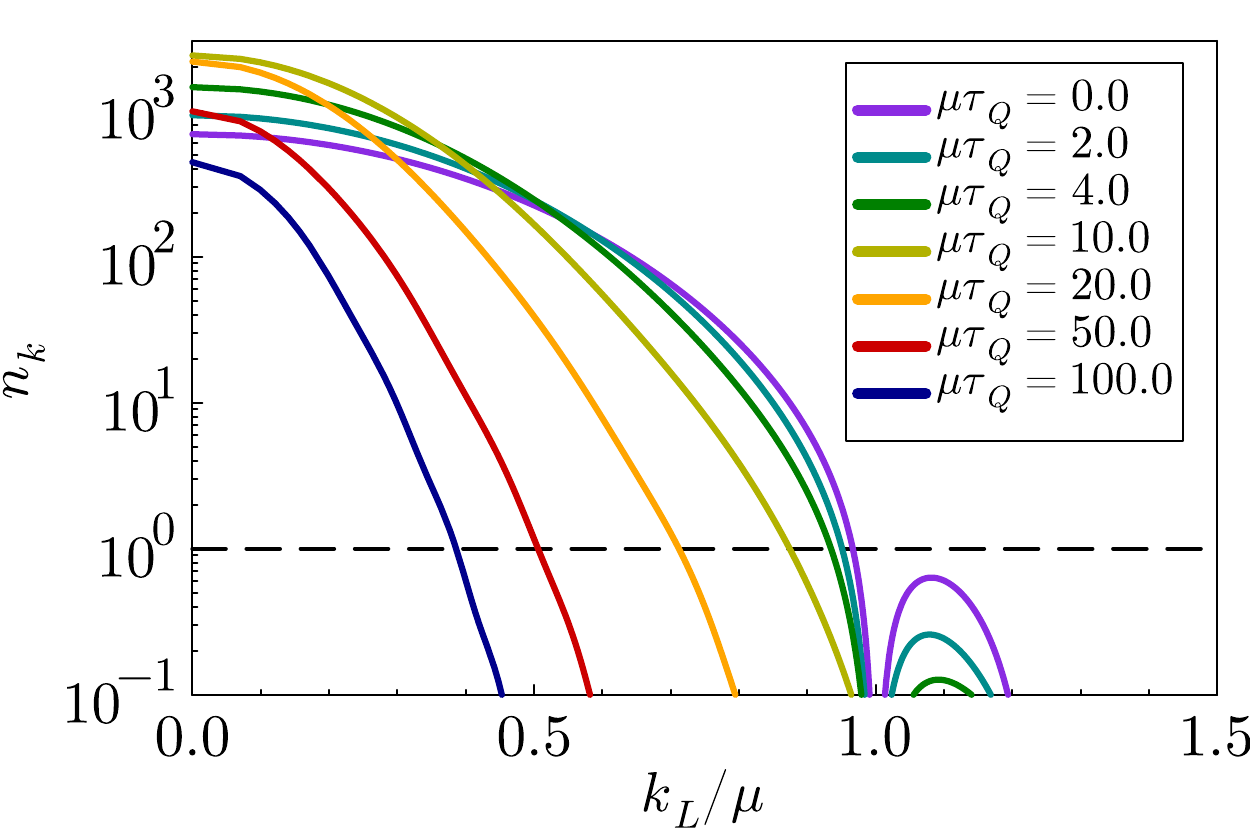}
  \caption{Left: Highest value for total particle numbers $n_{tot}$ for various quench times. Right: Particle spectrum in the LO approximation ($N=4$, $\lambda=1$) at 
  the time when the linear approximation breaks down ($\Delta F(t,t,x=0) = 0.2$). 
  Shown are spectra of unstable modes ($k_L/ \mu < 1$) for different values of quench times $\mu \tau_Q$. The spectra are calculated with a lattice of size $N_x^3=128^3$.}
  \label{fig:linearLOfinitequences}
\end{figure}

Figure \ref{fig:linearLOfinitequences} (left) shows the relation between quench time and the maximum total particle number that is achieved during evolution. Unsurprisingly, a fast quench leads to more overall particle production. 
As for the instantaneous quench, we show in figure \ref{fig:linearLOfinitequences} (right) the particle spectrum at the time when the free field description starts to fail.
We find that for slower quenches a smaller range of modes become classical ($n_k > 1$) at this time, and that for quenches slower than $\mu \tau_Q = 100$, it is essentially only the zero mode that grows large during the transition. The spectrum is also not completely monotonic in the quench time, although the total number of particles created is.
\begin{figure}[ht]
  \centering
  \includegraphics[width=0.5\textwidth]{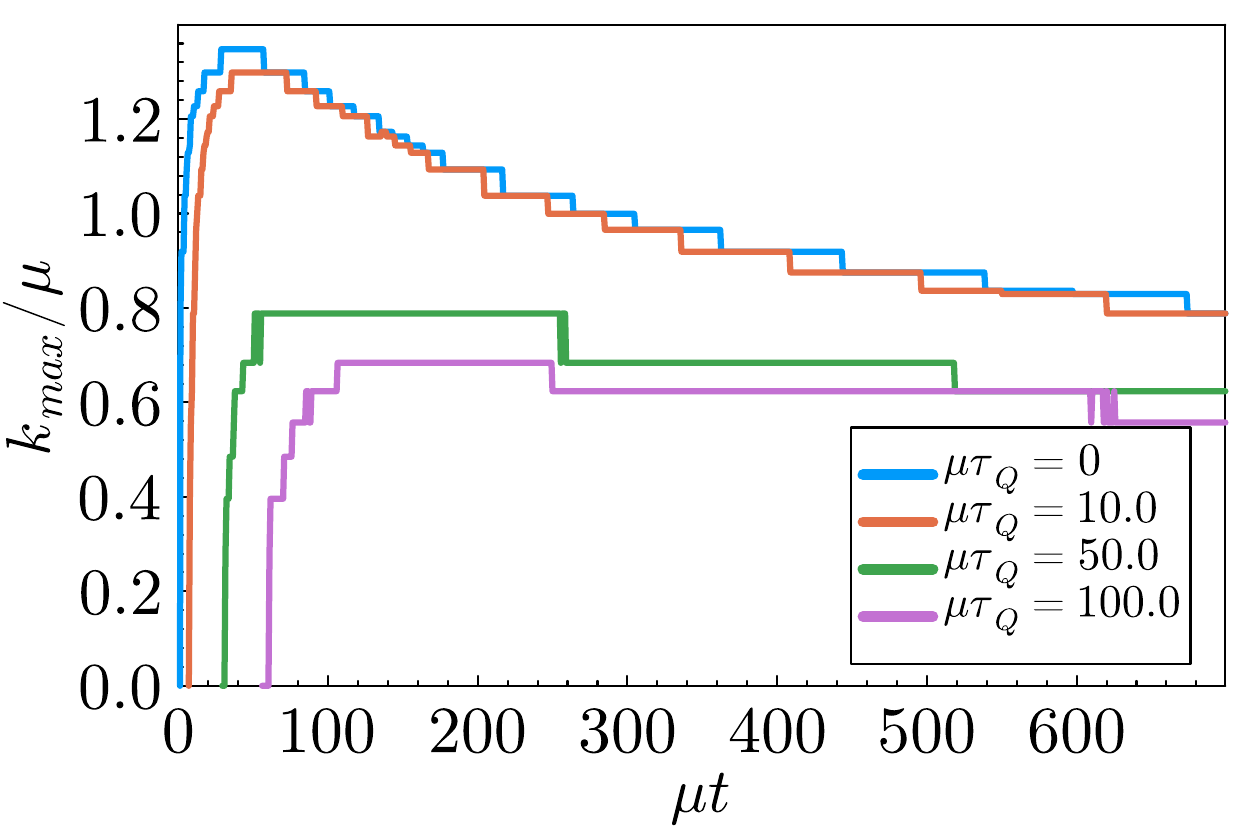}
  \caption{Time evolution of the maximum momentum with $n_k>1$ for different values of quench time. $N=4$, $\lambda=6$.}
  \label{fig:classicalityfinitequench}
\end{figure}
Finally, in figure \ref{fig:classicalityfinitequench}, we show the evolution of the classical range for different quench times. We see a separation into "fast" quenches $\mu \tau_Q < 20 $, where the classical range is large at first before it adjusts, and "slow" quenches $\mu \tau_Q > 50$ where the classical range is smaller, established later and changes on a longer timescale. 

\section{Conclusion}
\label{sec:conclusion}

Tachyonic preheating provides a mechanism to reheat the universe after inflation. During the process, a transient out-of-equilibrium state with very high occupation numbers in the IR is generated, which subsequently relaxes to an equilibrium reheating temperature $T_{\rm reh}$. Although the large occupation numbers allow an effective description in terms of classical field dynamics, quantum evolution equations are necessary to reach the correct equilibrium state and on the correct time scale\footnote{When not having access to full NLO dynamics, a semi-quantitative exponential relaxation from the LO post-transition spectrum towards equilibrium may be a useful alternative \cite{Kainulainen:2022lzp,Kainulainen:2021eki}.}.
We studied several aspects of this transition in a $O(N)$ scalar field theory with a time-dependent mass, using the 2PI-1/N formalism at NLO. 

We found that even for moderate couplings, the initial instability leads to IR occupation numbers well in the classical domain $n_k>1$ before non-linearities become important around $\mu t\simeq 2-4$, and direct comparison of quantum and classical 2PI approximations confirm that the early stages of the transition (say, $\mu t<50$) may be treated classically (LO dynamics agrees qualitatively, but occupation numbers tend to overshoot). Although this is as may be expected, it does imply that on this timescale the effect of all the unstable modes with low occupation is either negligible or at least does not introduce quantum corrections to the IR mode dynamics. During the initial stages of the evolution, we argued that the IR environment may be quantified through a classical effective temperature, which may be twice as high as the overall temperature. This could have implications for IR-dominated out-of-equilibrium processes taking place during this time.

As the system equilibrates kinetically on time scales $\mu \tau_{\rm kin}\simeq\mathcal{O}(100-300)$, however, classical evolution is no longer reliable. The quantum and classical evolution is substantially different, as energy from the initial zero-point fluctuations begins to leak from the UV into the IR, making classical occupation numbers too large. This is true also for "classical" modes with $n_k>1$. Hence from this time onwards we must rely on quantum 2PI-NLO evolution (or higher truncation, when available) or only consider modes with $n_k\gg 1$.

We proceeded to study the kinetic equilibration process in some detail, noting that full chemical equilibration is achieved at even later times somewhat beyond our current reach (see, however \cite{Arrizabalaga:2005tf,inprep}). The equilibration time-scale for moderate couplings is linear in $N$ and of order $\mu t_{\rm kin}\simeq 100-300$. In particular we pointed out that assuming an instantaneous redistribution of potential energy onto (in this case) $g^*=N-1$ massless non-interacting degrees of freedom fails to account for quantitatively significant effect of the equilibration stage, the massive mode(s), the interactions and the effective chemical potential, which lingers for $\mathcal{O}(>1000)$ after the transition. 

For application to the cosmological evolution, the equation of state of the system is established very early (around $\mu t=150$, see also \cite{Berges:2004ce}), but must be computed carefully. The system indeed behaves as a mixture of light Goldstone modes and a heavy Higgs mode.

2PI-simulations remain numerically challenging. Because tachyonic preheating involves very large occupation numbers, we made conservative choices in terms of lattice size, discretization and the length of the memory kernel. Technology exists for adding multiple scalar fields \cite{Aarts:2007ye,Tranberg:2013cka}, fermions \cite{Berges:2002wr,Shen:2020jya,Berges:2009bx}, Hubble expansion \cite{Tranberg:2008ae} and going to NNLO \cite{Aarts:2006cv,Aarts:2008wz} with robust renormalisation \cite{Berges:2004hn,Berges:2005hc}, and applications in cosmology are diverse \cite{Berges:2002cz,Gautier:2015pca,Serreau:2011fu}. Implementing all or most of these improvements for realistic models of reheating for larger volumes and the entire thermalisation process is daunting, but within reach. This would be a welcome supplement to the information provided by classical-statistical simulations of reheating and the post-inflationary evolution of the Universe.

\section*{Acknowledgements}
We thank Alexander Rothkopf for useful discussions and comments on the present manuscript.

\appendix
\section{Truncating the memory kernels}
\label{app:kernel}
\begin{figure}[ht]
  \includegraphics[width=0.5\textwidth]{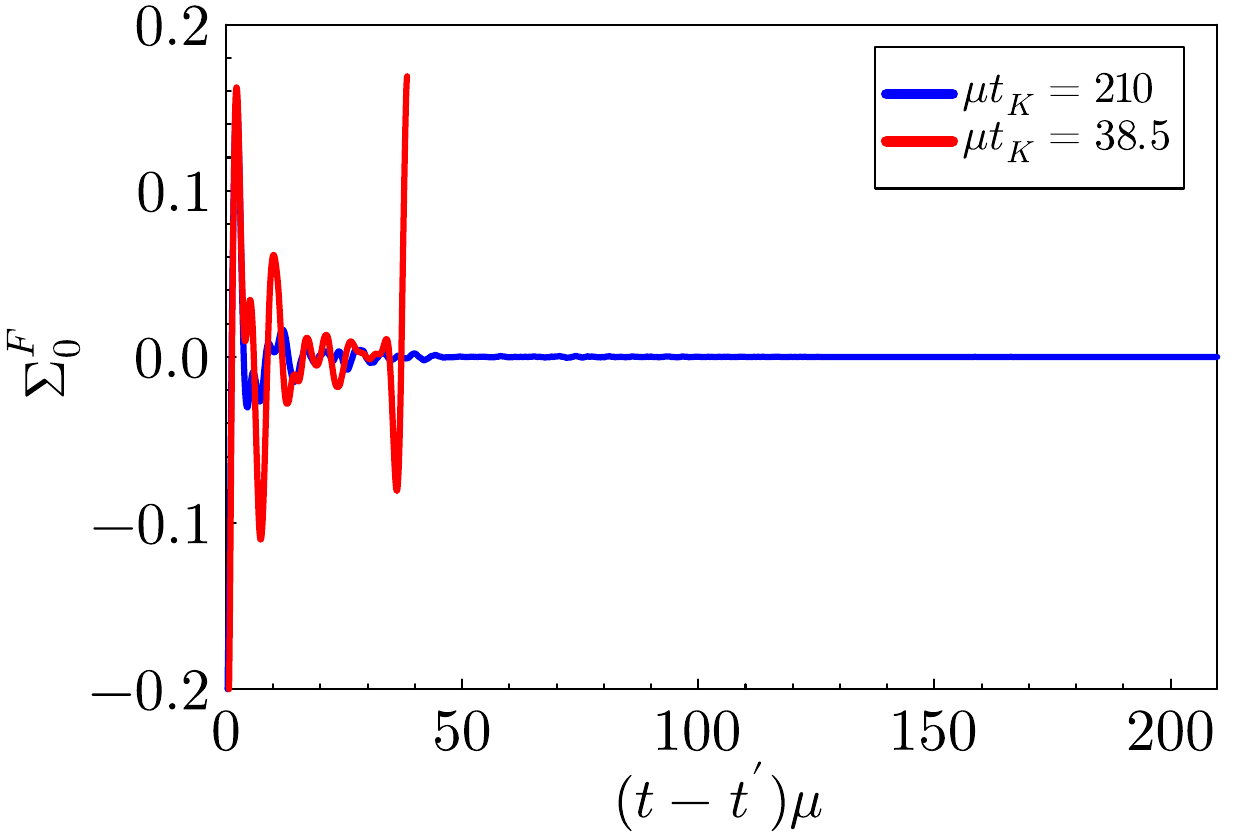}
  \includegraphics[width=0.5\textwidth]{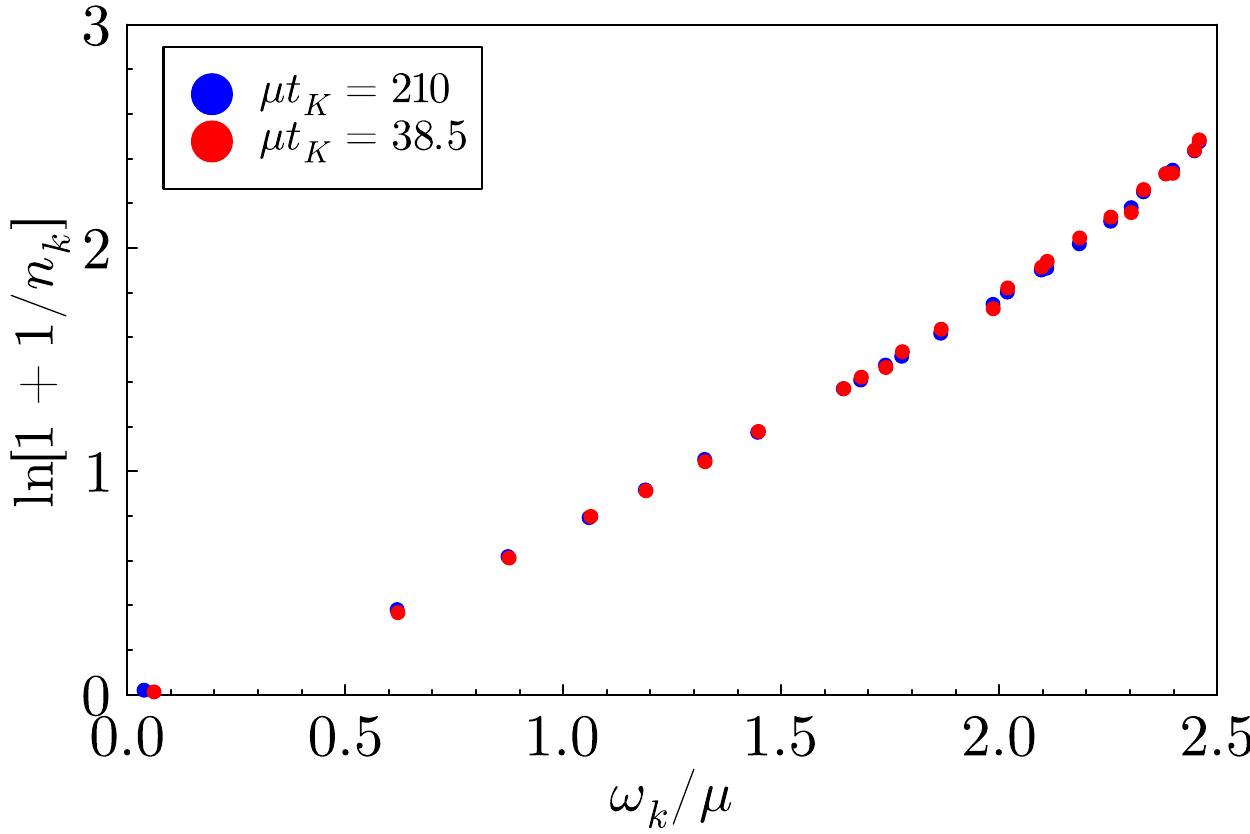}
  \caption{The effect of using reduced memory integrals of size $\mu t_k  = 38.5 $ and $200$ on the zero mode of $\Sigma_F$ (left), and the particle spectrum at $\mu t=700$ (right). $N_x=16$, $N=4$, $\lambda=6$.}
  \label{fig:SigmaFKernel}
\end{figure}

Throughout this work we used lattices with $N_x^3 = 32^3$ sites and a resolution $a\mu=0.7$. Giving a physical lattice size $(L\mu)^3$ of  $(22.4)^3$. This discretisation provides 833 distinct momentum values $(ak_L)^2$, of which 14 are unstable $(a k_L) < a\mu$. At our choice of time resolution, $dt = \frac{a_t}{a} = 0.1 $ a timescale of $\mu t=700 $ corresponds to 10.000 timesteps.
As mentioned in the main text, the equations of motions \cref{eq:eom} are integro-differential equations that are naturally solved by a simple Euler scheme. The computation time is heavily dependent on the lattice size and the time extent. 
In principle, the calculation of the memory integrals requires the entire history of $F(t,t^{\prime})$, $\rho(t,t^{\prime})$ which therefore needs to be kept in memory. 
On the other hand, the magnitude of contributions to the memory integrals become smaller far-in-the-past. In the interest of numerical efficiency, we may therefore put a cut-off on the history to reduce the required memory. Since much of the work is done computing Fast Fourier Transforms in the memory integrals, the runtime scales as the $N_x^3\log N_x\times t_K^2\times t_f$, where $t_f$ is the runtime and $t_K=a_t n_t$ is the length of the memory kernel.

\begin{figure}[ht]
  \centering
  \includegraphics[width=0.5\textwidth]{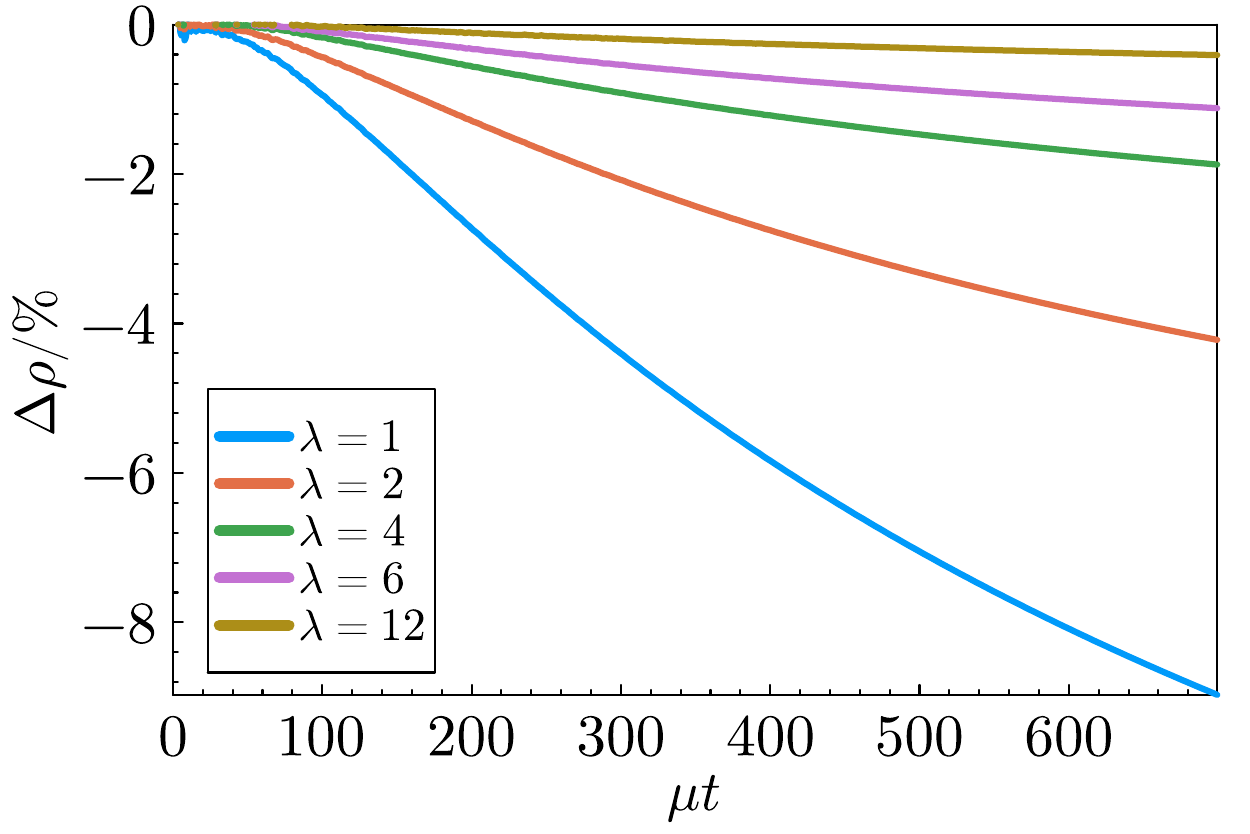}
  \caption{Energy loss for simulations caused by reduced memory kernels of size $\mu t_K=38.5$.}
  \label{fig:KernelEloss}
\end{figure}

In the following we show the effect of this procedure. We choose a lattice size of $N_x^3=16^3$ and compare simulations with a memory of $\mu t_K =200$ and $\mu t_K =38.5$ as done in \cite{Arrizabalaga:2004iw}. Figure \ref{fig:SigmaFKernel} shows the zero mode of the real part of the self-energy $\Sigma_F$ at $\mu t = 700$ where, to a good approximation, kinetic equilibrium has been reached. We see that the simulation with memory length of $\mu t_K =200$ shows the expected damping over time, whereas the simulation with $\mu t_K = 38.5$ shows a cut-off effect. 
We also found that if the memory kernel is too short, the simulations become unstable. Although figure \ref{fig:SigmaFKernel} might raise concerns about this approximation we found that the final states are the same for both, thus confirming previously obtained results \cite{Arrizabalaga:2004iw}. The final distributions at $\mu t = 700$ are shown in figure \ref{fig:SigmaFKernel}, and they are almost indistinguishable, showing that neglecting far-in-the-past quantities provides correct thermalization temperatures.
The interaction strength and occupation numbers impact the required kernel length, as large contributes to the memory integral need to be kept for longer. Keeping only a 
limited history causes an energy loss inversely proportional to the interaction strength (shown in figure \ref{fig:KernelEloss}), because smaller coupling leads to large occupation numbers.
The studied combinations of parameters required substantially larger physical memory length to complete correctly. All simulations have been performed with kernel length of $\mu t_K  =224$ and required $150$ GB of memory and about 50 hours of computation time on a computing cluster with 30 processing units.

\section{Chemical equilibration at late times on small lattices}
\label{App:16}

\begin{figure}[ht]
    \includegraphics[width=0.5\textwidth]{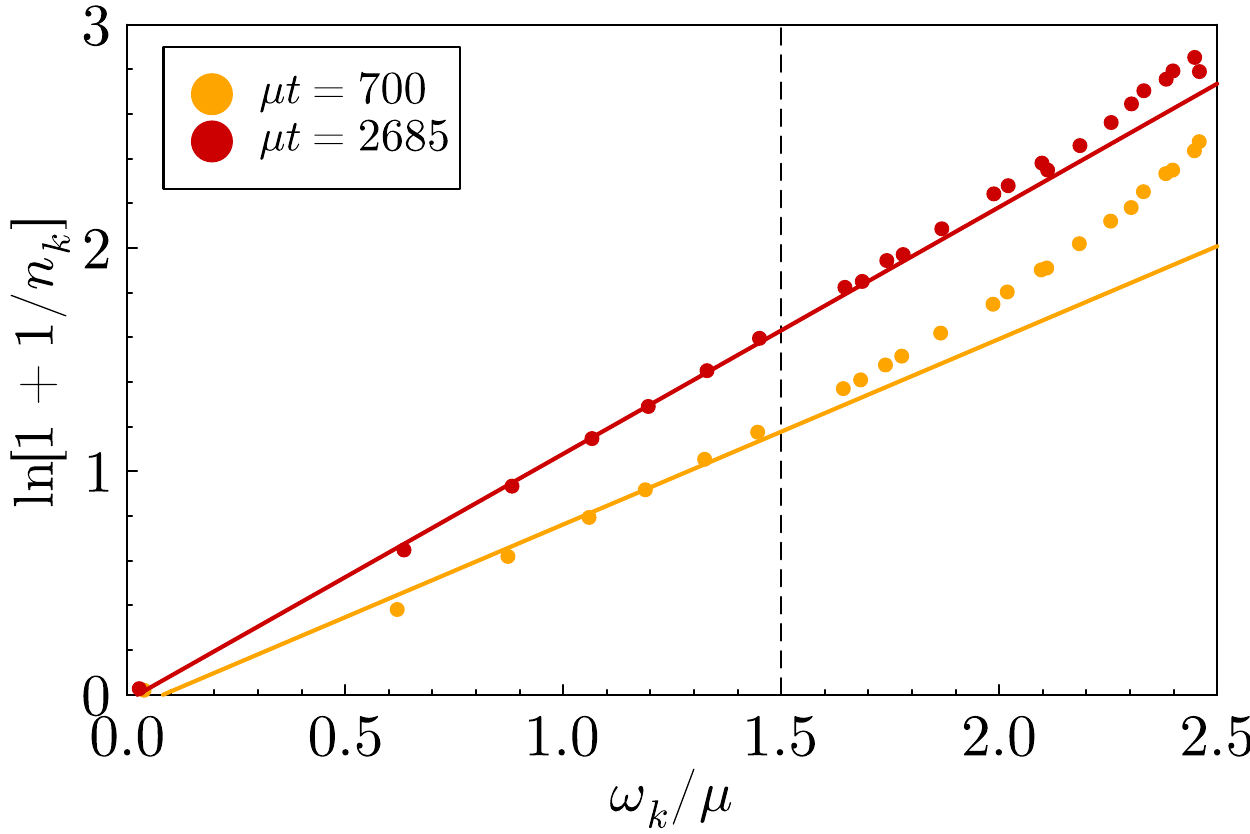}
    \includegraphics[width=0.5\textwidth]{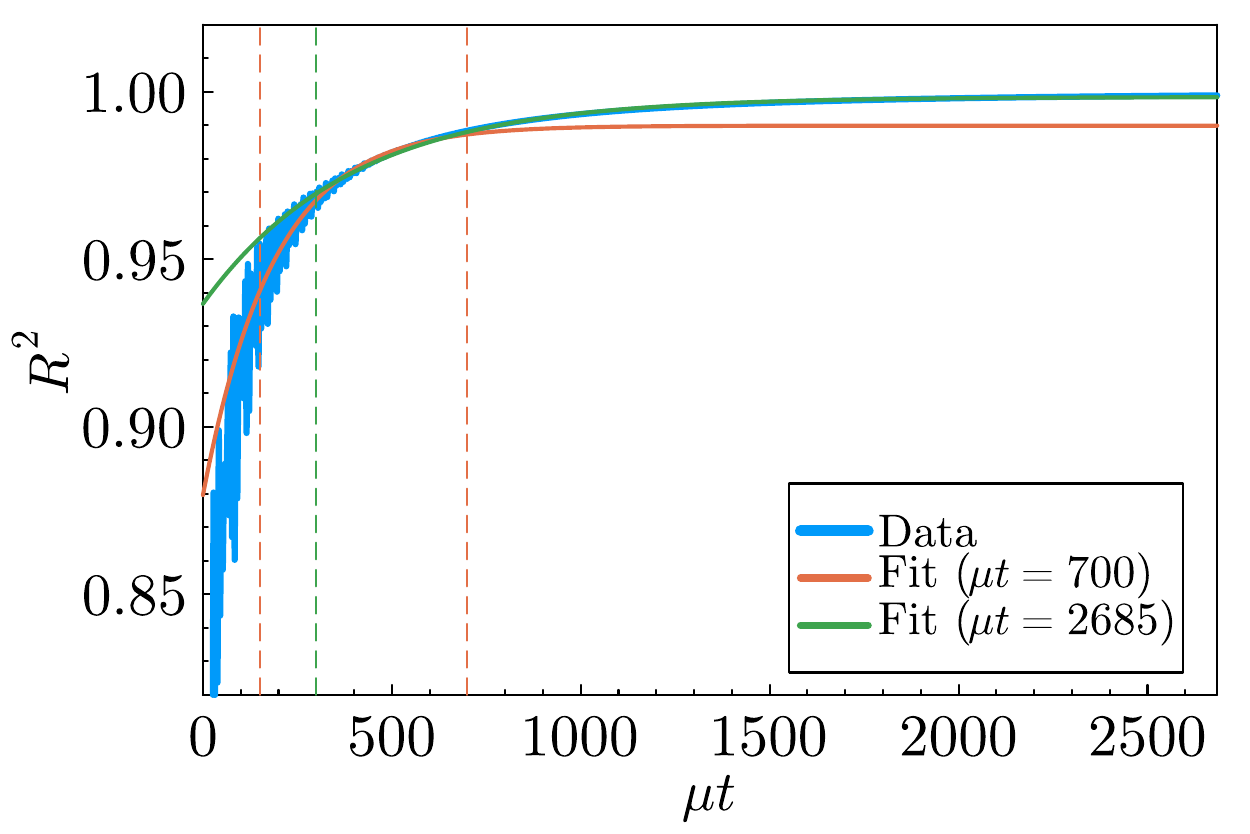}
  \caption{Left: Particle spectrum comparison after $\mu t=700$ and $\mu t=2685$ and corresponding fits to Bose-Einstein distributions. The latter shows not only temperatures and chemical potential closer to equilibrium but also that particles for $\omega_k/\mu>1.5$ are also thermal. Right: $R^2$ value, quantifying how close the state is to a thermal one, calculated for times up to $t\mu=2685$. We show fits, taking data into account for up to $\mu t=700$ and $\mu t=2685$.}
  \label{fig:SimulationlengthR2}
\end{figure}

\begin{figure}[ht]
    \includegraphics[width=0.5\textwidth]{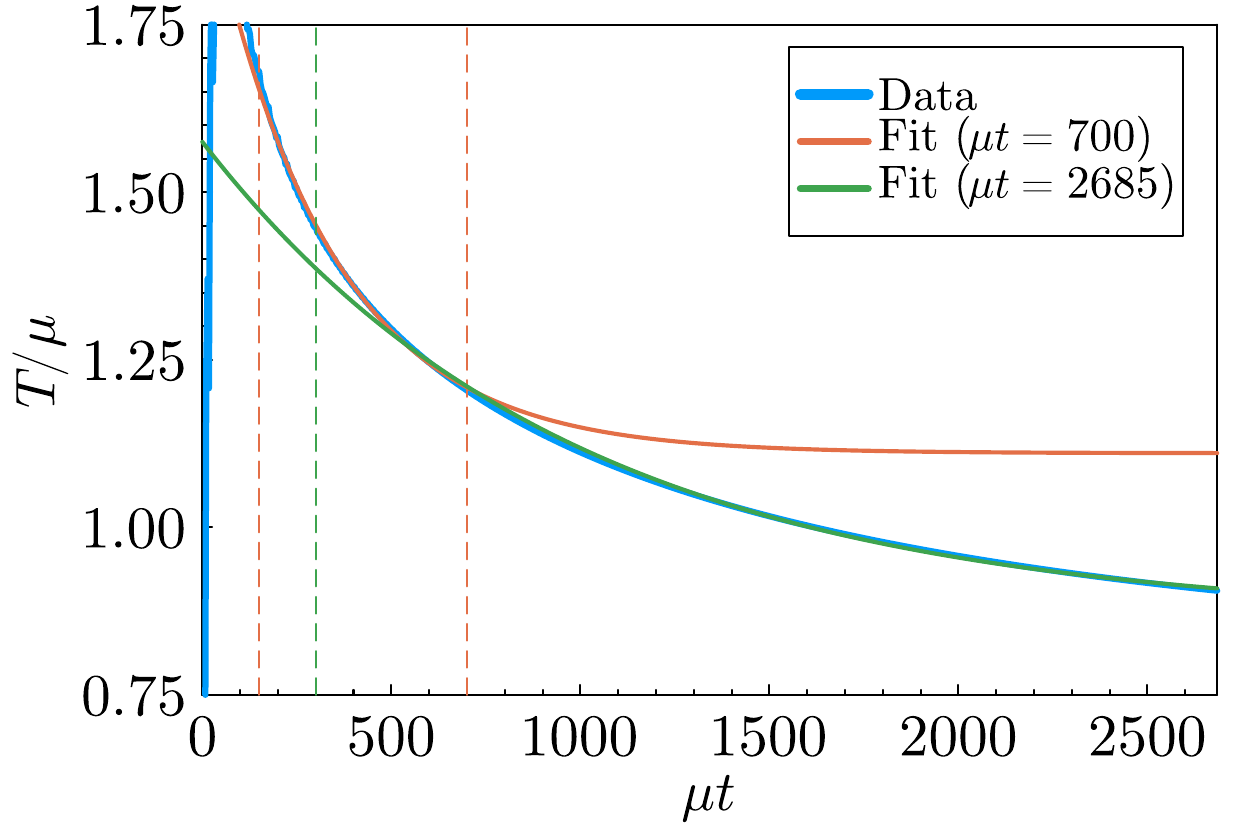}
    \includegraphics[width=0.5\textwidth]{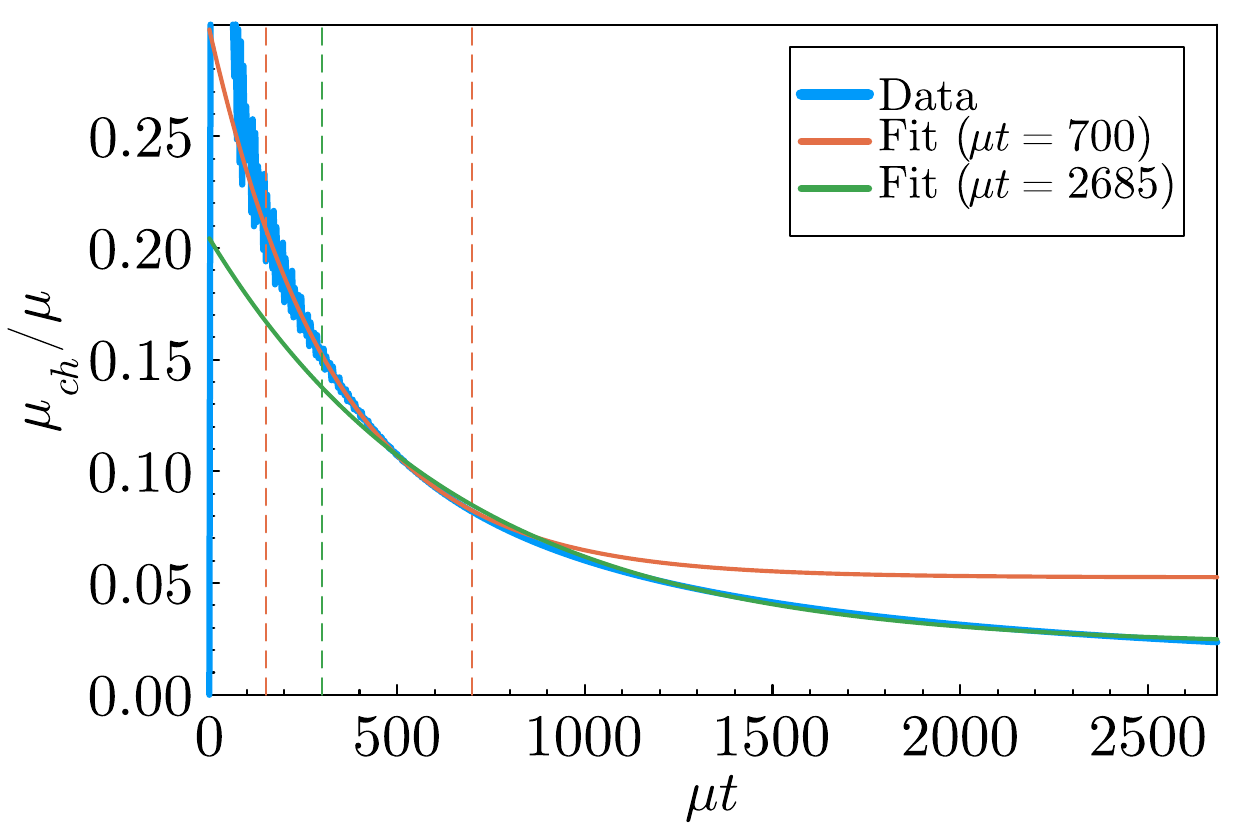}
  \caption{Evolution of the Bose-Einstein fit parameters $T$ and $\mu_{\rm ch}$ in time, and the corresponding fits. The dashed line indicates the fit region.}
  \label{fig:SimulationlengthTeff}
\end{figure}

The complexity of the algorithm allows us to calculate the dynamics only for times up to $\mu t = 700-1050$, on a lattice size of $N_x^3=32^3$.  As shown in the main text this allows to quantify equilibration properties. In this appendix we investigate the dependence of these estimates on having data for even longer times. Simulations of until $\mu t=2685$ are accessible for lattice sizes of $N_x^3 =16^3$.
  
In figure \ref{fig:SimulationlengthR2} (left) we show the particle distribution at $\mu t=700$ and $\mu t=2685$ and the corresponding fits to Bose-Einstein distributions. Note the reduced number of data points compared to simulations with $N_x^3=32^3$ lattice, e.g. figure \ref{fig:kineticequilibrationBEandR2}. As mentioned in the main text, particles in the IR equilibrate quicker than in the UV. In the figure we see that at $\mu t=2685$, particles for $\omega_k/\mu>1.5$ also align to the straight line, although the data points are still not included in determining the fit parameters. On the right of figure \ref{fig:SimulationlengthR2} we show the $R^2$ value that quantifies how closely the state has equilibrated (similar to figure \ref{fig:kineticequilibrationBEandR2} only that now equilibration fit values for times up to $\mu t=2685$ are available).
We perform the same analysis procedure as for $N_x^3=32^3$ lattices, namely discarding values for which $R^2<0.97$ and fitting the rest with eq. \ref{eq:R2fit}. When analysing data for up to $\mu t =2685$ we discard the first $\mu t =300$ data points. The figure shows both fits, taking data for up to $\mu t=700$ and up to $2685$. 
Taking data up to $\mu t=2685$ into account results in a kinetical equilibration time scale of $ \mu \tau_{kin} = 393$, compared to $272$. Therefore, taking only data up to $\mu t=700$ gives an equilibration scale that is about $50 \%$ smaller than the actual value. We note that the quoted error bars, which are not of statistical nature but quantify a range of parameters that are in agreement with the data drop from $7.6 \%$ to $0.6 \%$.
Figure \ref{fig:SimulationlengthTeff} shows the time evolution of the Bose-Einstein fit parameters $T$ (left) and $\mu$ (right) with the corresponding fits. For large values of $\mu t$ the fit that takes values up to $\mu t=700$ into account clearly deviates from the data. The reheating temperature extracted from the left plot is determined to be $0.86$, whereas the short simulation predicts $1.1$. Thus data up to $\mu t=700$ gives a temperature that is about $30\%$ too large. Again, errorbars drop from $6.1 \%$ to $0.6 \%$.
The chemical equilibration time scale, taken from the right plot in figure \ref{fig:SimulationlengthTeff} is $\mu \tau_{ch} = 660$, whereas the 
short simulation gives $333$. Thus, the short simulation gives only $50 \%$ of the actual value. The errorbars drop from $82 \%$ to $13 \%$.

\bibliography{biblio.bib}

\begin{thebibliography}{10}

\bibitem{PhysRevD.23.347}
Alan~H. Guth.
\newblock Inflationary universe: A possible solution to the horizon and
  flatness problems.
\newblock {\em Phys. Rev. D}, 23:347--356, Jan 1981.

\bibitem{Liddle:2000cg}
Andrew~R. Liddle and D.~H. Lyth.
\newblock {\em {Cosmological inflation and large scale structure}}.
\newblock 2000.

\bibitem{PhysRevLett.73.3195}
Lev Kofman, Andrei Linde, and Alexei~A. Starobinsky.
\newblock Reheating after inflation.
\newblock {\em Phys. Rev. Lett.}, 73:3195--3198, Dec 1994.

\bibitem{PhysRevLett.87.011601}
Gary Felder, Juan Garc\'{\i}a-Bellido, Patrick~B. Greene, Lev Kofman, Andrei
  Linde, and Igor Tkachev.
\newblock Dynamics of symmetry breaking and tachyonic preheating.
\newblock {\em Phys. Rev. Lett.}, 87:011601, Jun 2001.

\bibitem{PhysRevD.49.748}
Andrei Linde.
\newblock Hybrid inflation.
\newblock {\em Phys. Rev. D}, 49:748--754, Jan 1994.

\bibitem{Garcia-Bellido:2002fsq}
Juan Garcia-Bellido, Margarita Garcia~Perez, and Antonio Gonzalez-Arroyo.
\newblock {Symmetry breaking and false vacuum decay after hybrid inflation}.
\newblock {\em Phys. Rev. D}, 67:103501, 2003.

\bibitem{Smit:2002yg}
Jan Smit and Anders Tranberg.
\newblock {Chern-Simons number asymmetry from CP violation at electroweak
  tachyonic preheating}.
\newblock {\em JHEP}, 12:020, 2002.

\bibitem{Arrizabalaga:2004iw}
Alejandro Arrizabalaga, Jan Smit, and Anders Tranberg.
\newblock {Tachyonic preheating using 2PI-1/N dynamics and the classical
  approximation}.
\newblock {\em JHEP}, 10:017, 2004.

\bibitem{Aarts:1997kp}
Gert Aarts and Jan Smit.
\newblock {Classical approximation for time dependent quantum field theory:
  Diagrammatic analysis for hot scalar fields}.
\newblock {\em Nucl. Phys. B}, 511:451--478, 1998.

\bibitem{Dux:2022kuk}
Fr\'ed\'eric Dux, Adrien Florio, Juraj Klari\'c, Andrey Shkerin, and Inar
  Timiryasov.
\newblock {Preheating in Palatini Higgs inflation on the lattice}.
\newblock {\em JCAP}, 09:015, 2022.

\bibitem{Garcia:2023eol}
Marcos A.~G. Garcia and Mathias Pierre.
\newblock {Reheating after Inflaton Fragmentation}.
\newblock 6 2023.

\bibitem{Mahbub:2023faw}
Rafid Mahbub and Swagat~S. Mishra.
\newblock {Oscillon formation from preheating in asymmetric inflationary
  potentials}.
\newblock {\em Phys. Rev. D}, 108(6):063524, 2023.

\bibitem{Amin:2014eta}
Mustafa~A. Amin, Mark~P. Hertzberg, David~I. Kaiser, and Johanna Karouby.
\newblock {Nonperturbative Dynamics Of Reheating After Inflation: A Review}.
\newblock {\em Int. J. Mod. Phys. D}, 24:1530003, 2014.

\bibitem{Aarts:2001yn}
Gert Aarts and Juergen Berges.
\newblock {Classical aspects of quantum fields far from equilibrium}.
\newblock {\em Phys. Rev. Lett.}, 88:041603, 2002.

\bibitem{Tranberg:2008ae}
Anders Tranberg.
\newblock {Quantum field thermalization in expanding backgrounds}.
\newblock {\em JHEP}, 11:037, 2008.

\bibitem{Aarts:2001qa}
Gert Aarts and Juergen Berges.
\newblock {Nonequilibrium time evolution of the spectral function in quantum
  field theory}.
\newblock {\em Phys. Rev. D}, 64:105010, 2001.

\bibitem{Berges:2000ur}
Juergen Berges and Jurgen Cox.
\newblock {Thermalization of quantum fields from time reversal invariant
  evolution equations}.
\newblock {\em Phys. Lett. B}, 517:369--374, 2001.

\bibitem{Aarts:2002dj}
Gert Aarts, Daria Ahrensmeier, Rudolf Baier, Juergen Berges, and Julien
  Serreau.
\newblock {Far from equilibrium dynamics with broken symmetries from the 2PI -
  1/N expansion}.
\newblock {\em Phys. Rev. D}, 66:045008, 2002.

\bibitem{Guth}
Alan~H. Guth and So-Young Pi.
\newblock {The Quantum Mechanics of the Scalar Field in the New Inflationary
  Universe}.
\newblock {\em Phys. Rev. D}, 32:1899--1920, 1985.

\bibitem{PhysRevD.36.2474}
Erick~J. Weinberg and Aiqun Wu.
\newblock Understanding complex perturbative effective potentials.
\newblock {\em Phys. Rev. D}, 36:2474--2480, Oct 1987.

\bibitem{Calzetta}
E.~Calzetta.
\newblock {Spinodal Decomposition in Quantum Field Theory}.
\newblock {\em Annals Phys.}, 190:32--58, 1989.

\bibitem{Boyanovsky}
Daniel Boyanovsky.
\newblock {Quantum spinodal decomposition}.
\newblock {\em Phys. Rev. E}, 48:767--771, 1993.

\bibitem{Boyanovsky2}
D.~Boyanovsky, H.~J. de~Vega, and R.~Holman.
\newblock {Nonequilibrium evolution of scalar fields in FRW cosmologies I}.
\newblock {\em Phys. Rev. D}, 49:2769--2785, 1994.

\bibitem{Boyanovsky3}
D.~Boyanovsky, D.~Cormier, H.~J. de~Vega, R.~Holman, A.~Singh, and
  M.~Srednicki.
\newblock {Scalar field dynamics in Friedman-Robertson-Walker space-times}.
\newblock {\em Phys. Rev. D}, 56:1939--1957, 1997.

\bibitem{Salle}
Mischa Salle and Jan Smit.
\newblock {The Hartree ensemble approximation revisited: The Symmetric phase}.
\newblock {\em Phys. Rev. D}, 67:116006, 2003.

\bibitem{Salle2}
M.~Salle, Jan Smit, and Jeroen~C. Vink.
\newblock {Thermalization in a Hartree ensemble approximation to quantum field
  dynamics}.
\newblock {\em Phys. Rev. D}, 64:025016, 2001.

\bibitem{Tranberg:2022noe}
Anders Tranberg and Gerhard Ungersbaeck.
\newblock {Bubble nucleation and quantum initial conditions in classical
  statistical simulations}.
\newblock {\em JHEP}, 09:206, 2022.

\bibitem{Berges:2013lsa}
J.~Berges, K.~Boguslavski, S.~Schlichting, and R.~Venugopalan.
\newblock {Basin of attraction for turbulent thermalization and the range of
  validity of classical-statistical simulations}.
\newblock {\em JHEP}, 05:054, 2014.

\bibitem{Cornwall:1974vz}
John~M. Cornwall, R.~Jackiw, and E.~Tomboulis.
\newblock {Effective Action for Composite Operators}.
\newblock {\em Phys. Rev. D}, 10:2428--2445, 1974.

\bibitem{Aarts:2008wz}
Gert Aarts, Nathan Laurie, and Anders Tranberg.
\newblock {Effective convergence of the 2PI-1/N expansion for nonequilibrium
  quantum fields}.
\newblock {\em Phys. Rev. D}, 78:125028, 2008.

\bibitem{Arrizabalaga:2005tf}
Alejandro Arrizabalaga, Jan Smit, and Anders Tranberg.
\newblock {Equilibration in phi**4 theory in 3+1 dimensions}.
\newblock {\em Phys. Rev. D}, 72:025014, 2005.

\bibitem{Berges:2004hn}
J.~Berges, Sz. Borsanyi, U.~Reinosa, and J.~Serreau.
\newblock {Renormalized thermodynamics from the 2PI effective action}.
\newblock {\em Phys. Rev. D}, 71:105004, 2005.

\bibitem{Berges:2005hc}
Juergen Berges, Szabolcs Borsanyi, Urko Reinosa, and Julien Serreau.
\newblock {Nonperturbative renormalization for 2PI effective action
  techniques}.
\newblock {\em Annals Phys.}, 320:344--398, 2005.

\bibitem{Borsanyi}
Szabolcs Borsanyi and Urko Reinosa.
\newblock {Renormalised nonequilibrium quantum field theory: Scalar fields}.
\newblock {\em Phys. Rev. D}, 80:125029, 2009.

\bibitem{Millington:2020vkg}
Peter Millington, Zong-Gang Mou, Paul~M. Saffin, and Anders Tranberg.
\newblock {Statistics on Lefschetz thimbles: Bell/Leggett-Garg inequalities and
  the classical-statistical approximation}.
\newblock {\em JHEP}, 03:077, 2021.

\bibitem{Shen:2020jya}
Linda Shen, J\"urgen Berges, Jan~M. Pawlowski, and Alexander Rothkopf.
\newblock {Thermalization and dynamical spectral properties in the quark-meson
  model}.
\newblock {\em Phys. Rev. D}, 102(1):016012, 2020.

\bibitem{Berges:2004ce}
J.~Berges, S.~Borsanyi, and C.~Wetterich.
\newblock {Prethermalization}.
\newblock {\em Phys. Rev. Lett.}, 93:142002, 2004.

\bibitem{Kainulainen:2022lzp}
Kimmo Kainulainen, Olli Koskivaara, and Sami Nurmi.
\newblock {Tachyonic production of dark relics: a non-perturbative quantum
  study}.
\newblock {\em JHEP}, 04:043, 2023.

\bibitem{Kainulainen:2021eki}
Kimmo Kainulainen and Olli Koskivaara.
\newblock {Non-equilibrium dynamics of a scalar field with quantum
  backreaction}.
\newblock {\em JHEP}, 12:190, 2021.

\bibitem{inprep}
Anders Tranberg and Gerhard Ungersbaeck.
\newblock {In preparation}.

\bibitem{Aarts:2007ye}
Gert Aarts and Anders Tranberg.
\newblock {Thermal effects on slow-roll dynamics}.
\newblock {\em Phys. Rev. D}, 77:123521, 2008.

\bibitem{Tranberg:2013cka}
Anders Tranberg and David~J. Weir.
\newblock {On the quantum stability of Q-balls}.
\newblock {\em JHEP}, 04:184, 2014.

\bibitem{Berges:2002wr}
Juergen Berges, Szabolcs Borsanyi, and Julien Serreau.
\newblock {Thermalization of fermionic quantum fields}.
\newblock {\em Nucl. Phys. B}, 660:51--80, 2003.

\bibitem{Berges:2009bx}
Juergen Berges, Jens Pruschke, and Alexander Rothkopf.
\newblock {Instability-induced fermion production in quantum field theory}.
\newblock {\em Phys. Rev. D}, 80:023522, 2009.

\bibitem{Aarts:2006cv}
Gert Aarts and Anders Tranberg.
\newblock {Nonequilibrium dynamics in the O(N) model to next-to-next-to-leading
  order in the 1/N expansion}.
\newblock {\em Phys. Rev. D}, 74:025004, 2006.

\bibitem{Berges:2002cz}
Juergen Berges and Julien Serreau.
\newblock {Parametric resonance in quantum field theory}.
\newblock {\em Phys. Rev. Lett.}, 91:111601, 2003.

\bibitem{Gautier:2015pca}
Florian Gautier and Julien Serreau.
\newblock {Scalar field correlator in de Sitter space at next-to-leading order
  in a 1/N expansion}.
\newblock {\em Phys. Rev. D}, 92(10):105035, 2015.

\bibitem{Serreau:2011fu}
Julien Serreau.
\newblock {Effective potential for quantum scalar fields on a de Sitter
  geometry}.
\newblock {\em Phys. Rev. Lett.}, 107:191103, 2011.

\end{thebibliography}
\bibliographystyle{unsrt}

\end{document}